\documentclass{aa}
\usepackage[varg]{txfonts}
\usepackage{graphicx}
\usepackage{natbib}
\usepackage{booktabs}
\usepackage{morefloats}
\bibpunct{(}{)}{;}{a}{}{,}

\begin{document}

\title{Optical and near-infrared observations of SN 2011dh - The first 100 days.}

\author{M.~Ergon\inst{\ref{inst1}} \and J.~Sollerman\inst{\ref{inst1}} \and M.~Fraser\inst{\ref{inst2}} \and A.~Pastorello\inst{\ref{inst3}} \and S.~Taubenberger\inst{\ref{inst4}} \and N.~Elias-Rosa\inst{\ref{inst5}} \and M.~Bersten\inst{\ref{inst6}} \and A.~Jerkstrand\inst{\ref{inst2}} \and S.~Benetti\inst{\ref{inst3}} \and M.T.~Botticella\inst{\ref{inst7}} \and C.~Fransson\inst{\ref{inst1}} \and A.~Harutyunyan\inst{\ref{inst8}} \and R.~Kotak\inst{\ref{inst2}} \and S.~Smartt\inst{\ref{inst2}} \and S.~Valenti\inst{\ref{inst3}} \and F.~Bufano\inst{\ref{inst9},\ref{inst10}} \and E.~Cappellaro\inst{\ref{inst3}} \and M.~Fiaschi\inst{\ref{inst3}} \and A.~Howell\inst{\ref{inst11}} \and E.~Kankare\inst{\ref{inst12}} \and L.~Magill\inst{\ref{inst2},\ref{inst13}} \and S.~Mattila\inst{\ref{inst12}} \and J.~Maund\inst{\ref{inst2}} \and R.~Naves\inst{\ref{inst15}} \and P.~Ochner\inst{\ref{inst3}} \and J.~Ruiz\inst{\ref{inst14}} \and K.~Smith\inst{\ref{inst2}} \and L.~Tomasella\inst{\ref{inst3}} \and  M.~Turatto\inst{\ref{inst3}}}

\institute{The Oskar Klein Centre, Department of Astronomy, AlbaNova, Stockholm University, 106 91 Stockholm, Sweden 
\label{inst1}
\and Astrophysics Research Center, School of Mathematics and Physics, Queens University Belfast, Belfast, BT7 1NN, UK 
\label{inst2}
\and INAF, Osservatorio Astronomico di Padova, vicolo dell'Osservatorio n. 5, 35122 Padua, Italy 
\label{inst3}
\and Max-Planck-Institut für Astrophysik, Karl-Schwarzschild-Str. 1, D-85741 Garching, Germany 
\label{inst4}
\and Institut de Ci\`{e}ncies de l’Espai (IEEC-CSIC), Facultat de Ci\`{e}ncies, Campus UAB, E-08193 Bellaterra, Spain.
\label{inst5}
\and Kavli Institute for the Physics and Mathematics of the Universe (WPI), Todai Institutes for Advanced Study, University of Tokyo, 5-1-5 Kashiwanoha, Kashiwa, Chiba 277-8583, Japan
\label{inst6}
\and INAF-Osservatorio Astronomico di Capodimonte, Salita Moiariello, 16  80131 Napoli, Italy
\label{inst7}
\and Fundaci\'{o}n Galileo Galilei-INAF, Telescopio Nazionale Galileo, Rambla Jos\'{e} Ana Fern\'{a}ndez P\'{e}rez 7, 38712 Bre\~{n}a Baja, TF - Spain
\label{inst8}
\and INAF, Osservatorio Astrofisico di Catania, Via Santa Sofia, I-95123, Catania, Italy
\label{inst9}
\and Departamento de Ciencias Fisicas, Universidad Andres Bello, Av. Republica 252, Santiago, Chile
\label{inst10}
\and Las Cumbres Observatory Global Telescope Network, 6740 Cortona Dr., Suite 102, Goleta, CA 93117 
\label{inst11}
\and Finnish Centre for Astronomy with ESO (FINCA), University of Turku, V\"ais\"al\"antie 20, FI-21500 Piikki\"o, Finland
\label{inst12}
\and Isaac Newton Group, Apartado 321, E-38700 Santa Cruz de La Palma, Spain
\label{inst13}
\and Observatorio Montcabrer, C Jaume Balmes 24, Cabrils, Spain
\label{inst15}
\and Observatorio de C\'{a}ntabria, Ctra. de Rocamundo s/n, Valderredible, Cantabria, Spain
\label{inst14}}

\date{Submitted to Astronomy and Astrophysics}

\abstract{We present optical and near-infrared (NIR) photometry and spectroscopy of the Type IIb supernova (SN) 2011dh for the first 100 days. We complement our extensive dataset with SWIFT ultra-violet (UV) and Spitzer mid-infrared (MIR) data to build a UV to MIR bolometric lightcurve using both photometric and spectroscopic data. Hydrodynamical modelling of the SN based on this bolometric lightcurve have been presented in \citet{Ber12}. We find that the absorption minimum for the hydrogen lines is never seen below $\sim$11000 km s$^{-1}$ but approaches this value as the lines get weaker. This suggests that the interface between the helium core and hydrogen rich envelope is located near this velocity in agreement with the \citet{Ber12} He4R270 ejecta model. Spectral modelling of the hydrogen lines using this ejecta model supports the conclusion and we find a hydrogen mass of 0.01-0.04 M$_{\odot}$ to be consistent with the observed spectral evolution. We estimate that the photosphere reaches the helium core at 5-7 days whereas the helium lines appear between $\sim$10 and $\sim$15 days, close to the photosphere and then move outward in velocity until $\sim$40 days. This suggests that increasing non-thermal excitation due to decreasing optical depth for the $\gamma$-rays is driving the early evolution of these lines. The Spitzer 4.5 $\mu$m band shows a significant flux excess, which we attribute to CO fundamental band emission or a thermal dust echo although further work using late time data is needed. The distance and in particular the extinction, where we use spectral modelling to put further constraints, is discussed in some detail as well as the sensitivity of the hydrodynamical modelling to errors in these quantities. We also provide and discuss pre- and post-explosion observations of the SN site which shows a reduction by $\sim$75 percent in flux at the position of the yellow supergiant coincident with SN 2011dh. The $B$, $V$ and $r$ band decline rates of 0.0073, 0.0090 and 0.0053 mag day$^{-1}$ respectively are consistent with the remaining flux being emitted by the SN. Hence we find that the star was indeed the progenitor of SN 2011dh as previously suggested by \citet{Mau11} and which is also consistent with the results from the hydrodynamical modelling.}


\titlerunning{SN 2011dh - The first 100 days.}
\authorrunning{M. Ergon et al.}
\maketitle

\defcitealias{Ber12}{B12}
\defcitealias{Arc11}{A11}
\defcitealias{Mau11}{M11}
\defcitealias{Tsv12}{T12}
\defcitealias{Vin12}{V12}
\defcitealias{Sch98}{S98}
\defcitealias{Sch11}{SF11}
\defcitealias{Mun97}{MZ97}
\defcitealias{Poz12}{P12}
\defcitealias{Val11}{V11}
\defcitealias{Bes12}{BM12}
\defcitealias{Pas09}{P09}
\defcitealias{Val11}{V11}
\defcitealias{Cho11}{C11}
\defcitealias{Mar13}{M13}
\defcitealias{Dyk13b}{D13}
\defcitealias{Sah13}{S13}

\section{Introduction}

Core-collapse (CC) supernovae (SNe) are caused by the gravitational collapse of the core in massive stars. The diversity of the events that we observe reflects the diversity of the progenitor stars and their surrounding circumstellar media (CSM). In particular, the extent to which the star has lost its hydrogen envelope has a profound impact on the observed properties of the SN. Through the presence or absence of hydrogen lines in their spectra these SNe are classified as Type II or Type I, respectively. The ejecta mass of Type I SNe tends to be smaller and thus the diffusion time shorter and the expansion velocity higher. The designation IIb is used for SNe which show a spectral transition from Type II (with hydrogen) at early times to Type Ib (without hydrogen but with helium) at later times. These SNe are thought to arise from stars that have lost most, but not all, of their hydrogen envelope. The prime example of such a SN is 1993J, where the progenitor star was a yellow (extended) supergiant proposed to have lost most of its hydrogen envelope through interaction with its blue (compact) companion star \citep{Pod93,Mau04,Sta09}. As Type IIb SNe are surprisingly common given the brief period single stars spend in the appropriate state, binary stars have been suggested as the main production channel - but the issue remains unresolved. Bright and nearby Type IIb SNe are rare but detection of the progenitor star in archival pre-explosion images and, when the SN has faded, a search for the companion star is feasible. By comparison of the magnitude and colour of the progenitor star to predictions from stellar evolutionary models, basic properties such as the initial mass can be estimated \citep{Sma09}. High quality multi-wavelength monitoring of these SNe followed by detailed modelling of the data is crucial to improve our understanding of Type IIb SNe and their progenitor stars. This paper presents the first 100 days of the extensive optical and NIR dataset we have obtained for such a SN, the Type IIb 2011dh. Detailed hydrodynamical modelling of the SN using these data have been presented in \citet[hereafter \citetalias{Ber12}]{Ber12} and identification and analysis of the plausible progenitor star in \citet[hereafter \citetalias{Mau11}]{Mau11}. The remaining data and further modelling will be presented in forthcoming papers.

\subsection{Supernova 2011dh}

SN 2011dh was discovered by A. Riou on 2011 May 31.893 UT \citep{Gri11} in the nearby galaxy M51 at a distance of about 8 Mpc (Sect.~\ref{s_distance}). The latest non-detection reported in the literature is by Palomar Transient Factory (PTF) from May 31.275 UT \citep[hereafter \citetalias{Arc11}]{Arc11}. In this paper we adopt May 31.5 UT as the epoch of explosion and the phase of the SN will be expressed relative to this date throughout the paper.

The host galaxy M51, also known as the Whirlpool galaxy, was the first galaxy for which the spiral structure was observed \citep{Ros50} and is frequently observed. Thus it is not surprising that excellent pre-explosion data were available in the Hubble Space Telescope (HST) archive. In \citetalias{Mau11} we used these data to identify a yellow (extended) supergiant progenitor candidate which, by comparison to stellar evolutionary models, corresponds to a star of $13 \pm 3$ M$_\odot$ initial mass. A similar analysis by \citet{Dyk11} estimated an initial mass between 17 and 19 M$_\odot$, the difference mainly stemming from the different method used to identify the evolutionary track in the HR-diagram. Recent HST \citep{Dyk13b} and Nordic Optical Telescope (NOT) \citep[this paper]{Erg13} observations show that the yellow supergiant is now gone and indeed was the progenitor of SN 2011dh. We discuss this issue in Sect.~\ref{s_prog_dis} and provide details of the NOT observations in Appendix~\ref{a_prog_obs}.

The SN has been extensively monitored from X-ray to radio wavelengths by several teams. Optical and NIR photometry and spectroscopy, mainly from the first 50 days, have been published by \citetalias{Arc11}, \citetalias{Mau11}, \citet[hereafter \citetalias{Tsv12}]{Tsv12}, \citet[hereafter \citetalias{Vin12}]{Vin12}, \citet[hereafter \citetalias{Mar13}]{Mar13}, \citet[hereafter \citetalias{Dyk13b}]{Dyk13b} and \citet[hereafter \citetalias{Sah13}]{Sah13}. Radio and millimeter observations have been published by \citet{Mar11}, \citet{Kra12}, \citet{Bie12}, \citet{Sod12} and \citet{Hor12} and X-ray observations by \citet{Sod12}, \citet{Sas12} and \citet{Cam12}. The SN has been monitored in the ultraviolet (UV) using SWIFT, in the mid-infrared (MIR) using Spitzer and at sub-millimeter wavelengths using Herschel. In this paper we will focus on the UV to MIR emission.

The nature of the progenitor star is an issue of great interest and there has been some debate in the literature. Using approximate models \citetalias{Arc11} argued that the SN cooled too fast and \citet{Sod12} that the speed of the shock was too high to be consistent with an extended progenitor. However, in \citetalias{Ber12} we have used detailed hydrodynamical modelling to show that a 3.3-4 M$_\odot$ helium core with an attached thin and extended hydrogen envelope well reproduces the early photometric evolution and is also consistent with the temperature inferred from early spectra. The findings in \citetalias{Ber12} are in good agreement with those in \citetalias{Mau11} and the issue now seems to be settled by the disappearance of the yellow supergiant. See also \citet{Mae12} for a discussion of the assumptions made in \citet{Sod12}.

The presence of a companion star (as for SN 1993J) or not is another issue of great interest. As shown in \citet{Ben12} a binary interaction scenario that reproduces the observed and modelled properties of the yellow supergiant is certainly possible. Furthermore, the prediction of a blue (compact) companion star would be possible to confirm using HST observations, preferably in the UV where the star would be at its brightest.

The paper is organized as follows. In Sections \ref{s_distance} and \ref{s_extinction} we discuss the distance and extinction, in Sect.~\ref{s_obs} we present the observations and describe the reduction and calibration procedures, in Sect. \ref{s_analysis} we analyse the observations and calculate the bolometric lightcurve, in Sect. \ref{s_sn_comp} we compare the observations to other SNe and in Sect.~\ref{s_discussion} we provide a discussion, mainly related to the hydrodynamical modelling in \citetalias{Ber12} and the disappearance of the progenitor. Finally, we conclude and summarize the paper in Sect.~\ref{s_conclusions}. In Appendix \ref{a_phot_cal} we provide details on the calibration of the photometry and in Appendix \ref{a_prog_obs} we provide details on the progenitor observations.

\subsection{Distance}
\label{s_distance}

In Table \ref{t_d} we list all estimates for the distance to M51 we have found in the literature. As the sample is reasonably large and as it is not clear how to judge the reliability of the individual estimates we will simply use a median and the 16 and 84 percentiles to estimate the distance and the corresponding error bars. This gives a distance of 7.8$^{+1.1}_{-0.9}$ Mpc which we will use throughout this paper.

\begin{table*}[tb]
\caption{Distance to M51. Literature values.}
\begin{center}
\begin{tabular}{lll}
\toprule
Distance & Method & Reference \\
(Mpc) & & \\
\midrule
9.60 $\pm{0.80}$ & Size of HII regions & \citet{San74} \\
6.91 $\pm{0.67}$ & Young stellar clusters  & \citet{Geo90} \\
8.39 $\pm{0.60}$ & Planetary nebula luminosity function & \citet{Fel97} \\
7.62 $\pm{0.60}$ & Planetary nebula luminosity function & \citet{Cia02} \\
7.66 $\pm{1.01}$ & Surface brightness fluctuations & \citet{Ton01} \\
7.59 $\pm{1.02}$ & Expanding photosphere method (SN 2005cs) &\citet{Tak06} \\
6.36 $\pm{1.30}$ & Type IIP SN standard candle method (SN 2005cs) & \citet{Tak06} \\
8.90 $\pm{0.50}$ & Spectral expanding photosphere method (SN 2005cs) & \citet{Des08} \\
6.92 & Type Ic SN properties (SN 1994I) & \citet{Iwa94} \\
7.90 $\pm{0.70}$ & Spectral expanding photosphere method (SN 2005cs) & \citet{Bar07} \\
6.02 $\pm{1.92}$ & Spectral expanding photosphere method (SN 1994I) & \citet{Bar96} \\
8.36 & Type IIP SN standard candle method (SN 2005cs) & \citet{Poz09} \\
9.30 & Tully-Fisher relation & \citet{Tul88} \\
8.40 $\pm{0.7}$ & Expanding photosphere method (SNe 2005cs and 2011dh) & \citet{Vin12} \\
\bottomrule
\end{tabular}
\end{center}
\label{t_d}
\end{table*}

\subsection{Extinction}
\label{s_extinction}

The interstellar line-of-sight extinction towards SN 2011dh within the Milky Way as given by the extinction maps presented by \citet[hereafter \citetalias{Sch98}]{Sch98} and recently recalibrated by \citet[hereafter \citetalias{Sch11}]{Sch11} is $E$($B$-$V$)$_\mathrm{MW}$=0.031 mag. Here and in the following the extinction within the Milky Way, the host galaxy and in total will be subscripted "MW", "H" and "T" respectively and, except where otherwise stated, refer to the interstellar line-of-sight extinction towards the SN. The extinction within host galaxies is generally difficult to estimate. One class of methods used are empirical relations between the equivalent widths of the interstellar \ion{Na}{i} D absorption lines and $E$($B$-$V$). Relations calibrated to the extinction within other galaxies as the one by \citet{Tur03} are based on low resolution spectroscopy and as demonstrated by \citet{Poz11} the scatter is very large. Relations based on high or medium resolution spectroscopy as the ones by \citet[hereafter \citetalias{Mun97}]{Mun97} and \citet[hereafter \citetalias{Poz12}]{Poz12} show a surprisingly small scatter but are calibrated to the extinction within the Milky Way. Nevertheless, given the line of sight nature of the method and the rough similarity between M51 and the Milky Way we will use these for an estimate of the extinction within M51. \citet{Rit12} presented high-resolution spectroscopy of SN 2011dh resolving 8 \ion{Na}{i} D components near the M51 recession velocity. The total widths of the \ion{Na}{i} D$_2$ and D$_1$ lines were $180.1 \pm 5.0$ and $106.2 \pm 5.1$ m\AA~respectively. Using the \citetalias{Mun97} relations and summing the calculated extinction for all individual components (see discussions in \citetalias{Mun97} and \citetalias{Poz12}) we get $E$($B$-$V$)$_\mathrm{H}$=0.05 mag. Using the \citetalias{Poz12} relations for the total equivalent widths we get $E$($B$-$V$)$_\mathrm{H}$=0.03 mag. Taking the average of these two values and adding the extinction within the Milky Way (see above) gives $E$($B$-$V$)$_\mathrm{T}$=0.07 mag. Such a low extinction is supported by estimates from X-rays \citep{Cam12} and the progenitor Spectral Energy Distribution (SED) \citepalias{Mau11} and we will use this value throughout the paper. The stellar population analysis done by \citet{Mur11} suggests a somewhat higher extinction ($E$($B$-$V$)$_\mathrm{T}$=0.14 mag). We will adopt that value and the extinction within the Milky Way as our upper and lower error bars giving $E$($B$-$V$)$_\mathrm{T}$=0.07$^{+0.07}_{-0.04}$ mag. Further constraints on the extinction from the SN itself and comparisons to other SNe is discussed in Sect. \ref{s_extinction_rev}. To calculate the extinction as a function of wavelength we have used the reddening law of \citet{Car89} and R$_V$=3.1. For broad-band photometry the extinction was calculated at the mean energy wavelength of the filters. In this paper we will consequently use the definitions from \citet[hereafter \citetalias{Bes12}]{Bes12} for the mean energy wavelength and other photometric quantities.

\section{Observations}
\label{s_obs}

\subsection{Software}
\label{s_software}

Two different software packages have been used for 2-D reductions, measurements and calibrations of the data. The {\sc iraf} based {\sc quba} pipeline \citep[hereafter \citetalias{Val11}]{Val11} and another {\sc iraf} based package developed during this work which we will refer to as the {\sc sne} pipeline. This package has been developed with the particular aim to provide the high level of automation needed for large sets of data.

\subsection{Imaging}
\label{s_obs_image}

An extensive campaign of optical and NIR imaging was initiated for SN 2011dh shortly after discovery using a multitude of different instruments. Data have been obtained with the Liverpool Telescope (LT), the Nordic Optical Telescope (NOT), Telescopio Nazionale (TNG), Telescopio Carlos Sanchez (TCS), the Calar Alto 3.5m and 2.2m telescopes, the Faulkes Telescope North (FTN), the Asiago 67/92cm Schmidt and 1.82m Copernico telescopes, the William Herschel Telescope (WHT), the Large Binocular Telescope (LBT) and Telescopi Joan Oro (TJO). Amateur observations obtained at the Cantabria and Montcabrer observatories have also been included. The major contributors were the LT, the NOT, the TCS and the TNG. The dataset includes 85 epochs of optical imaging and 23 epochs of NIR imaging for the first 100 days and have been obtained thanks to a broad collaboration of European observers.

\subsubsection{Reductions and calibration}
\label{s_obs_image_red_cal}

The optical raw data were reduced with the {\sc quba} pipeline except for the LT data for which the automatic telescope pipeline reductions have been used. 

The NIR raw data were reduced with the {\sc sne} pipeline except for UKIRT data for which the reductions provided by CASU have been used. Except for the standard procedures the pipeline has support for second pass sky subtraction using an object mask, correction for field distortion and unsharp masking. Correction for field distortion is necessary to allow co-addition of images with large dithering shifts and has been applied to the TNG data. Unsharp masking removes large scales structures (e.g. the host galaxy) in the images to facilitate the construction of a master sky in the case of large scale structure overlap. Given the (usually) small fields of view and the large size of the host galaxy this technique has been applied to all data where separate sky frames were not obtained. 

Photometry was performed with the {\sc sne} pipeline. We have used aperture photometry on the reference stars as well as the SN using a relatively small aperture (1.5$-$2.0 times the FWHM). A mild ($>$0.1 mag error) rejection of the reference stars as well as a mild (3 $\sigma$) rejection of the calculated zero points were also used. Both measurement and calibration errors were propagated using standard formulae. To ensure that the photometry is free from background contamination we have, as a test, template-subtracted the NOT and LT data sets using a {\sc hotpants}\footnote{http://www.astro.washington.edu/users/becker/hotpants.html} based tool provided by the {\sc sne} pipeline and late-time ($\sim$200 days) SN subtracted images. The contamination was negligible in all bands which is not surprising as the SN is still bright compared to the background at $\sim$100 days.

The optical and NIR photometry was calibrated to the Johnson-Cousins (JC), Sloan Digital Sky Survey (SDSS) and 2 Micron All Sky Survey (2MASS) systems using reference stars in the SN field in turn calibrated using standard fields. The calibration procedure is described in detail in Appendix~\ref{a_phot_cal} where we also discuss the related uncertainties. The photometry was transformed to the standard systems using S-corrections \citep{Str02} except for the JC $U$ and SDSS $u$ bands which were transformed using linear colour-terms. We find the calibration to be accurate to within five percent in all bands, except for the early (0-40 days) NOT $U$ band observations, which show a systematic offset of $\sim$20 percent, possibly due to the lack of S-corrections in this band. Comparisons to S-corrected SWIFT JC photometry as well as the photometry published in \citetalias{Arc11}, \citetalias{Vin12}, \citetalias{Tsv12}, \citetalias{Mar13}, \citetalias{Dyk13b} and \citetalias{Sah13} supports this conclusion although some datasets show differences in the 15-30 percent range in some bands. Note that we have used JC-like $UBVRI$ filters and SDSS-like $gz$ filters at NOT whereas we have used JC-like $BV$ filters and SDSS-like $ugriz$ filters at LT and FTN. The JC-like $URI$ and SDSS-like $uri$ photometry were then tied to both the JC and SDSS systems to produce full sets of JC and SDSS photometry.

\subsubsection{Space Telescope Observations}
\label{s_obs_image_space}

We have also performed photometry on the Spitzer 3.6 and 4.5 $\mu$m imaging\footnote{Obtained through the DDT program by G. Helou.} and the SWIFT optical and UV imaging. 

For the Spitzer imaging we performed aperture photometry using the {\sc sne} pipeline and the zero points and standard aperture provided in the IRAC Instrument Handbook to calculate magnitudes in the natural (energy flux based) Vega system of IRAC. The Spitzer images were template subtracted using a {\sc hotpants} based tool provided by the {\sc sne} pipeline and templates constructed from archive images. Comparing with photometry on the original images, the background contamination was less than five percent in all bands.

For the SWIFT imaging we performed aperture photometry using the {\sc uvotsource} tool provided by the {\sc heasoft} package and the standard aperture of 5 arcsec to calculate magnitudes in the natural (photon count based) Vega system of UVOT. Observations were combined using the {\sc uvotimsum} tool provided by the {\sc heasoft} package and after day 5 subsequently combined in sequences of three to increase the signal-to-noise ratio (SNR). The SWIFT UV images were template subtracted using a {\sc hotpants} based tool provided by the {\sc sne} pipeline and templates constructed from archive images ($UVW1$) and SN subtracted late-time ($\sim$80 days) images ($UVM2$ and $UVW2$). Comparing with photometry on the original images, the background contamination was negligible in the $UVW1$ band whereas the $UVM2$ and $UVW2$ bands were severely affected, differing by more than a magnitude at late times. Our SWIFT photometry is in good agreement with that published in \citetalias{Mar13} except in the $UVM2$ and $UVW2$ bands after $\sim$10 days, which is expected since \citetalias{Mar13} did not perform template subtraction. Our SWIFT photometry is also in good agreement with that published in \citetalias{Arc11} (given in the natural AB system of UVOT) except in the $UVM2$ and $UVW2$ bands after $\sim$30 days, the differences probably arising from differences in the template subtraction.

\subsubsection{Results}
\label{s_obs_image_results}

\begin{figure}[tb]
\includegraphics[width=0.48\textwidth,angle=0]{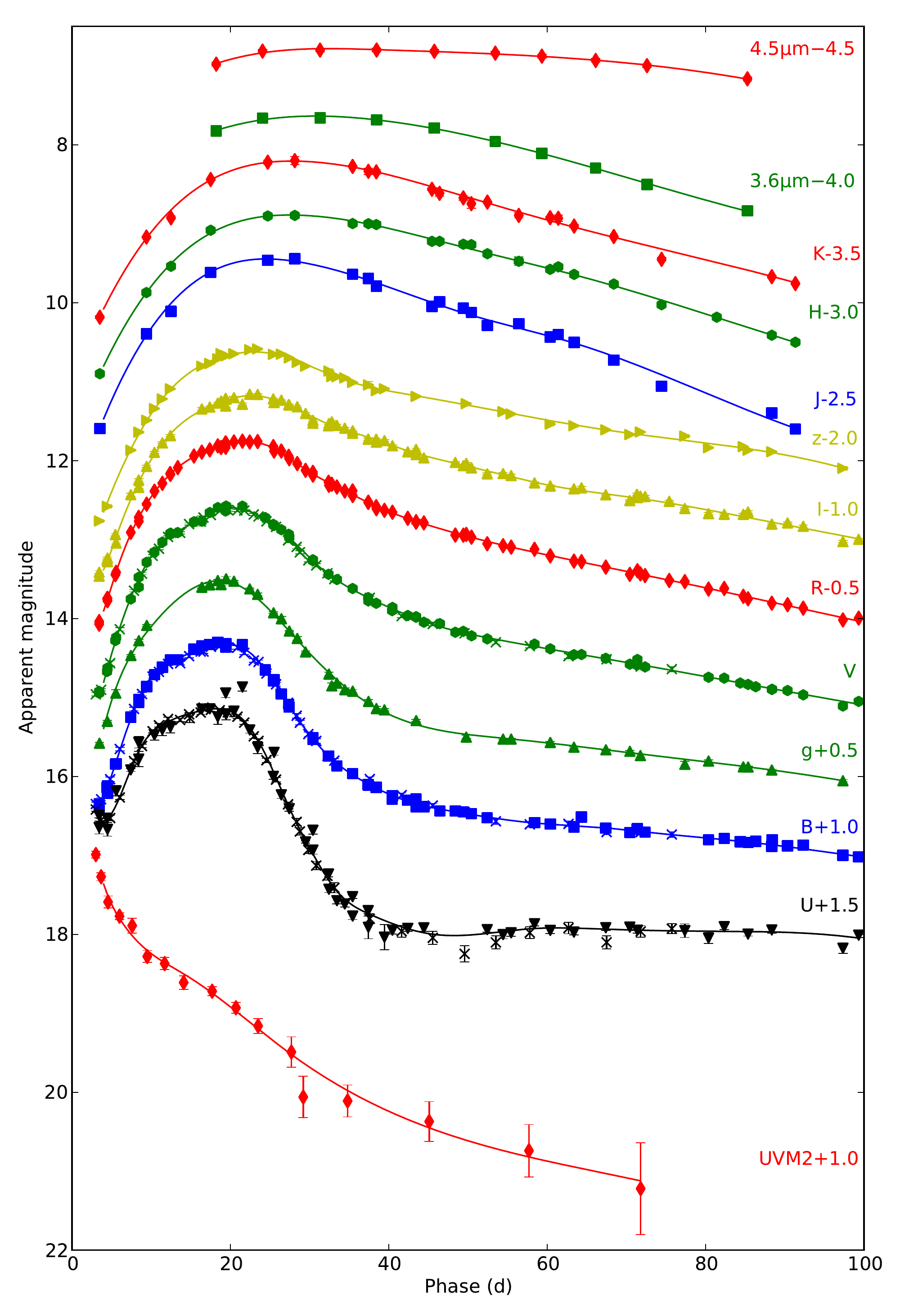}
\caption{Photometric evolution of SN 2011dh in the UV, optical, NIR and MIR. For clarity each band have been shifted in magnitude. Each lightcurve have been annotated with the name of the band and the shift applied. We also show the S-corrected SWIFT JC photometry (crosses) and cubic spline fits (solid lines).}
\label{f_uv_opt_nir_mir}
\end{figure}

The S-corrected optical (including SWIFT JC) and NIR magnitudes and their corresponding errors are listed in Tables \ref{t_jc}, \ref{t_jc_swift}, \ref{t_sloan} and \ref{t_nir} and the JC $UBVRI$, SDSS $gz$ and 2MASS $JHK$ magnitudes shown in Fig. \ref{f_uv_opt_nir_mir}. The Spitzer 3.6 and 4.5 $\mu$m magnitudes and their corresponding errors are listed in Table~\ref{t_mir} and shown in Fig.~\ref{f_uv_opt_nir_mir}. The SWIFT UV magnitudes and their corresponding errors are listed in Table~\ref{t_uv} and the SWIFT $UVM2$ magnitudes shown in Fig.~\ref{f_uv_opt_nir_mir}. As discussed in Appendix~\ref{a_phot_cal}, because of the red tail of the filters and the strong blueward slope of the SN spectrum, the $UVW1$ and $UVW2$ lightcurves do not reflect the evolution of the spectrum at their mean energy wavelengths. These bands will therefore be excluded from any subsequent discussion and the calculation of the bolometric lightcurve in Sect.~\ref{s_bol_lightcurve}. Figure~\ref{f_uv_opt_nir_mir} also shows cubic spline fits using 3-5 point knot separation, error weighting and a 5 percent error floor. The standard deviation around the fitted splines is less than 5 percent and mostly less than a few percent except for the SWIFT $UVM2$ band for which the standard deviation is between 5 and 10 percent on the tail. All calculations in Sect. \ref{s_analysis}, including the bolometric lightcurve, are based on these spline fits. In these calculations the errors have been estimated as the standard deviation around the fitted splines and then propagated.

\subsection{Spectroscopy}
\label{s_obs_spec}

An extensive campaign of optical and NIR spectroscopic observations was initiated for SN 2011dh shortly after discovery with data obtained from a multitude of telescopes. Data have been obtained with the NOT, the TNG, the WHT, the Calar Alto 2.2m telescope, the Asiago 1.82m Copernico and 1.22m Galileo telescopes and the LBT. The major contributors were the NOT and the TNG. Details of all spectroscopic observations, the telescope and instrument used, epoch and instrument characteristics are given in Table~\ref{t_speclog}. The dataset includes 55 optical spectra obtained at 26 epochs and 18 NIR spectra obtained at 10 epochs for the first 100 days.

\subsubsection{Reductions and calibration}
\label{s_obs_spec_red_cal}

The optical and NIR raw data were reduced using the {\sc quba} and {\sc sne} pipelines respectively. Flats for NOT Grisms 4 and 5 were spatially shifted, typically by one pixel, to minimize the fringing in the reduced data.

The flux of optical and NIR spectra was extracted using the {\sc quba} and {\sc sne} pipelines respectively. A large aperture and error weighting was used to reduce the wavelength dependent effect on the size of the PSF in the spatial direction. No corrections were done for this effect in the dispersion direction. The slit was always (initially) vertically aligned so the position of the PSF in the dispersion direction should not vary much.

The optical spectra were flux calibrated using the {\sc quba} pipeline. A sensitivity function was derived using a spectroscopic standard star and corrected for the relative atmospheric extinction using tabulated values for each site. Telluric absorption was removed using a normalized absorption profile derived from the standard star. The significant second order contamination present in NOT Grism 4 spectra was corrected for using the method presented in \citet{Sta07}. The optical spectra were wavelength calibrated using arc lamp spectra and cross-correlated and shifted to match sky-lines.

The NIR spectra were flux calibrated and the telluric absorption removed with the {\sc sne} pipeline. A sensitivity function was derived using solar or Vega analogue standard stars selected from the Hipparchos catalogue and spectra of the sun and Vega. The interstellar extinction of the standards have been estimated from Hipparchos $BV$ photometry and corrected for when necessary. The NIR spectra were wavelength calibrated using arc lamp spectra and cross-correlated and shifted to match sky-lines.

Finally, the absolute flux scale of all spectra has been calibrated against interpolated photometry using a least square fit to all bands for which the mean energy wavelength is at least half an equivalent width within the spectral range.

\subsubsection{Results}
\label{s_obs_spec_results}

\begin{figure*}[tb]
\includegraphics[width=1.0\textwidth,angle=0]{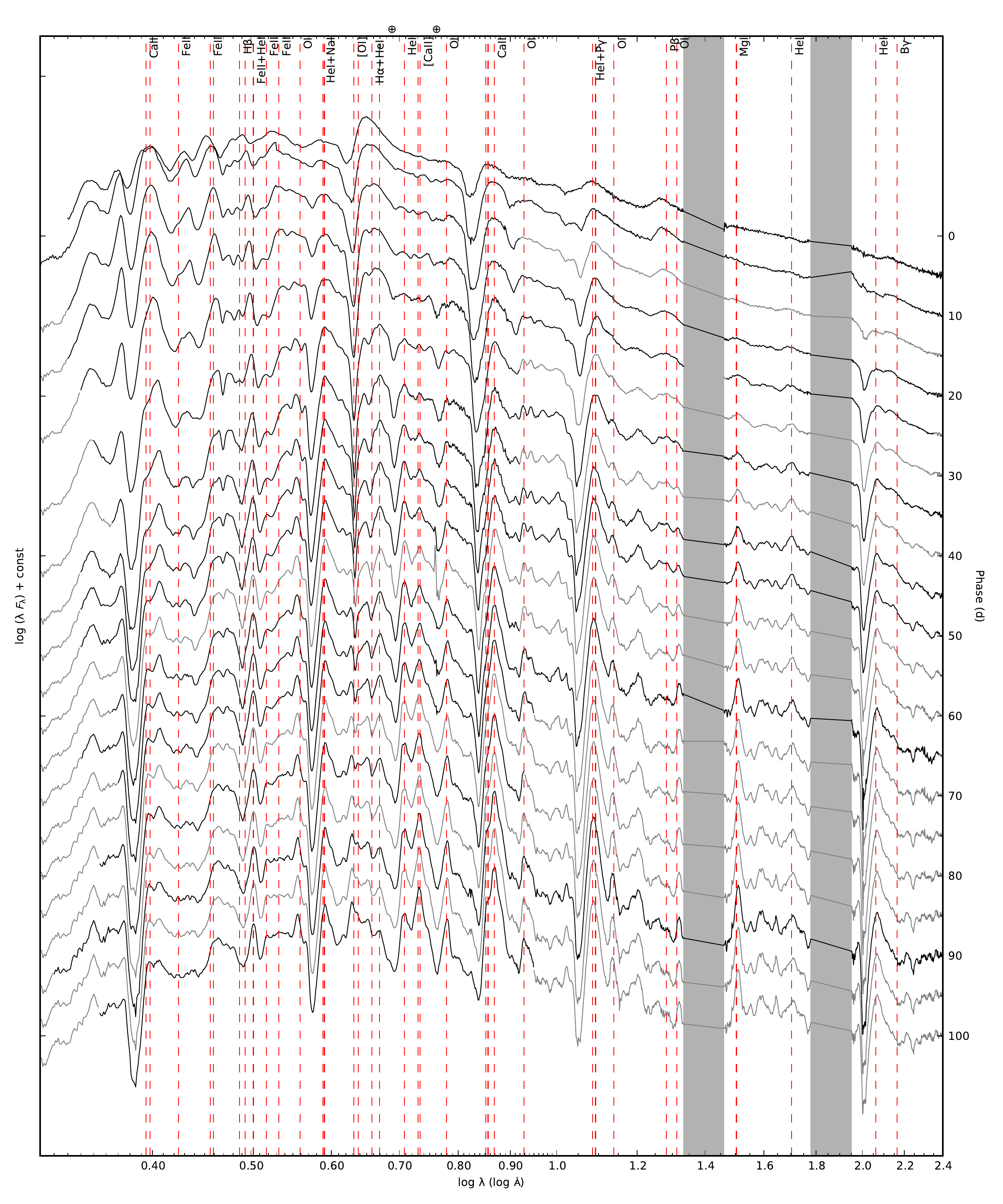}
\caption{Optical and NIR (interpolated) spectral evolution for SN 2011dh for days 5$-$100 with a 5-day sampling. Telluric absorption bands are marked with a $\oplus$ symbol in the optical and shown as grey regions in the NIR.}
\label{f_spec_evo_opt_NIR}
\end{figure*}

\begin{figure*}[p]
\includegraphics[width=1.0\textwidth,angle=0]{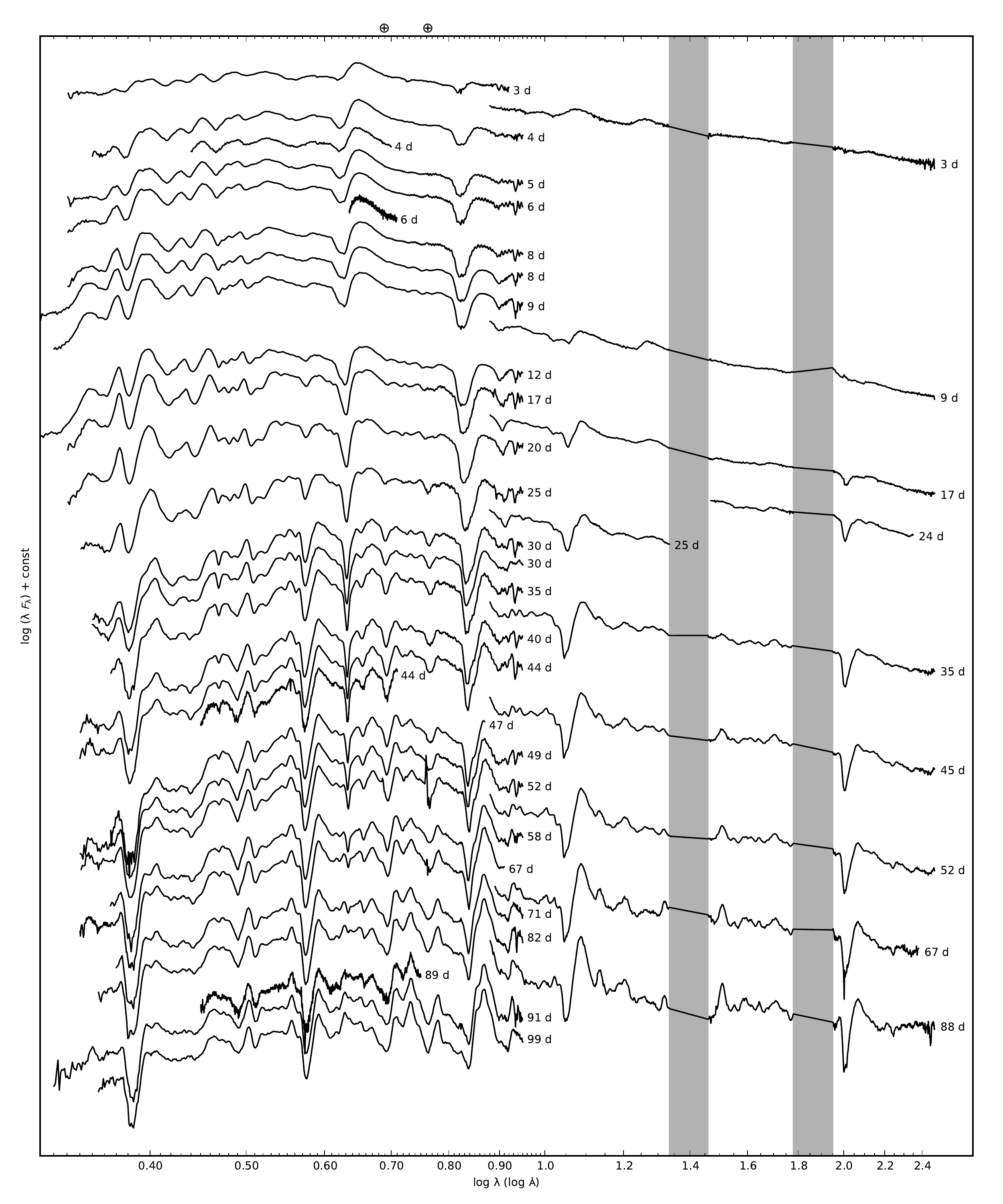}
\caption{Sequence of the observed spectra for SN 2011dh. Spectra obtained on the same night using the same telescope and instrument have been combined and each spectra have been labelled with the phase of the SN. Telluric absorption bands are marked with a $\oplus$ symbol in the optical and shown as grey regions in the NIR.}
\label{f_spec_evo_opt_NIR_trad}
\end{figure*}

All reduced, extracted and calibrated spectra will be made available for download from the Weizmann Interactive Supernova data REPository\footnote{http://www.weizmann.ac.il/astrophysics/wiserep/} (WISeREP) \citep{Yar12}.  Figure~\ref{f_spec_evo_opt_NIR_trad} shows the sequence of observed spectra where those obtained on the same night using the same telescope and instrument have been combined. For clarity, and as is motivated by the frequent sampling of spectra, all subsequent figures in this and the following sections are based on time-interpolations of the spectral sequence. Interpolated spectra separated more than half the sampling time from observed spectra are displayed in shaded colour and should be taken with some care whereas interpolated spectra displayed in full colour are usually more or less indistinguishable from observed spectra. To further visualize the evolution, the spectra have been aligned to a time axis at the right border of the panels. The interpolations were done as follows. First all spectra were re-sampled to a common wavelength dispersion. Then, for each interpolation epoch the spectra closest in time before and after the epoch were identified resulting in one or more wavelength ranges and associated pre- and post-epoch spectra. For each wavelength range the pre- and post-epoch spectra were then linearly interpolated and finally scaled and smoothly averaged using a 500 \AA~overlap range. Spectra interpolated using this method were also used in the calculations of the bolometric lightcurve (Sect.~\ref{s_bol_lightcurve}) and S-corrections (Appendix~\ref{a_phot_cal}). Figure~\ref{f_spec_evo_opt_NIR} shows the interpolated optical and NIR spectral evolution of SN 2011dh for days 5$-$100 with a 5-day sampling. All spectra in this and subsequent figures spectra have been corrected for redshift and interstellar extinction.

\section{Analysis}
\label{s_analysis}

\subsection{Photometric evolution}
\label{s_analysis_phot}

Absolute magnitudes were calculated as $M_i=m_i-\mu-A_i$, where $m_{i}$ is the apparent magnitude in band $i$, $\mu$ the distance modulus and $A_i$ the interstellar absorption at the mean energy wavelength of band $i$. The systematic errors stemming from this approximation (as determined from synthetic photometry) is less than a few percent and can be safely ignored. The systematic errors stemming from the uncertainty in distance (Sect.~\ref{s_distance}) and extinction (Sect.~\ref{s_extinction}) on the other hand are at the 30 percent level and this should be kept in mind in the subsequent discussions. All bands except the SWIFT $UVM2$ band show a similar evolution (the Spitzer MIR imaging did not start until day 20) with a strong initial increase from day 3 to the peak followed by a decrease down to a tail with a roughly linear decline rate. The maximum occurs at increasingly later times for redder bands. The drop from the maximum down to the tail is more pronounced for bluer bands and is not seen for bands redder than $z$. Both these trends are reflections of the strong decrease in temperature seen between 10 and 40 days (Fig.~\ref{f_bb_T_evo}). The tail decline rates are highest for the reddest bands and almost zero for the bluest bands. It is interesting to note that the Spitzer 4.5 $\mu$m band breaks this pattern and shows a markedly slower decline than the 3.6 $\mu$m and the NIR bands. Warm dust or CO fundamental band emission are two possible explanations (Sect.~\ref{s_45_micron_excess}). The times and absolute magnitudes of the maximum as well as the tail decline rates at 60 days are listed in Table~\ref{t_lc_char} as measured from cubic spline fits (Fig.~\ref{f_uv_opt_nir_mir}).

Early time data for the first three days have been published in A11 and T12 and show a strong decline in the $g$, $V$ and $R$ bands. This initial decline phase ends at about the same time as our observations begins. 

\begin{table}[tb]
\caption{Times and absolute magnitudes of the maximum and tail decline rates at 60 days as measured from cubic spline fits.}
\begin{center}
\begin{tabular}{llll}
\hline\hline \\ [-1.5ex]
Band & Maximum & Absolute magnitude & Decline rate \\ [0.5ex]
& (days) & (mag) & (mag day$^{-1}$)\\
\hline \\ [-1.5ex]
$UVM2$ & ... & ... & 0.019 \\
$U$ & 18.30 & -16.16 & -0.002 \\
$B$ & 18.88 & -16.43 & 0.007 \\
$V$ & 20.22 & -17.08 & 0.018 \\
$R$ & 21.86 & -17.38 & 0.021 \\
$I$ & 22.62 & -17.41 & 0.020 \\
$J$ & 24.45 & -17.58 & 0.027 \\
$H$ & 27.42 & -17.61 & 0.025 \\
$K$ & 27.71 & -17.78 & 0.027 \\
3.6 $\mu$m & 31.55 & -17.83 & 0.028 \\
4.5 $\mu$m & 39.62 & -18.18 & 0.009 \\ [0.5ex]
\hline
\end{tabular}

\end{center}
\label{t_lc_char}
\end{table}

\subsection{Colour evolution and blackbody fits}
\label{s_analysis_colour}

\begin{figure}
\includegraphics[width=0.48\textwidth,angle=0]{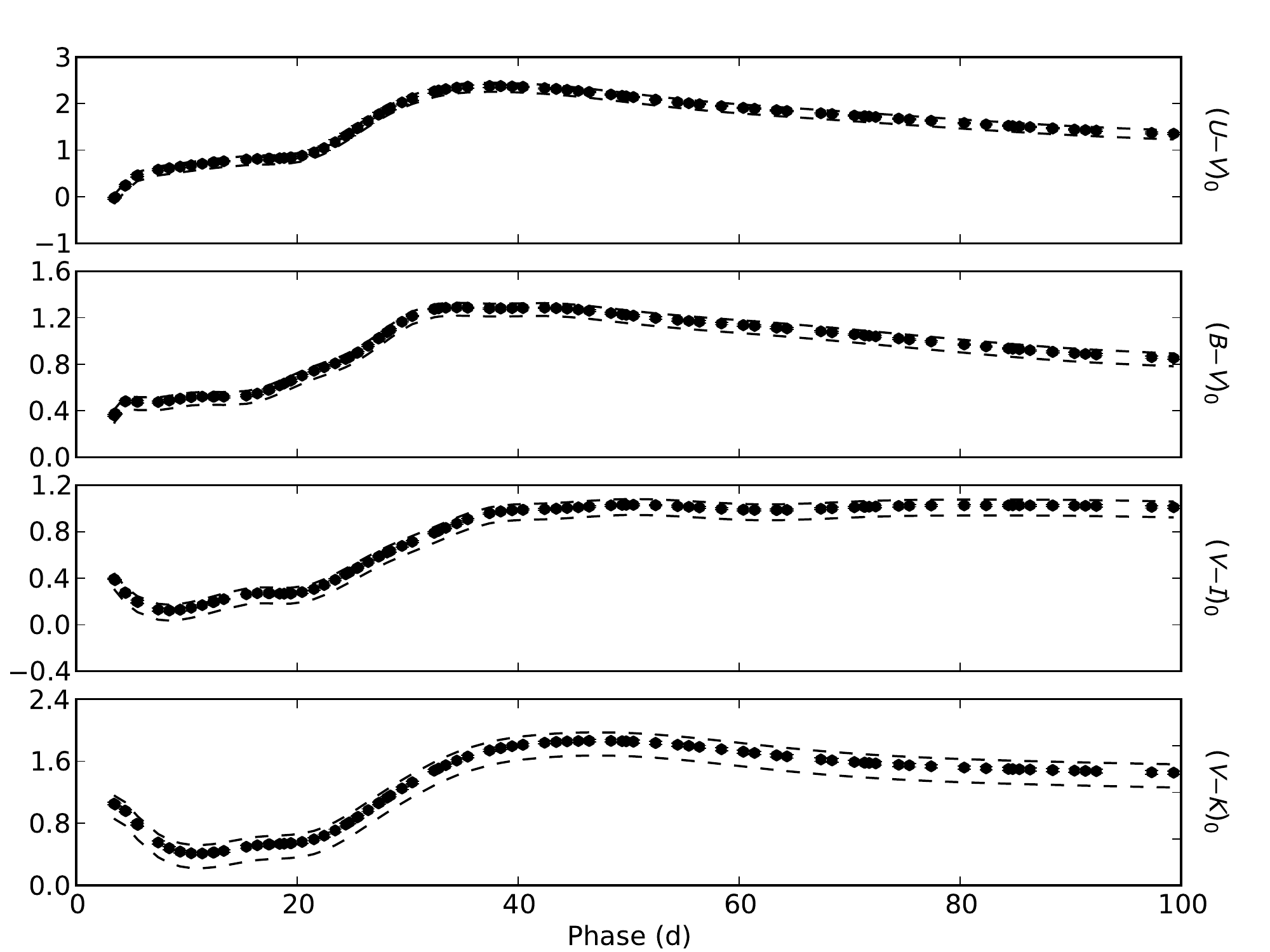}
\caption{$U$-$V$, $B$-$V$, $V$-$I$ and $V$-$K$ intrinsic colour evolution for SN 2011dh for the adopted extinction (black dots). The upper and lower error bars for the systematic error arising from extinction (black dashed lines) are also shown.}
\label{f_colour_evo}
\end{figure}

\begin{figure}
\includegraphics[width=0.48\textwidth,angle=0]{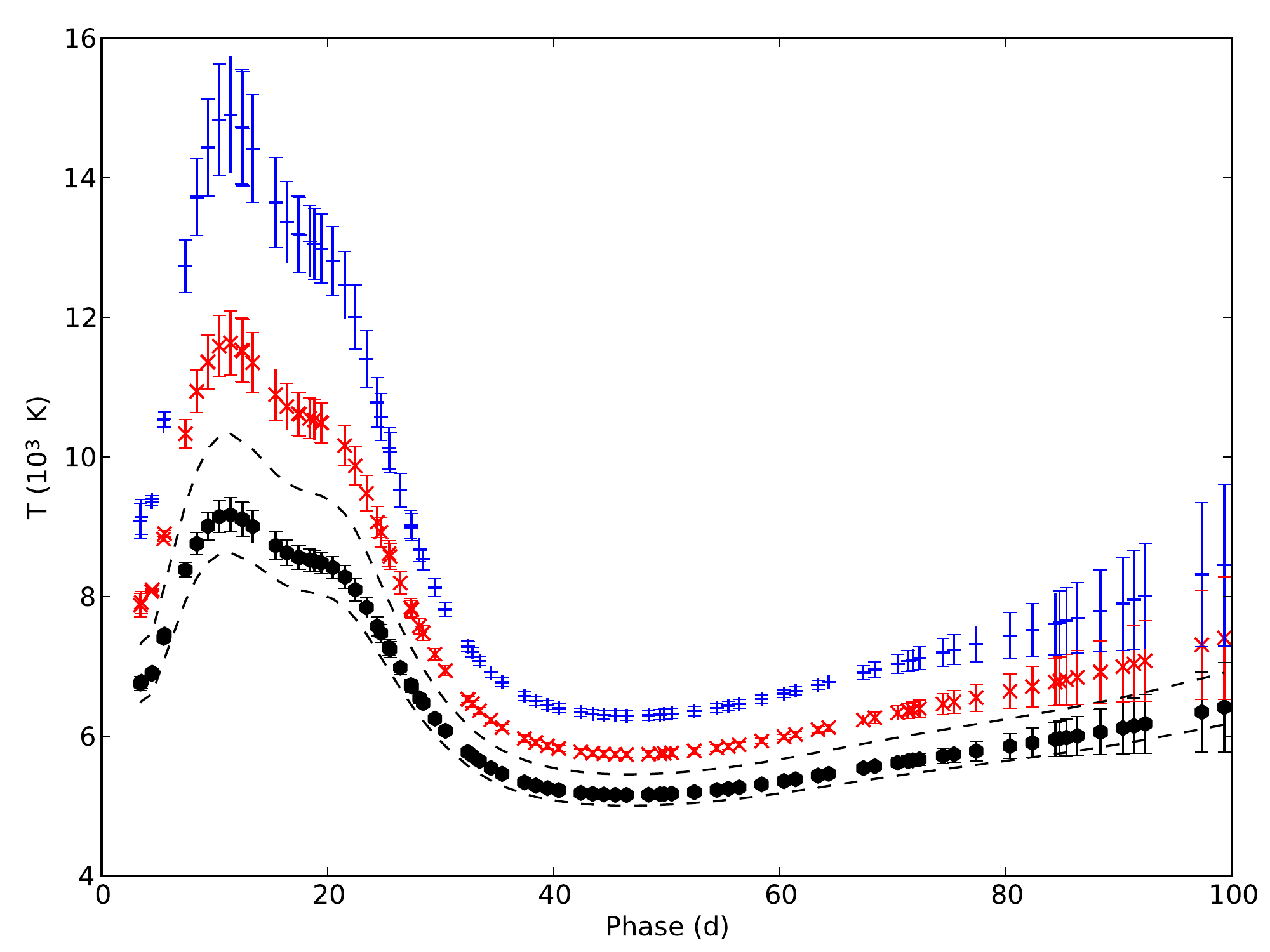}
\caption{Evolution of the blackbody temperature for SN 2011dh as inferred from fits to the $V$, $I$, $J$, $H$ and $K$ bands for the adopted extinction (black dots). The upper and lower error bars for the systematic error arising from extinction (black dashed lines) and two higher extinction scenarios, $E$($B$-$V$)$_\mathrm{T}$=0.2 mag (red crosses) and $E$($B$-$V$)$_\mathrm{T}$=0.3 mag (blue pluses), discussed in Sect. \ref{s_extinction_rev}, are also shown.}
\label{f_bb_T_evo}
\end{figure}

\begin{figure}
\includegraphics[width=0.48\textwidth,angle=0]{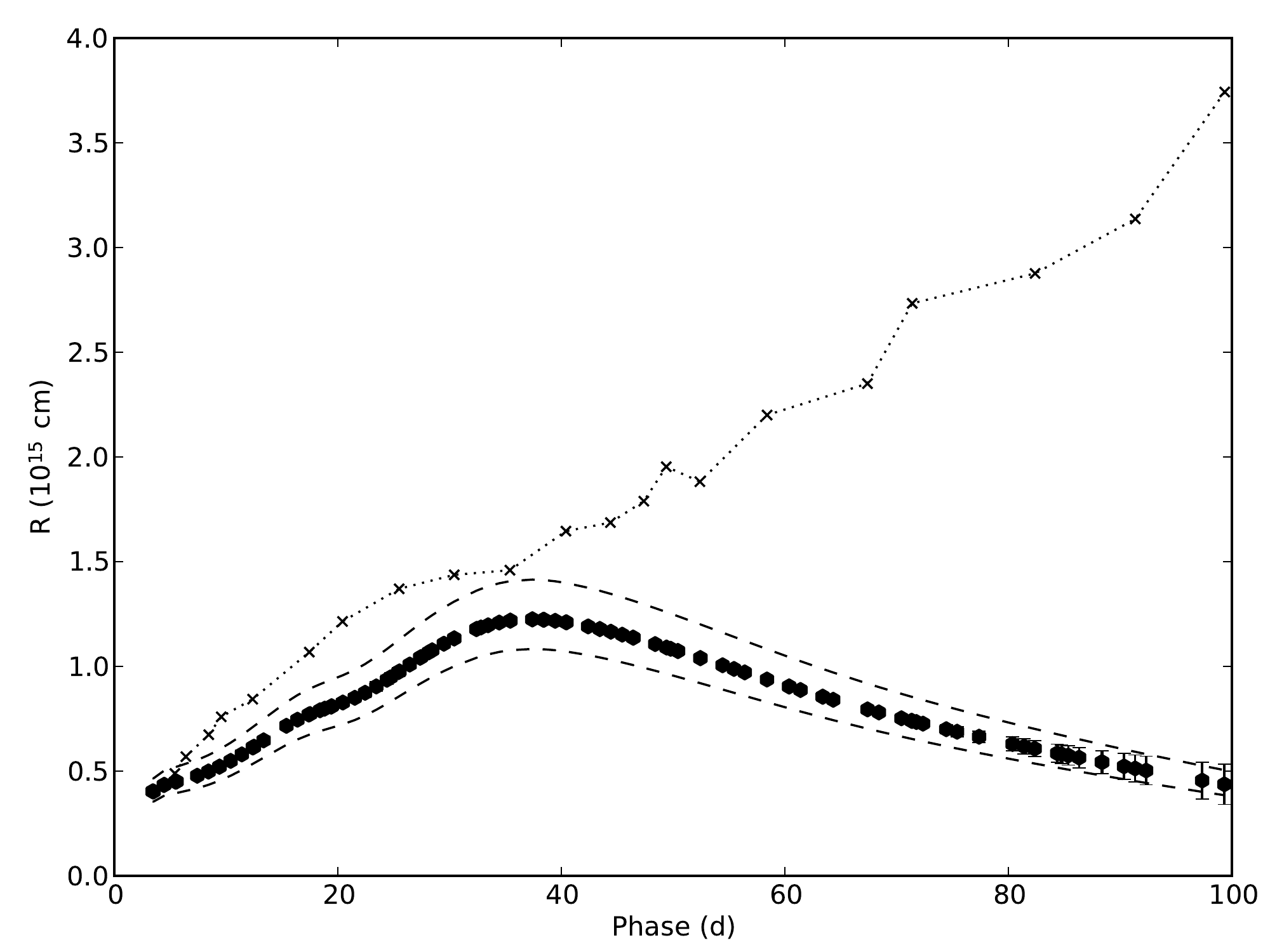}
\caption{Evolution of blackbody radius for SN 2011dh as inferred from fits to the $V$, $I$, $J$, $H$ and $K$ bands for the adopted extinction. The upper and lower error bars for the systematic error arising from extinction and distance (black dashed lines) and the radius corresponding to the P-Cygni minimum of the \ion{Fe}{ii} 5169 \AA~line (black dotted line) are also shown.}
\label{f_bb_R_evo}
\end{figure}

Figure~\ref{f_colour_evo} shows the intrinsic $U$-$V$, $B$-$V$, $V$-$I$ and $V$-$K$ colour evolution of SN 2011dh given the adopted extinction. Initially we see a quite strong blueward trend in the $V$-$I$ and $V$-$K$ colours reaching a minimum at $\sim$10 days which is not reflected in the $U$-$V$ and $B$-$V$ colours. Subsequently all colours redden reaching a maximum at $\sim$40 days for the $U$-$V$ and $B$-$V$ colours and $\sim$50 days for the $V$-$I$ and $V$-$K$ colours followed by a slow blueward trend for all colours. Figures \ref{f_bb_T_evo} and \ref{f_bb_R_evo} show the evolution of blackbody temperature and radius as inferred from fits to the $V$, $I$, $J$, $H$ and $K$ bands given the adopted extinction. As discussed in Sect. \ref{s_analysis_spec}, the flux in bands blueward of $V$ is strongly reduced by the line opacity in this region, in particular between 10 and 30 days. Therefore we have excluded these bands from the fits whereas the $R$ band has been excluded to avoid influence from H$\alpha$ emission at early times. Note that the temperature and radius obtained correspond to the surface of thermalization rather than the photosphere (total optical depth $\sim$1) and lose physical meaning when the ejecta become optically thin in the continuum. The evolution of the $V$-$I$ and $V$-$K$ colours is reflected in the evolution of the blackbody temperature, initially increasing from $\sim$7000 K at 3 days to a maximum of $\sim$9000 K at 8 days, subsequently decreasing to a minimum of $\sim$5000 K at $\sim$50 days followed by a slow increase. The blackbody radius shows an almost linear increase from ~$\sim$0.4 $\times$10$^{15}$ cm to a maximum of $\sim$1.2 $\times$10$^{15}$ cm and a subsequent almost linear decrease. 

\begin{figure}
\includegraphics[width=0.48\textwidth,angle=0]{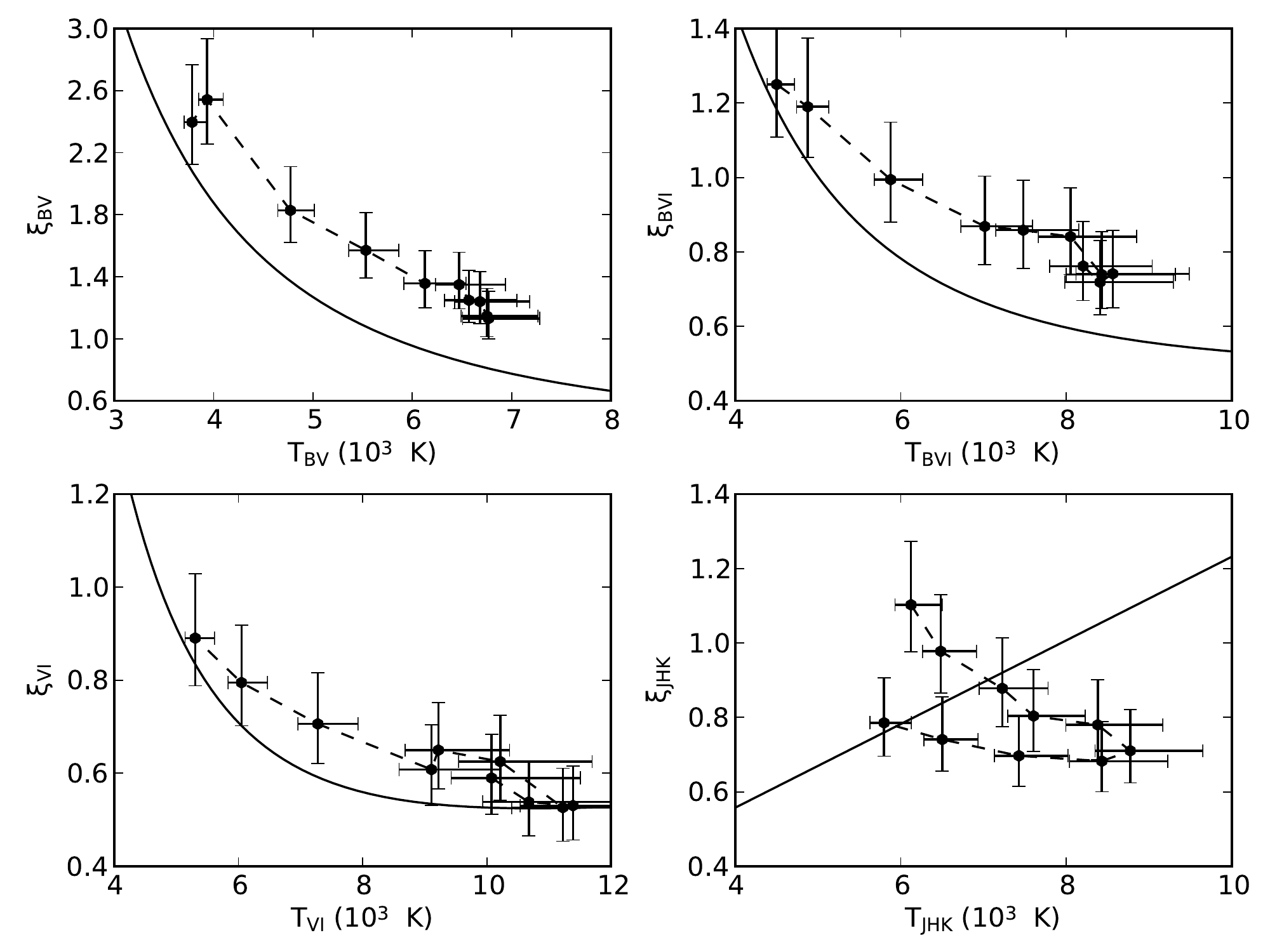}
\caption{Dilution factors as a function of colour temperature as inferred from blackbody fits (dots) compared to the dilution factors for Type IIP SNe determined by \citet{Des05} (solid lines) for the $B$ and $V$ (upper left panel), the $B$, $V$ and $I$ (upper right panel), the $V$ and $I$ (lower left panel) and the $J$, $H$ and $K$ (lower right panel) bands. In all panels we also show the upper and lower error bars for the systematic error arising from the extinction and distance.}
\label{f_bb_T_xi}
\end{figure}

In Fig.~\ref{f_bb_R_evo} we also show the radius corresponding to the P-Cygni minimum of the \ion{Fe}{ii} 5169 \AA~line. Interpreting this (Sect.~\ref{s_analysis_spec}) as the photospheric radius and the blackbody radius as the thermalization radius we see a fairly consistent evolution between 8 and 40 days corresponding to a dilution factor (ratio of photospheric and blackbody radius) increasing from $\sim$0.7 to $\sim$0.8 as the temperature decreases. The figure also suggests that such an interpretation breaks down for later epochs. Dilution factors for Type IIP SNe have been discussed extensively in the literature because of their importance for the EPM method \citep{Des05} but are not well known for Type IIb SNe. In Fig~\ref{f_bb_T_xi} we show dilution factors as a function of colour temperature as inferred from blackbody fits compared to the $\xi_{BV}$, $\xi_{BVI}$, $\xi_{VI}$ and $\xi_{JHK}$ dilution factors determined for Type IIP SNe using NLTE modelling by \citet{Des05}. The $VI$ and $JHK$ dilution factors are $\sim$10 percent higher and $\sim$10 percent lower on average as compared to $\xi_{VI}$ and $\xi_{JHK}$ respectively. If free-free absorption is dominating the absorptive opacity in the NIR but not in the optical, this is naively consistent with the lower charge density for helium core composition as compared to the hydrogen envelope composition of Type IIP SNe. The $BV$ and $BVI$ dilution factors are $\sim$25 and $\sim$40 percent higher on average as compared to $\xi_{BVI}$ and $\xi_{BV}$ respectively. The main reason for this is likely a stronger flux deficit (caused by a higher line opacity) in the $B$ band as compared to Type IIP SNe for a given thermalization temperature. \citetalias{Vin12} argue for higher values of the dilution factors as compared to Type IIP SNe because of the lower charge density and, as they point out, \citet{Bar95} have used NLTE modelling of SN 1993J to determine a $BV$ dilution factor $\sim$60 percent higher than for Type IIP SNe. This is similar to our (observational) result although in our interpretation this is rather due to a stronger flux deficit in the $B$ band. In the end \citetalias{Vin12} chose a value of 1.0 for their $BVRI$ dilution factor which is $\sim$10 percent higher than our average value, the difference explained by the $\sim$10 percent longer distance they derive. Dilution factors can never be observationally determined with better accuracy than the distance is known and NLTE modelling of Type IIb SNe is probably needed to accurately determine these. We find dilution factors involving bands redwards $B$, in particular, the $VI$ dilution factor most promising for future use in the EPM method applied to Type IIb SNe.

\subsection {Bolometric evolution}
\label{s_bol_lightcurve}

To calculate the pseudo-bolometric lightcurve of SN 2011dh we have used a combination of two different methods. One, which we will refer to as the spectroscopic method, for wavelength regions with spectral information and one, which we will refer to as the photometric method, for wavelength regions without. The prefix pseudo here refers to the fact that a true bolometric lightcurve should be integrated over all wavelengths. We do not assume anything about the flux in wavelength regions not covered by data but discuss this issue at the end of the section.

When using the spectroscopic method we divide the wavelength region into sub-regions corresponding to each photometric band. For each epoch of photometry in each of the sub-regions a bolometric correction $BC_{i}=M_{\mathrm{bol},i}^{\mathrm{syn}}-M_{i}^{\mathrm{syn}}$ is determined. Here $M_{i}^{\mathrm{syn}}$ and  $M_{\mathrm{bol},i}^{\mathrm{syn}}$ are the absolute and bolometric magnitudes respectively, as determined from synthetic photometry and integration of the sub-region flux per wavelength using observed spectra. The bolometric magnitude in the region $M_{\mathrm{bol}}=-2.5 \log \sum 10^{-0.4(M_{i}+BC_{i})}$ is then calculated as the sum over all sub-regions, where $M_{i}$ is the absolute magnitude as determined from observed photometry. Spectra are linearly interpolated to match each epoch of photometry as described in Sect.~\ref{s_obs_spec_results}. This method makes use of both spectral and photometric information and is well motivated as long as the spectral sampling is good.

When using the photometric method we log-linearly interpolate the flux per wavelength between the mean energy wavelengths of the filters. This is done under the constraint that the synthetic absolute magnitudes as determined from the interpolated SED equals the absolute magnitudes as determined from observed photometry. The solution is found by a simple iterative scheme. The total flux in the region is then calculated by integration of the interpolated flux per wavelength.

The absolute magnitudes in each band were calculated using cubic spline fits as described in Sect.~\ref{s_obs_image_results}, which is justified by the frequent sampling in all bands. When necessary, as for the SWIFT UV and Spitzer MIR magnitudes, extrapolations were done assuming a constant colour. The filter response functions and zeropoints used to represent the different photometric systems are discussed in Appendix~\ref{a_phot_cal}.

\begin{figure}[tb]
\includegraphics[width=0.48\textwidth,angle=0]{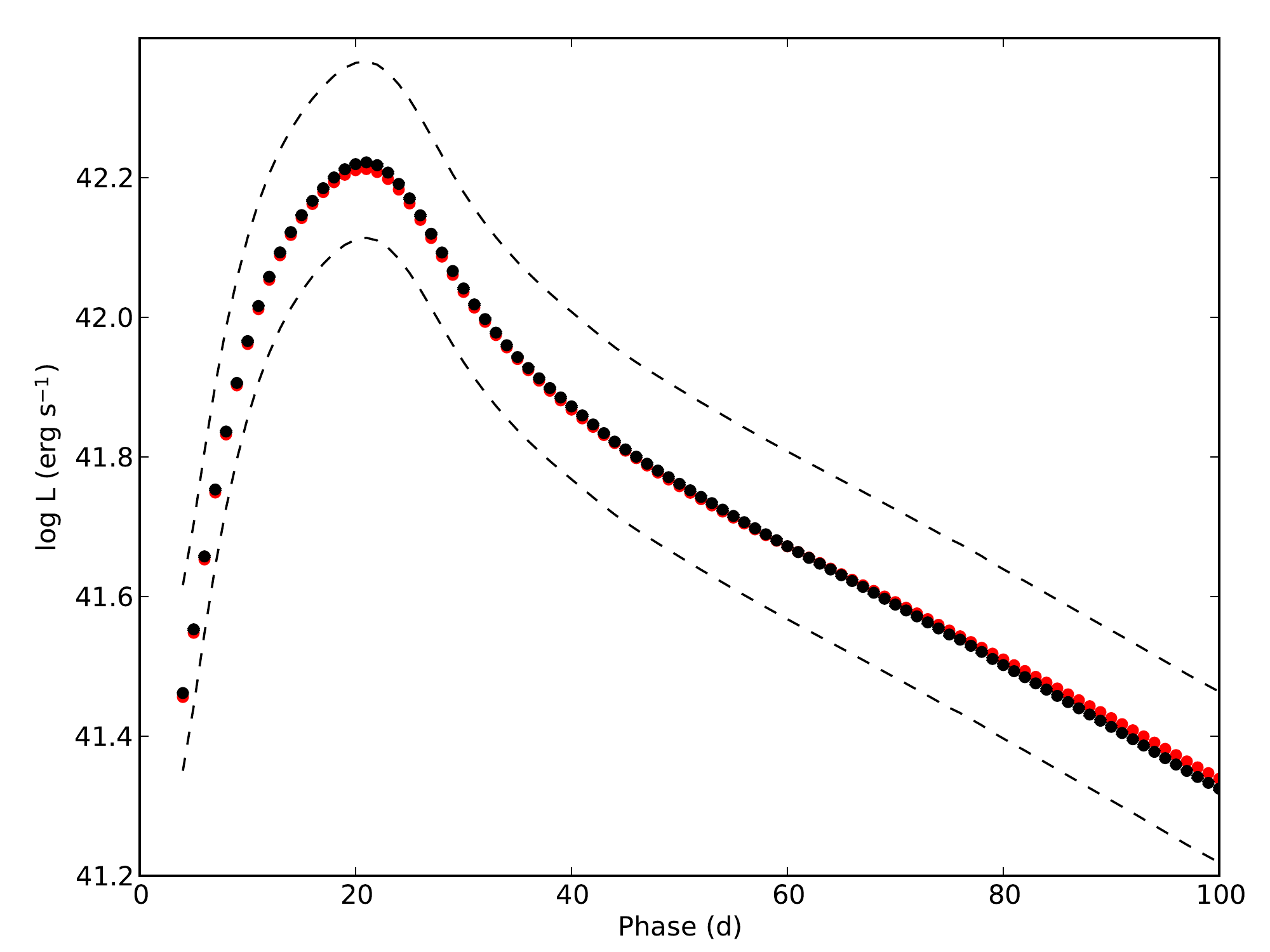}
\caption{Pseudo-bolometric UV to MIR lightcurve for SN 2011dh calculated with the spectroscopic (black dots) and photometric (red dots) method. The upper and lower error bars for the systematic error arising from extinction and distance (black dashed lines) are also shown.}
\label{f_UV_MIR_bol}
\end{figure}

For SN 2011dh we have optical and NIR spectra with good sampling between 3 and 100 days and we have used the spectroscopic method in the $U$ to $K$ region and the photometric method in the UV and MIR regions. The pseudo-bolometric UV to MIR (1900-50000 \AA) lightcurve of SN 2011dh is shown in Fig.~\ref{f_UV_MIR_bol} and listed in Table~\ref{t_UV_MIR_bol} for reference. These data together with the photospheric velocity as estimated in Sect.~\ref{s_analysis_spec} provide the observational basis for the hydrodynamical modelling of SN 2011dh presented in B12. For comparison we also show the pseudo-bolometric lightcurve calculated using the photometric method only. The difference is small but, as expected, increases slowly when the spectrum evolves to become more line dominated. The bolometric lightcurve shows the characteristics common to Type I and Type IIb SNe with a rise to peak luminosity followed by a decline phase and a subsequent tail phase with a roughly linear decline rate (Sect.~\ref{s_physics_sn_IIb}). The maximum occurs at 20.9 days at a pseudo-bolometric luminosity of 16.67$\pm{0.05}^{+6.59}_{-3.67}$$\times$10$^{41}$ ergs s$^{-1}$, where the second error bars give the systematic error arising from the distance and extinction. The tail decline rates are 0.033, 0.021, 0.022 and 0.020 mag day$^{-1}$ at 40, 60, 80 and 100 days respectively.

\begin{figure}[tb]
\includegraphics[width=0.48\textwidth,angle=0]{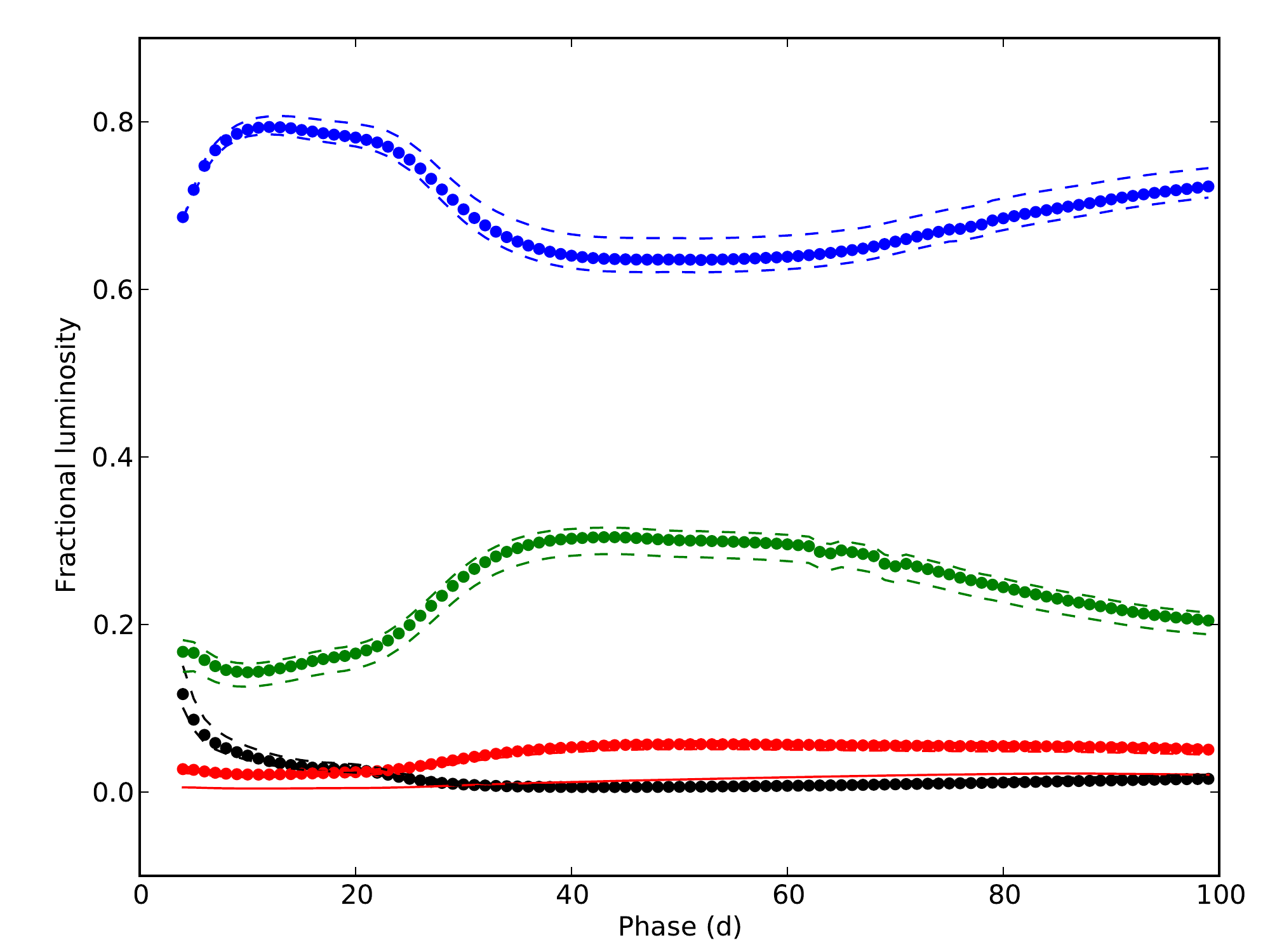}
\caption{Fractional UV (black dots), optical (blue dots), NIR (green dots) and MIR (red dots) luminosity for SN 2011dh. The upper and lower error bars for the systematic error arising from extinction (dashed lines) and the fractional Rayleigh-Jeans luminosity redwards of 4.5 $\mu$m (red solid line) are also shown.}
\label{f_bol_frac}
\end{figure}

Figure~\ref{f_bol_frac} shows the fractional luminosity in the UV (1900-3300 \AA), optical (3300-10000 \AA), NIR (10000-24000 \AA) and MIR (24000-50000 \AA) regions respectively. The optical flux dominates and varies between $\sim$75 and $\sim$60 percent whereas the NIR flux varies between $\sim$15 and $\sim$30 percent. The UV flux initially amounts to $\sim$10 percent, decreasing to $\sim$1 percent at the beginning of the tail and onwards. The MIR flux initially amounts to $\sim$1 percent, increasing to $\sim$5 percent at the beginning of the tail and onwards. The evolution of the fractional luminosities mainly reflects the evolution of the temperature (Fig. \ref{f_bb_T_evo}) although we expect the UV to be quite sensitive to the evolution of the line opacity (Sect. \ref{s_analysis_spec}).

\begin{figure}[tb]
\includegraphics[width=0.48\textwidth,angle=0]{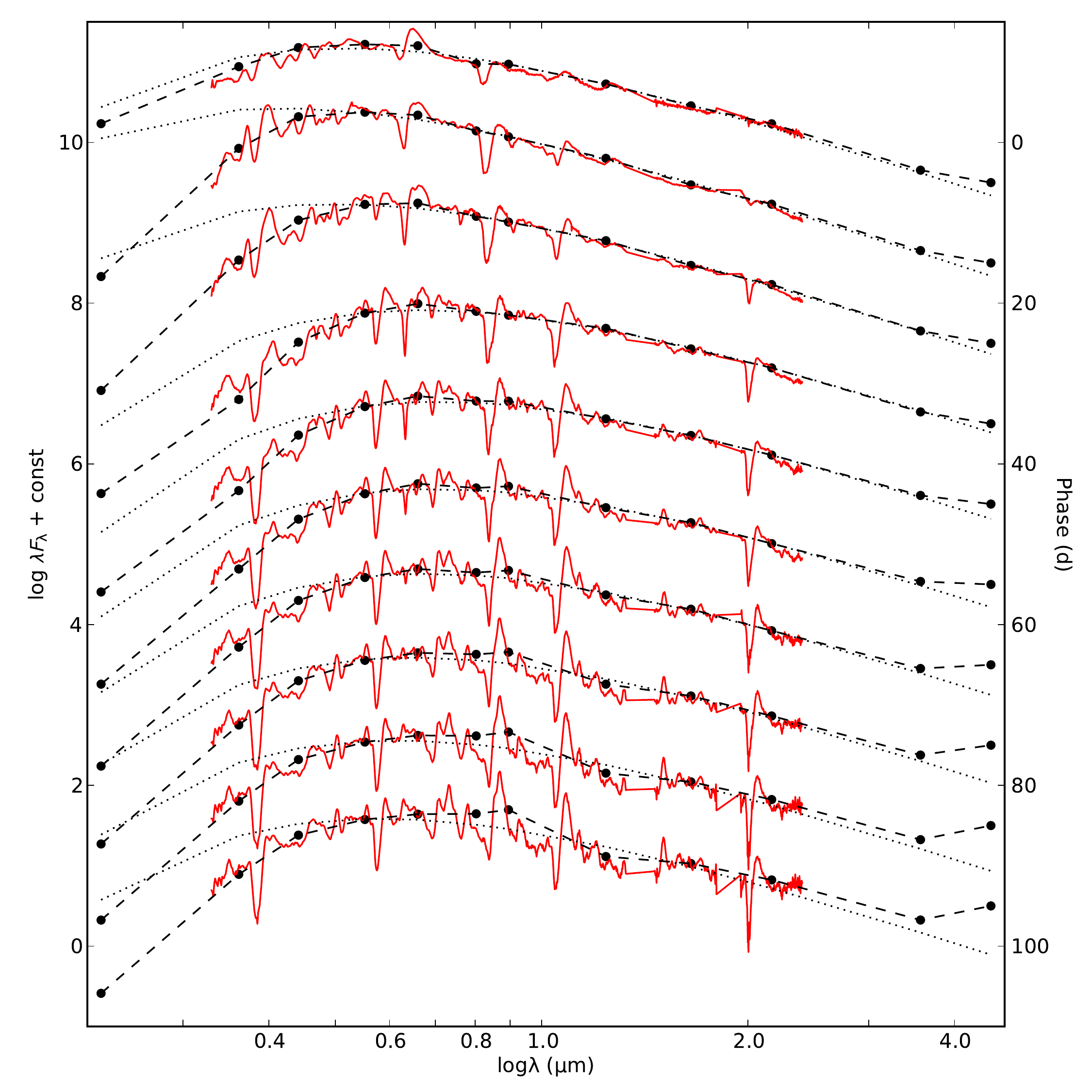}
\caption{The evolution of the SED as calculated with the photometric method (black dots and dashed lines) overplotted with the blackbody fits discussed in Sect.~\ref{s_analysis_colour} (black dotted lines) as well as the observed spectra interpolated as described in Sect. \ref{s_obs_spec_results} (red solid lines).}
\label{f_sed_evo}
\end{figure}

Figure~\ref{f_sed_evo} shows the evolution of the SED as calculated with the photometric method overplotted with the blackbody fits discussed in Sect.~\ref{s_analysis_colour} as well as the observed spectra interpolated as described in Sect.~\ref{s_obs_spec_results}. The strong blueward slope in the UV region (except for the first few days) suggests that the flux bluewards of the $UVM2$ band is negligible. The flux redwards of 4.5 $\mu$m could be approximated with a Rayleigh-Jeans tail or a model spectrum. As shown in Fig. \ref{f_bol_frac} the fractional Rayleigh-Jeans luminosity redwards of 4.5 $\mu$m is at the percent level. Note again the excess at 4.5 $\mu$m that develops between 50 and 100 days. Whereas the other bands redward of $V$ are well approximated by the blackbody fits the flux at 4.5 $\mu$m is a factor of $\sim$5 in excess at 100 days. Note also the strong reduction of the flux as compared to the fitted blackbodies in bands blueward of $V$ between 10 and 30 days (Sect.~\ref{s_analysis_spec}).

\subsection{Spectroscopic evolution}
\label{s_analysis_spec}

\begin{figure}[tb]
\includegraphics[width=0.48\textwidth,angle=0]{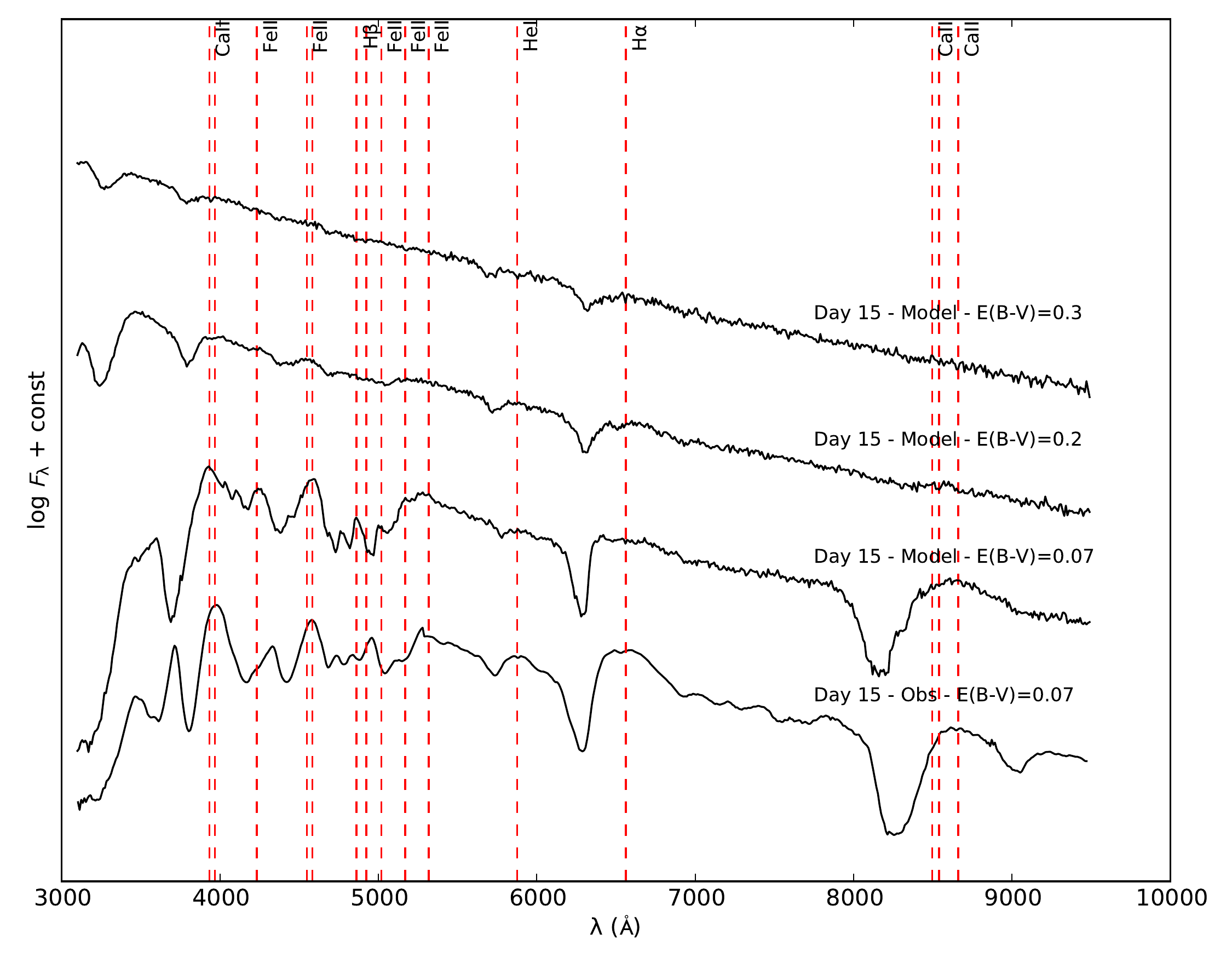}
\caption{Modelled and observed optical spectrum at 15 days. Lines identified by their optical depth being $\gtrsim$1 have been marked at their rest wavelength. We also show model spectra for two higher extinction scenarios, $E$($B$$-$$V$)$_\mathrm{T}$=0.2 mag and $E$($B$$-$$V$)$_\mathrm{T}$=0.3 mag, discussed in Sect. \ref{s_extinction_rev}.}
\label{f_spec_model_comp}
\end{figure}

We have used a SN atmosphere code implementing the method presented by \citet{Maz93} and \citet{Abb85} and the \citetalias{Ber12} He4R270 ejecta model with all elements except hydrogen and helium replaced with solar abundances to aid in identification of lines and some qualitative analysis of the spectra. The factor $\xi$ in eq. 15 in \citet{Abb85} has been set to one which might lead to overestimates of the line absorption in the optically thick limit. The Monte-Carlo based method treats line and electron scattering in the nebular approximation where the ionization fractions and level populations of bound states are determined by the radiation field approximated as a diluted blackbody parametrized by a radiation temperature. Line emission will be underestimated as the contribution from recombination is not included whereas line absorption is better reproduced. Following \citet{Maz93}, for each epoch we have determined the temperature for the blackbody emitting surface from fits to the $V$, $I$, $J$, $H$ and $K$ bands and iterated the radius until the observed luminosity was achieved. Note that, except for the temperature peak between $\sim$10 and $\sim$20 days, the \ion{He}{i} lines cannot be reproduced by the model as non-thermal excitation from the ground state is needed to populate the higher levels \citep{Luc91}. For a quantitative analysis a NLTE-treatment solving the rate equations is necessary, in particular with respect to non-thermal excitations and ionizations. Figure~\ref{f_spec_model_comp} shows a comparison between model and observed spectra at 15 days where we also have marked the rest wavelengths of lines identified by their optical depth being $\gtrsim$1. The atmosphere model is appropriate at early times when the approximation of a blackbody emitting surface is justified and we do not use it for phases later than $\sim$30 days. To aid in line-identifications at later times we use preliminary results from NLTE spectral modelling of the SN spectrum at 100 days to be presented in Jerkstrand et al. 2013 (in preparation). The details of this code have been presented in \citet{Jer11,Jer12}. Both the atmosphere and NTLE code uses the same atomic data as described in these papers. The lines identified by the atmosphere modelling, the NLTE modelling or both are discussed below and have been marked in Fig. \ref{f_spec_evo_opt_NIR}.

\begin{figure}[tb]
\includegraphics[width=0.48\textwidth,angle=0]{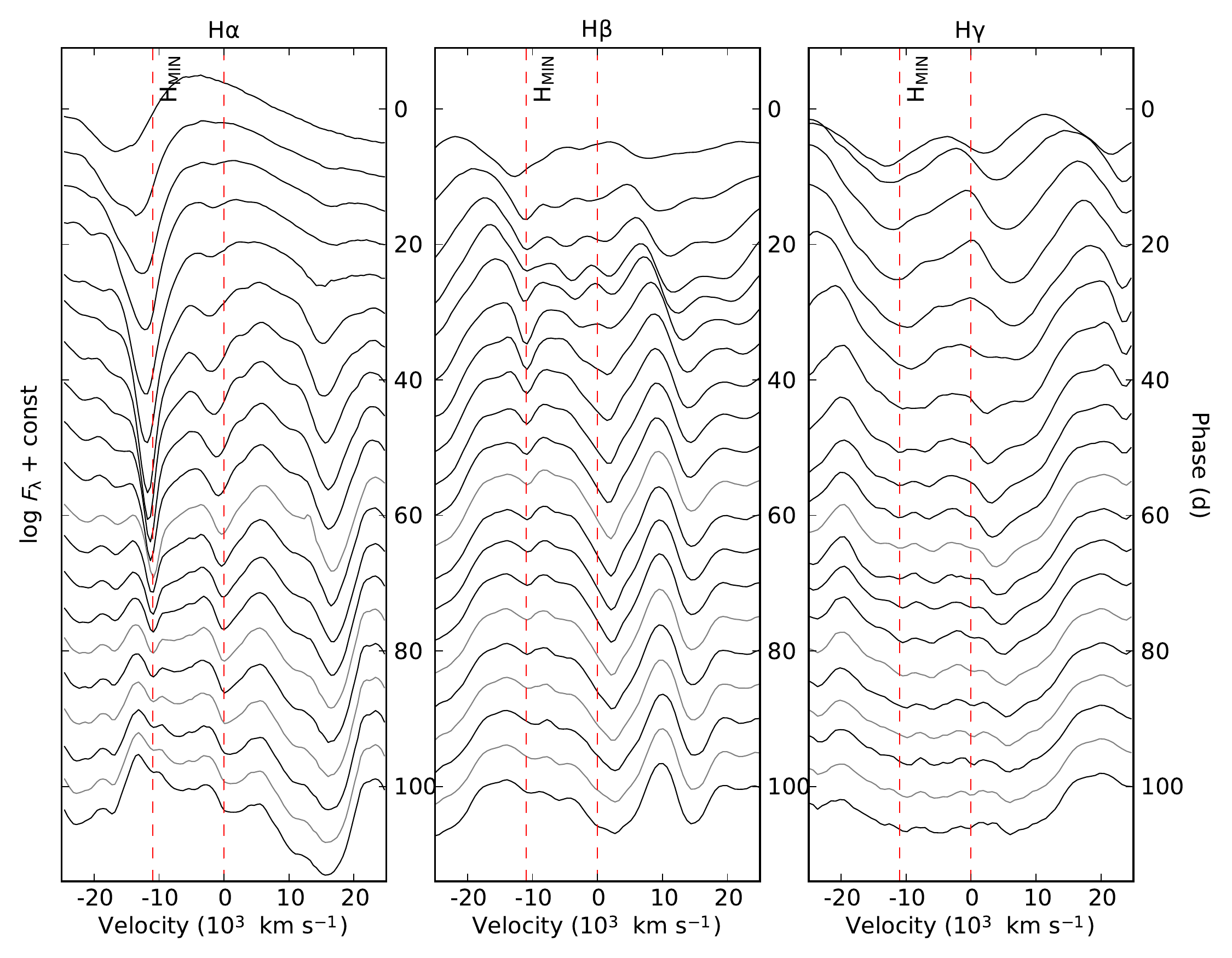}
\caption{Closeup of (interpolated) spectral evolution centred on the H$\alpha$ (left panel), H$\beta$ (middle panel) and H$\gamma$ (right panel) lines. All panels in this and the following figure show the minimum velocity for the H$\alpha$ absorption minimum (marked H$_{\mathrm{MIN}}$) interpreted as the interface between the helium core and hydrogen envelope.}
\label{f_spec_evo_H}
\end{figure}

The transition of the spectra from hydrogen (Type II) to helium (Type Ib) dominated starts at $\sim$10 days with the appearance of the \ion{He}{i} 5876 and 10830 \AA~lines and ends at $\sim$80 days with the disappearance of the H$\alpha$ line. This transition is likely determined by the photosphere reaching the helium core, the ejecta gradually becoming optically thin to the $\gamma$-rays and eventually to the hydrogen lines. At 3 days the hydrogen signature in the spectrum is strong and we identify the Balmer series $\alpha$$-$$\gamma$, Paschen series $\alpha$$-$$\gamma$ as well as Bracket $\gamma$ using the atmosphere modelling. H$\alpha$ shows a strong P-Cygni profile, extending in absorption to at least $\sim$25000 km s$^{-1}$, which gradually disappears in emission but stays strong in absorption until $\sim$50 days. Most other hydrogen lines fade rather quickly and have disappeared at $\sim$30 days. Weak absorption in H$\alpha$ and H$\beta$ remains until $\sim$80 days. Figure~\ref{f_spec_evo_H} shows closeups of the evolution centred on the hydrogen Balmer lines. Note that the absorption minimum for H$\alpha$ as well as H$\beta$ is never seen below $\sim$11000 km s$^{-1}$ but approaches this value as the lines get weaker (see also Fig.~\ref{f_vel_evo_p_cygni}). This suggests that a transition in the ejecta from helium core to hydrogen rich envelope material occurs at this velocity. Atmosphere modelling of the hydrogen lines using the \citetalias{Ber12} He4R270 ejecta model with all elements except hydrogen and helium replaced with solar abundances well reproduce the observed evolution of the absorption minima and the minimum velocity coincides with the model interface between the helium core and hydrogen rich envelope at $\sim$11500 km s$^{-1}$. The good agreement with the observed minimum velocity gives further support to the \citetalias{Ber12} ejecta model. \citetalias{Mar13} estimated hydrogen to be absent below $\sim$12000 km s$^{-1}$ by fitting a {\sc synow} \citep{Bra03} model spectrum to the observed spectrum at 11 days. We find the behaviour of the hydrogen lines in the weak limit to provide a better constraint and conclude that the interface between the helium core and hydrogen rich envelope is likely to be located at $\sim$11000~km~s$^{-1}$. By varying the fraction of hydrogen in the envelope we find a hydrogen mass of 0.01-0.04 M$_{\odot}$, in agreement with the 0.02 M$_{\odot}$ in the original model, to be consistent with the observed evolution of the hydrogen lines. \citetalias{Arc11} used spectral modelling similar to the one in this paper, but with a NLTE treatment of hydrogen and helium, to estimate the hydrogen mass to 0.024 M$_{\odot}$.

\begin{figure}[tb]
\includegraphics[width=0.48\textwidth,angle=0]{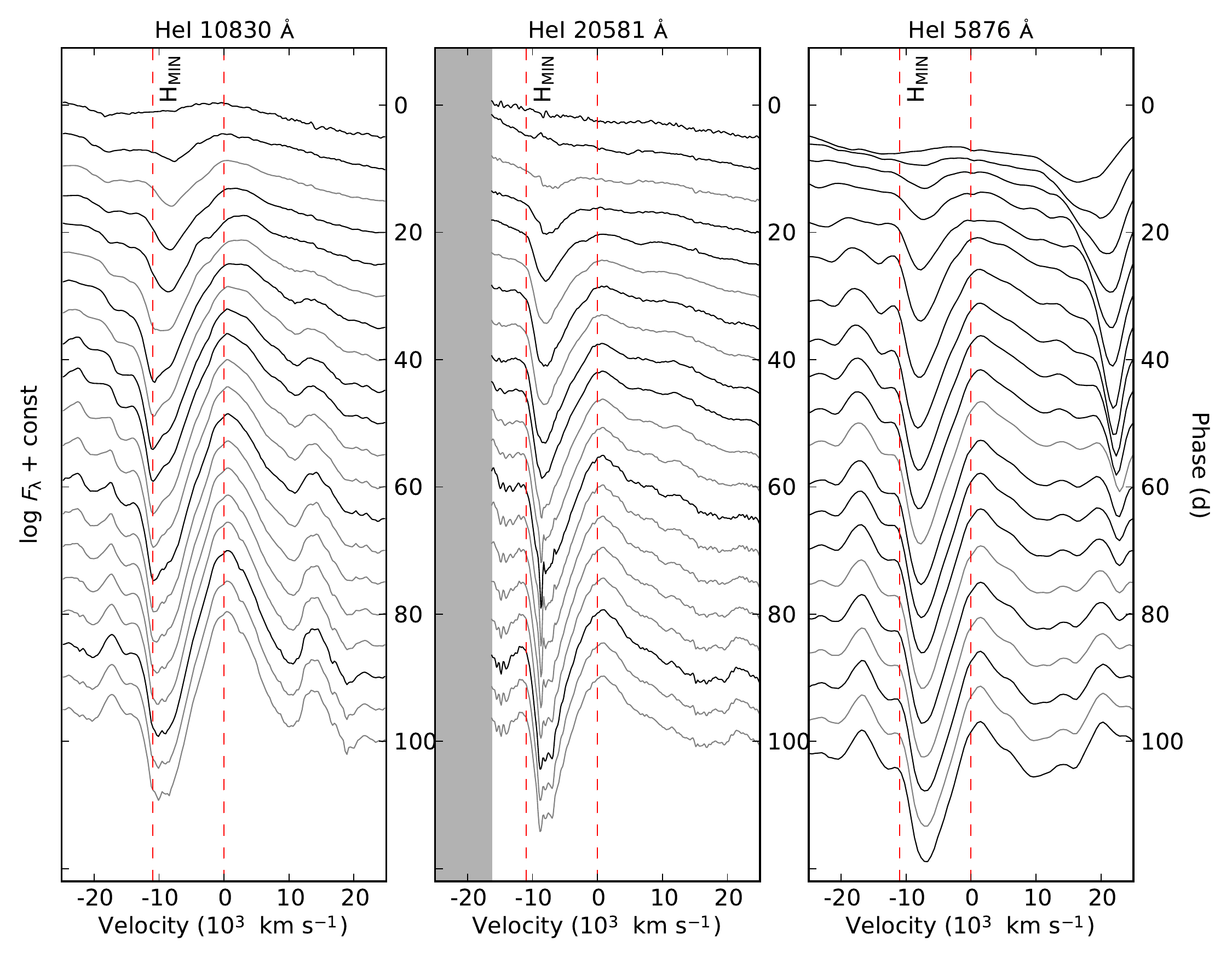}
\caption{Closeup of (interpolated) spectral evolution centred on the \ion{He}{i} 10830 \AA~(left panel), \ion{He}{i} 20581 \AA~(middle panel) and \ion{He}{i} 5876 \AA~(right panel) lines.}
\label{f_spec_evo_He}
\end{figure}

\begin{figure}[tb]
\includegraphics[width=0.48\textwidth,angle=0]{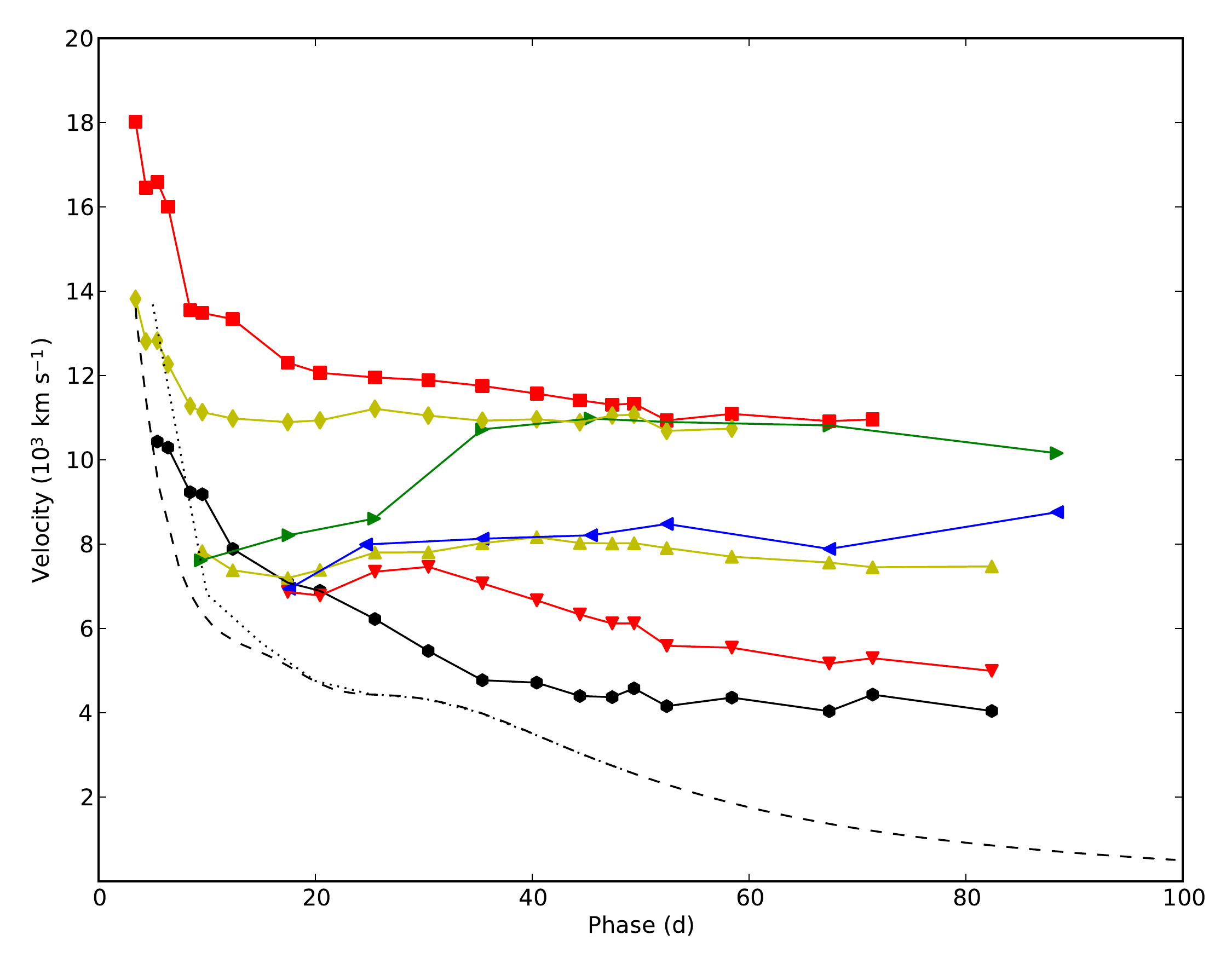}
\caption{Velocity evolution of the absorption minimum of the \ion{Fe}{ii} 5169 \AA~(black circles), \ion{He}{i} 5876 \AA~(yellow upward triangles), \ion{He}{i} 6678 \AA~(red downward triangles), \ion{He}{i} 10830 \AA~(green rightward triangles), \ion{He}{i} 20581 \AA~(blue leftward triangles), H$\alpha$~(red squares) and H$\beta$~(yellow diamonds) lines as automatically measured from the spectral sequence. For comparison we also show the velocity corresponding to the blackbody radius as determined from fits to the photometry (black dashed line) and as iteratively determined by the atmosphere modelling (black dotted line).}
\label{f_vel_evo_p_cygni}
\end{figure}

The \ion{He}{i} lines appears in the spectra between $\sim$10 (\ion{He}{i} 10830 and 5876 \AA) and $\sim$15 (\ion{He}{i} 6678, 7065 and 20581 \AA) days. Later on we see the 5016 and 17002 \AA~lines emerge as well. As mentioned the atmosphere modelling does not well reproduce the \ion{He}{i} lines but those identified here are present in the model spectrum with optical depths of 0.1$-$5 during the temperature peak between $\sim$10 and $\sim$20 days. Increasing the \ion{He}{i} excitation fraction to mimic the non-thermal excitation reproduce the \ion{He}{i} lines and their relative strengths reasonably well. At 100 days, all \ion{He}{i} lines, except \ion{He}{i} 17002 \AA, are present and identified by the NLTE modelling. Given the low ionization potential of \ion{Na}{i} and the high temperatures we find it unlikely that \ion{He}{i} 5876 is blended with \ion{Na}{i} 5890/5896 at early times. Using the atmosphere modelling we find a very low ion fraction of \ion{Na}{i} (<10$^{-7}$) and the optical depth for \ion{Na}{i} 5890/5896 to be negligible during the first 30 days. Using the NLTE modelling at 100 days we find emission to arise primarily from \ion{Na}{i} 5890/5896 and absorption to be a blend. \ion{He}{i} 10830 is likely to be blended with Paschen $\gamma$ at early times and \ion{He}{i} 5016 \AA~is likely to be blended with \ion{Fe}{i} 5018 \AA. Figure~\ref{f_spec_evo_He} shows a closeup of the evolution centred on the \ion{He}{i} lines. Helium absorption is mainly seen below the $\sim$11000 km s$^{-1}$ attributed to the interface between the helium core and the hydrogen rich envelope although \ion{He}{i} 10830 \AA~absorption extends beyond this velocity and also shows a narrow dip close to it between $\sim$30 and $\sim$60 days. We may speculate that this dip is caused by a denser shell of material close to the interface as was produced in explosion modelling of SN 1993J \citep[e.g.][]{Woo94}. Whereas the fading and disappearance of the hydrogen lines are driven by the decreasing density and temperature of the envelope the appearance and growth of the helium lines is likely to be more complex. \citetalias{Mar13} suggest that the helium lines appear because the photosphere reaches the helium core. However, Fig.~\ref{f_vel_evo_p_cygni} (see below) suggests that the photosphere reaches the helium core at 5-7 days whereas the helium lines appear later, at lower velocities, close to the region where we expect the continuum photosphere to be located and then move outwards in velocity until $\sim$40 days. This rather suggest the appearance and subsequent evolution to be driven by increasing non-thermal excitation due to the decreasing optical depth for the $\gamma$-rays. For the line optical depth at a given velocity (using the \citet{Sob57} approximation) we have $\tau \propto t^{-2} x_{l}$, where $x_{l}$ is the fraction of \ion{He}{i} in the lower state. As the temperature decreases after $\sim$10 days and the ion fraction of \ion{He}{i} is high according to the modelling, we would expect the line optical depth at a given velocity to decrease if non-thermal excitation was not important. Detailed modelling including a treatment of non-thermal excitation of the helium lines is needed to better understand the behaviour of the \ion{He}{i} lines.

Except for \ion{H}{i} and \ion{He}{i} we also identify lines from \ion{Ca}{ii}, \ion{Fe}{ii}, \ion{O}{i}, \ion{Mg}{i} and \ion{Na}{i} in the spectra. The \ion{Ca}{ii} 3934/3968 \AA~and 8498/8542/8662 \AA~lines are present throughout the evolution showing strong P-Cygni profiles and are identified by both the atmosphere and NLTE modelling whereas the [\ion{Ca}{ii}] 7291/7323 \AA~line is identified by the NLTE modelling at 100 days. The \ion{O}{i} 5577, 7774, 9263, 11300, and 13164 \AA~lines are all identified by the NLTE modelling at 100 days. The atmosphere modelling does not reproduce the \ion{O}{i} lines at early times but the \ion{O}{i} 7774 \AA~line seems to appear already at $\sim$25 days and the other lines between $\sim$30 and $\sim$50 days. The NLTE modelling also identifies the emerging [\ion{O}{i}] 6300/6364 lines at 100 days. The \ion{Mg}{i} 15040 \AA~line is identified by the NLTE modelling at 100 days and seem to emerge at $\sim$40 days. As mentioned above, we identify the \ion{Na}{i} 5890/5896 \AA~lines in emission and blended in absorption with the \ion{He}{i} 5876 \AA~line at 100 days using the NLTE modelling. In the region 4000$-$5500 \AA, we identify numerous \ion{Fe}{ii} lines using the atmosphere modelling, the most prominent being \ion{Fe}{ii} 4233, 4549, 4584, 4924, 5018, 5169 and 5317 \AA. These lines are present already at $\sim$5 days and most of them persist to at least 50 days. As mentioned in Sect. \ref{s_bol_lightcurve} and as can be seen in Fig.~\ref{f_sed_evo} there is a strong reduction of the flux bluewards of 5000 \AA~between $\sim$10 and $\sim$30 days. This well known behaviour, which is also reproduced by the modelling, is caused by an increased line opacity from a large number of metal ion (e.g. \ion{Fe}{ii} and \ion{Cr}{ii}) lines. This explains the initial redward trend in the $U$-$V$ and $B$-$V$ colours contrary to the blueward trend in $V$-$I$ and $V$-$K$ caused by the increasing temperature (see Fig.~\ref{f_colour_evo}). Judging from Fig.~\ref{f_sed_evo} the reduction of the flux is considerably reduced after $\sim$30 days.

Figure~\ref{f_vel_evo_p_cygni} shows the evolution of the absorption minimum for a number of lines as determined from the spectral sequence. These were measured by a simple automatic centring algorithm where the spectra were first smoothed down to 500 km s$^{-1}$ and the absorption minimum then traced through the interpolated spectral sequence and evaluated at the dates of observation. We also show the velocity corresponding to the blackbody radius as determined from fits to the photometry and as iteratively determined by the atmosphere modelling. Because of backscattering, the model blackbody radius is larger than the fitted. It is reasonable to expect that the photosphere is located somewhere between the blackbody surface and the region where the line with the lowest velocity is formed. This line is the \ion{Fe}{ii} 5169 \AA~line which was used in \citetalias{Ber12} to estimate the photospheric velocities. \citet{Des05} have used NLTE modelling to show that the absorption minimum of the \ion{Fe}{ii} 5169 \AA~line is a good estimator for the photospheric velocity in Type IIP SNe but it is not clear that this also apply to Type IIb SNe. Thus we have to consider the possibility that the photospheric velocities could be overestimated with up to 50 percent and in section \ref{s_error_b12} we will discuss how such an error would effect the results in \citetalias{Ber12}.

\section{Comparison to other SNe}
\label{s_sn_comp}

In this section we compare the observations of SN 2011dh to the well observed Type IIb SNe 1993J and 2008ax. In order to do this we need to estimate their distance and extinction. This will be done without assuming similarity among the SNe and in analogy with SN 2011dh we will use high-resolution spectroscopy of the \ion{Na}{i} D and \ion{K}{i} 7699 \AA~interstellar absorption lines to estimate the extinction. In the end of the section we will investigate what difference an assumption of similarity among the SNe will make.

\subsection{SN 1993J}
\label{s_sn_1993J}

SN 1993J which occurred in M81 is one of the best observed SNe ever and the nature of this SN and its progenitor star is quite well understood. \citet{Shi94} and \citet{Woo94} used hydrodynamical modelling to show that a progenitor star with an initial mass of 12$-$15 $M_{\odot}$ with an extended (not specified) but low mass (0.2$-$0.9 $M_{\odot}$) hydrogen envelope reproduces the observed bolometric lightcurve. This was confirmed by the more detailed modelling of \citet{Bli98}. Progenitor observations were presented in \citet{Mau04} while \citet{Sta09} used stellar evolutionary models to show that a progenitor star with an initial mass of 15$-$17 $M_{\odot}$ with an extended but low mass hydrogen envelope, stripped through mass transfer to a companion star, reproduces the observed progenitor luminosity and effective temperature. Photometric and spectroscopic data for SN 1993J were taken from \citet{Lew94}, \citet{Ric96}, \citet{Mat02}, \citet{Wad97} and IAU circulars. 

The distance to M81 is well constrained by Cepheid measurements, the mean and standard deviation of all such measurements listed in the NASA/IPAC Extragalactic Database (NED) being 3.62$\pm{0.22}$ Mpc, which we will adopt. The extinction within the Milky Way as given by the \citetalias{Sch98} extinction maps recalibrated by \citetalias{Sch11} is $E$($B$-$V$)$_\mathrm{MW}$=0.07 mag. \citet{Ric94} discuss the extinction in some detail and suggest a total $E$($B$-$V$)$_\mathrm{T}$ between 0.08 and 0.32 mag. High-resolution spectroscopy of the \ion{Na}{i} D lines was presented in \citet{Bow94}. Given the rough similarity between M81 and the Milky Way we will use the \citetalias{Mun97} and \citetalias{Poz12} relations to estimate the extinction within M81. \citet{Bow94} resolve a system of components near the M81 recession velocity and another one near zero velocity. There is also a third system which the authors attribute to extragalactic dust in the M81/M82 interacting system. The individual components of all three systems are quite heavily blended. As it is not clear whether the third system belongs to the Milky Way or M81, we calculate the extinction for all the three systems with the \citetalias{Mun97} and \citetalias{Poz12} relations and sum to get estimates of the total extinction. The \citetalias{Mun97} relation gives $E$($B$-$V$)$_\mathrm{T}$=0.28 mag and the \citetalias{Poz12} relations $E$($B$-$V$)$_\mathrm{T}$=0.17 mag (on average). Given that each system clearly consists of multiple components the \citetalias{Mun97} relation rather provides an upper limit (see discussion in \citetalias{Mun97}) and we will adopt the lower value given by the \citetalias{Poz12} relations. Adopting the higher value given by the \citetalias{Mun97} relation and the extinction within the Milky Way as upper and lower error limits we then get $E$($B$-$V$)$_\mathrm{T}$=0.17$^{+0.11}_{-0.10}$ mag.

\subsection{SN 2008ax}
\label{s_sn_2008ax}

SN 2008ax is another well observed Type IIb SN but the nature of this SN and its progenitor star is not as well understood as for SN 1993J. \citet{Tsv09} used the hydrodynamical code STELLA \citep{Bli98} to show that a progenitor star with an initial mass of 13 $M_{\odot}$ with an extended (600 $R_{\odot}$) and low mass (not specified) hydrogen envelope well reproduces the $UBVRI$ lightcurves except for the first few days. Progenitor observations were presented in \citet{Cro08} but the conclusions about the nature of the progenitor star were not clear. Photometric and spectroscopic data for SN 2008ax were taken from \citet{Pas08}, \citet{Rom09}, \citet{Tsv09}, \citet{Tau11} and \citet[hereafter \citetalias{Cho11}]{Cho11}.

The distance to the host galaxy NGC 4490 is not very well known. We have found only three measurements in the literature \citep{Tul88,Ter02,The07}. Taking the median and standard deviation of these and the Virgo, Great Attractor and Shapley corrected kinematic distance as given by NED we get 9.38$\pm{0.85}$ Mpc which we will adopt. The extinction within the Milky Way as given by the \citetalias{Sch98} extinction maps recalibrated by \citetalias{Sch11} is $E$($B$-$V$)$_\mathrm{MW}$=0.02 mag. High resolution spectroscopy of the \ion{Na}{i} D and \ion{K}{i} 7699 \AA~lines were presented in \citetalias{Cho11}. The host galaxy NGC 4490 is a quite irregular galaxy so it is not clear if relations calibrated to the Milky Way are applicable. However, as we have no alternative, we will use the \citetalias{Mun97} relations to estimate the extinction within NGC 4490. The \citetalias{Cho11} spectra show blended multiple components of the \ion{Na}{i} D$_2$ line most of which are clearly saturated. We measure the total equivalent width to 1.0 \AA~which using the linear (unsaturated) part of the \citetalias{Mun97} relation corresponds to a lower limit of $E$($B$-$V$)$_\mathrm{H}$$>$0.25 mag. As the \ion{Na}{i} D$_2$ lines are saturated we cannot use these to derive a useful upper limit. \citetalias{Cho11} measures the total equivalent width of the \ion{K}{i} 7699 \AA~line components to 0.142 \AA~which using the corresponding \citetalias{Mun97} relation gives $E$($B$-$V$)$_\mathrm{H}$=0.54 mag. Adding the extinction within the Milky way and adopting the lower limit from the \citetalias{Mun97} \ion{Na}{i} D$_2$ relation and the extinction corresponding to the bluest SN colours allowed for a blackbody (Sect. \ref{s_extinction_rev}) as the lower and upper error limits we then get $E$($B$-$V$)$_\mathrm{T}$=0.56$^{+0.14}_{-0.29}$ mag.

\subsection{Comparison}
\label{s_sn_comp_comp}

The absolute magnitudes in each band were calculated using cubic spline fits as described in Sect. \ref{s_obs_image_results} and, when neccesary, extrapolated assuming constant colour.

\begin{figure}[tb]
\includegraphics[width=0.48\textwidth,angle=0]{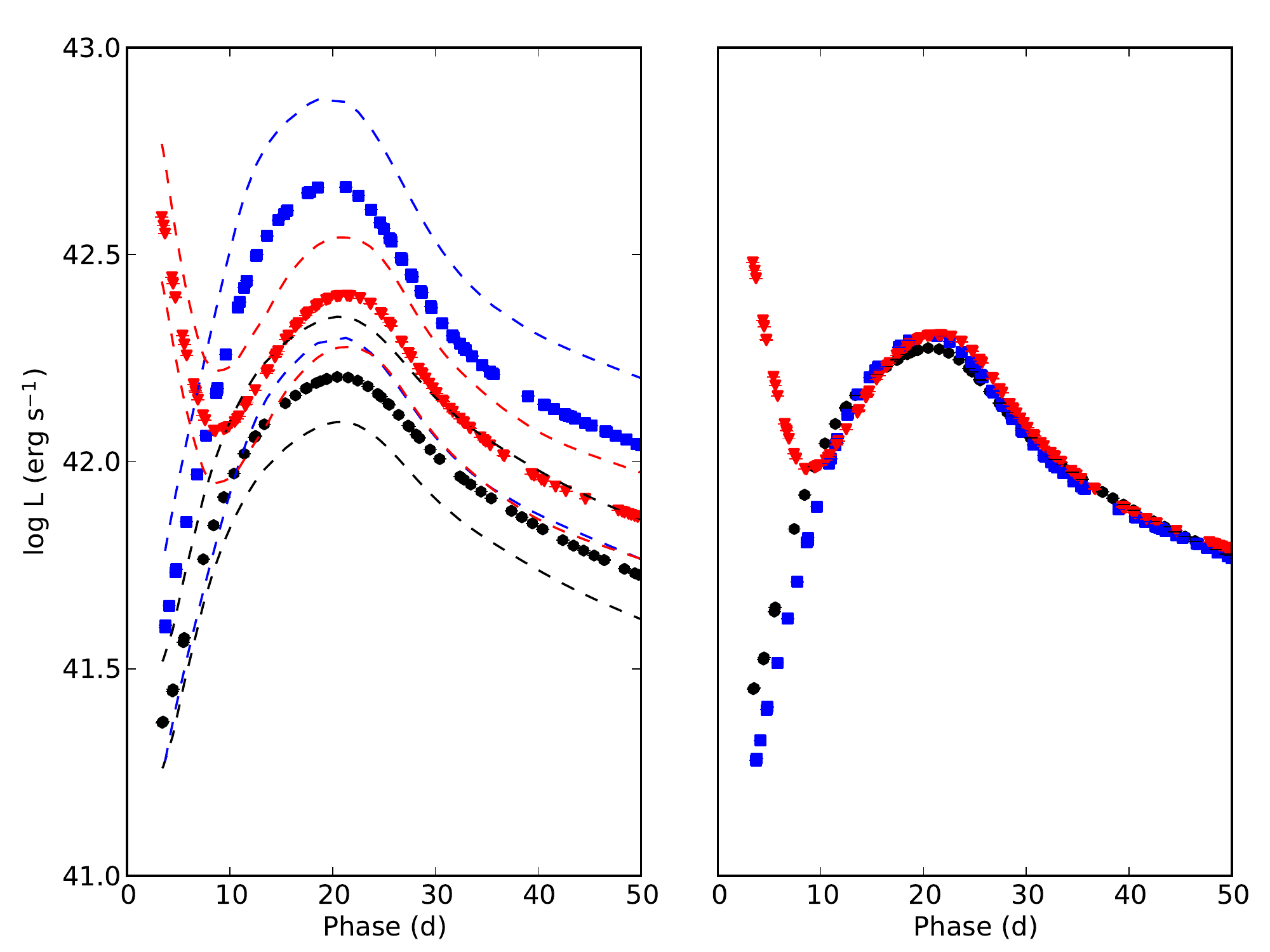}
\caption{Pseudo-bolometric $U$ to $K$ lightcurve for SN 2011dh (black circles) as compared to SNe 1993J (red triangles) and 2008ax (blue squares) for the adopted extinctions (left panel) and for a revised scenario where we have set $E$($B$-$V$)$_\mathrm{T}$ to 0.14, 0.09 and 0.27 mag for SNe 2011dh, 1993J and 2008ax respectively (right panel). In the left panel we also show the systematic error arising from the distance and extinction (dashed lines).}
\label{f_UK_bol_comp_comb}
\end{figure}

The left panel of Fig.~\ref{f_UK_bol_comp_comb} shows the pseudo-bolometric $U$ to $K$ (3000-24000 \AA) lightcurves of SNe 2011dh, 1993J and 2008ax as calculated with the photometric method (Sect.~\ref{s_bol_lightcurve}). The absolute magnitudes have been calculated using cubic spline fits as described in Sect.~\ref{s_obs_image_results} and extrapolated assuming constant colour. Except for the first few days the shape is similar and they all show the characteristics common to Type I and IIb SNe lightcurves (Sect.~\ref{s_physics_sn_IIb}). As shown in \citetalias{Ber12} the differences during the first few days could be explained by differences in the radius and mass of the hydrogen envelope. Given the adopted distances and extinctions SN 2011dh is fainter than SN 1993J which, in turn, is fainter than SN 2008ax. The peak luminosity occurs at similar times but the peak-to-tail luminosity ratio for SN 2011dh is smaller than for SN 1993J which, in turn, is smaller than for SN 2008ax.

\begin{figure}[tb]
\includegraphics[width=0.48\textwidth,angle=0]{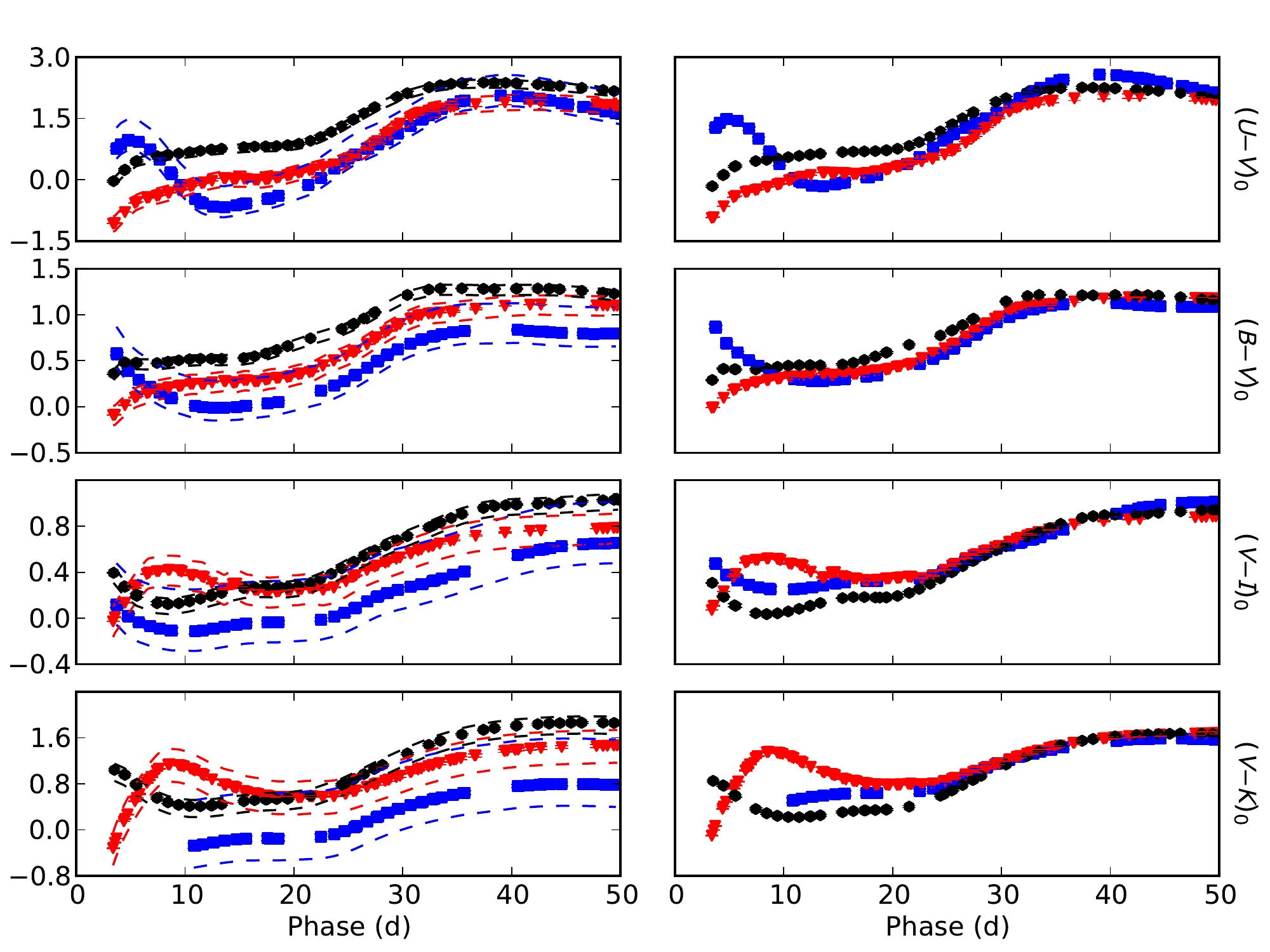}
\caption{Colour evolution of SN 2011dh (black circles) as compared to SNe 1993J (red triangles) and 2008ax (blue squares) for the adopted extinctions (left panel) and for a revised scenario where we have set $E$($B$-$V$)$_\mathrm{T}$ to 0.14, 0.09 and 0.27 mag for SNe 2011dh, 1993J and 2008ax respectively (right panel). In the left panel we also show the systematic error arising from the extinction (dashed lines).}
\label{f_colour_evo_comp_comb}
\end{figure}

The left panel of Fig.~\ref{f_colour_evo_comp_comb} shows the colour evolution for the three SNe. The absolute magnitudes have been calculated using cubic spline fits as described in Sect.~\ref{s_obs_image_results}. As for the lightcurves, the shape is quite different for the first few days which could again be explained by differences in the radius and mass of the hydrogen envelope. The shape of the subsequent evolution is quite similar with a blueward trend in the $V$-$I$ and $V$-$K$ colours (corresponding to increasing temperature) during the rise to peak luminosity and then a redward trend in all colours (corresponding to decreasing temperature) to a colour maximum at 40$-$50 days and a subsequent slow blueward trend. Given the adopted extinction, SN 2011dh is redder than SN 1993J which, in turn, is redder than SN 2008ax. 

\begin{figure}[tb]
\includegraphics[width=0.48\textwidth,angle=0]{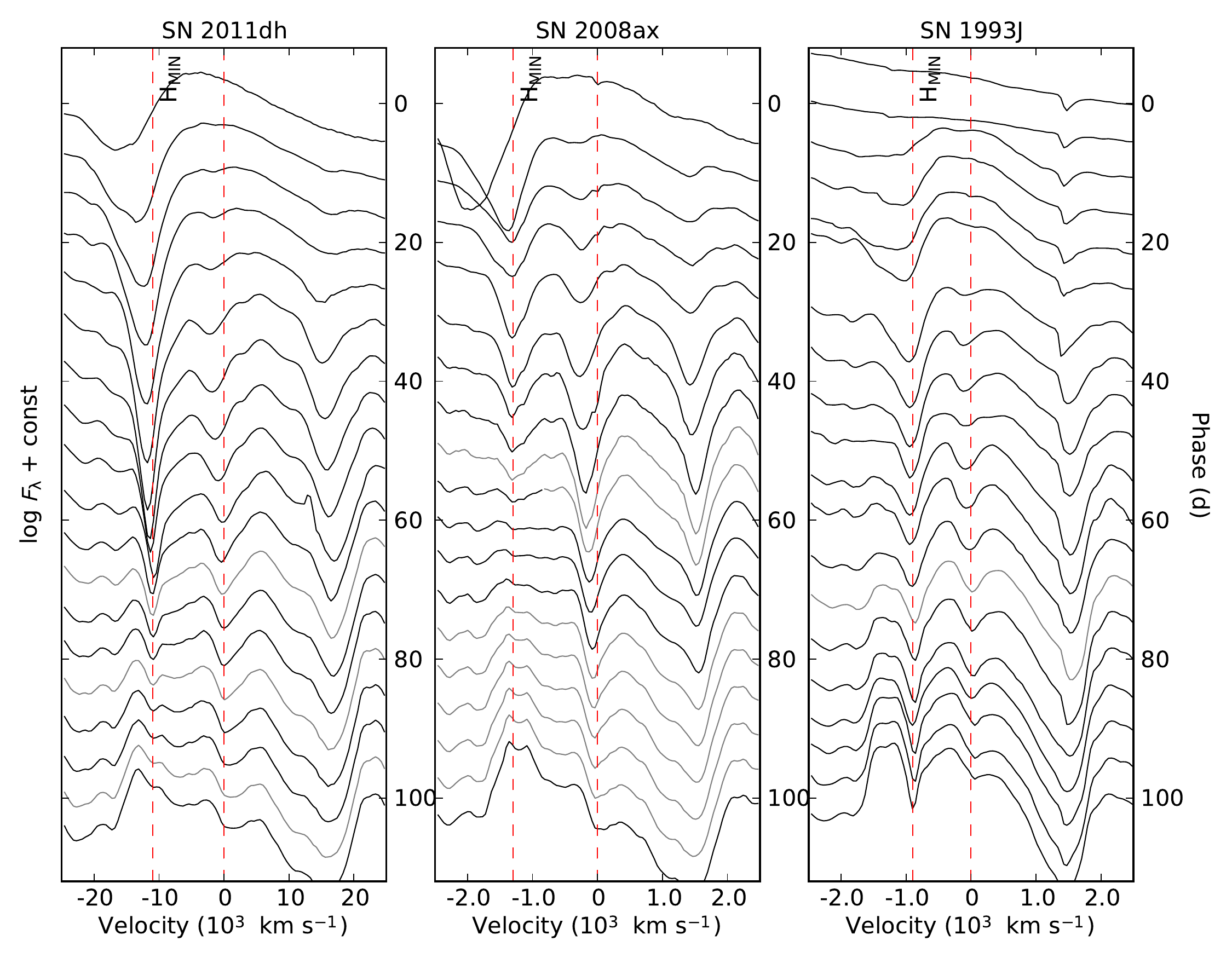}
\caption{The (interpolated) evolution of the H$\alpha$ line for SN 2011dh (left panel) as compared to SNe 2008ax (middle panel) and 1993J (right panel). All panels in this and the following figure show the minimum velocity for the H$\alpha$ absorption minimum (marked H$_{\mathrm{MIN}}$) interpreted as the interface between the helium core and hydrogen envelope.}
\label{f_spec_evo_Ha_comp}
\end{figure}

\begin{figure}[tb]
\includegraphics[width=0.48\textwidth,angle=0]{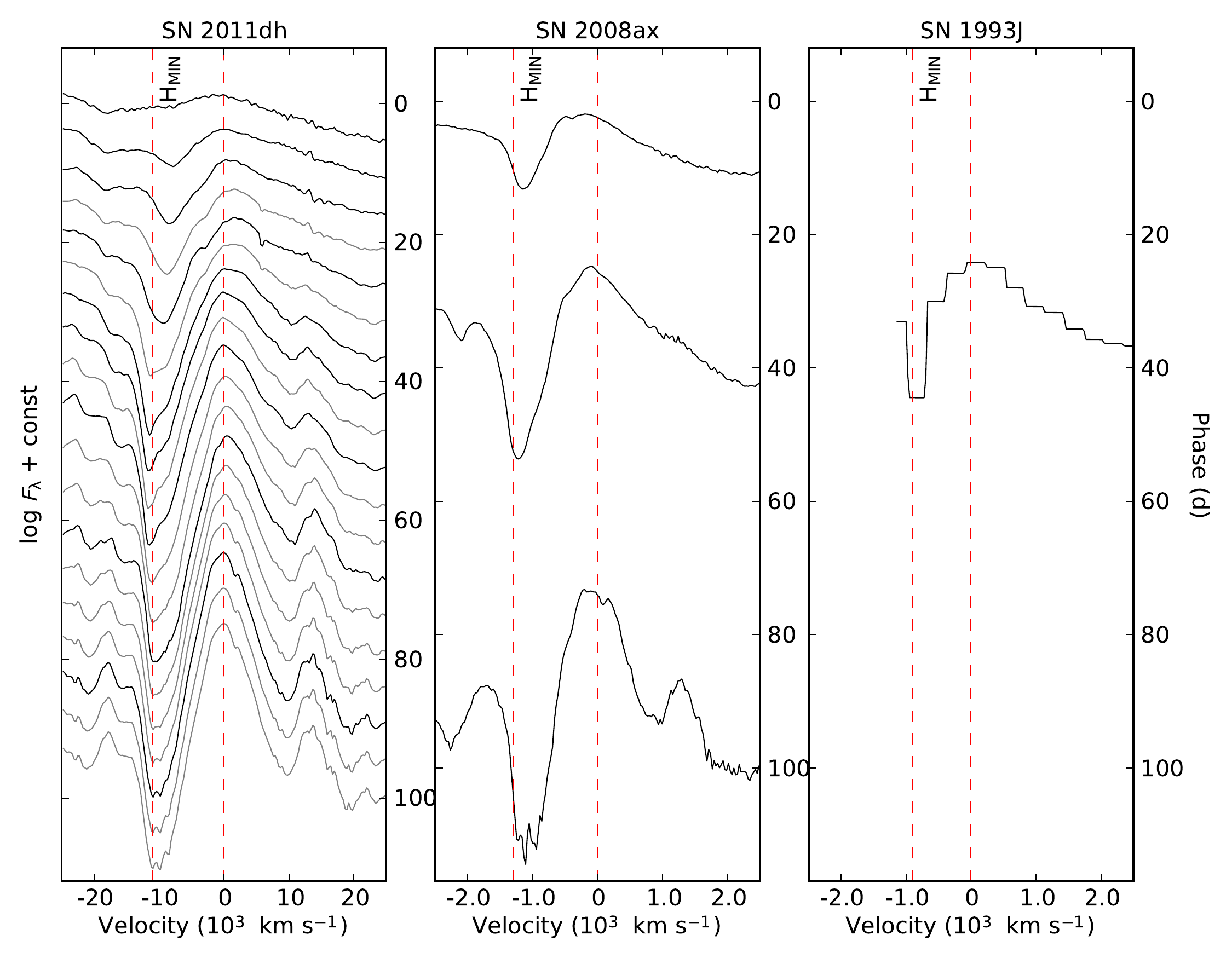}
\caption{The (interpolated) evolution of the \ion{He}{i} 10830 \AA~line for SN 2011dh (left panel) as compared to SNe 2008ax (middle panel) and 1993J (right panel). Given the sparse data available for SNe 1993J and 2008ax we show observed spectra for these SNe.}
\label{f_spec_evo_He_10833_comp}
\end{figure}

\begin{figure}[tb]
\includegraphics[width=0.48\textwidth,angle=0]{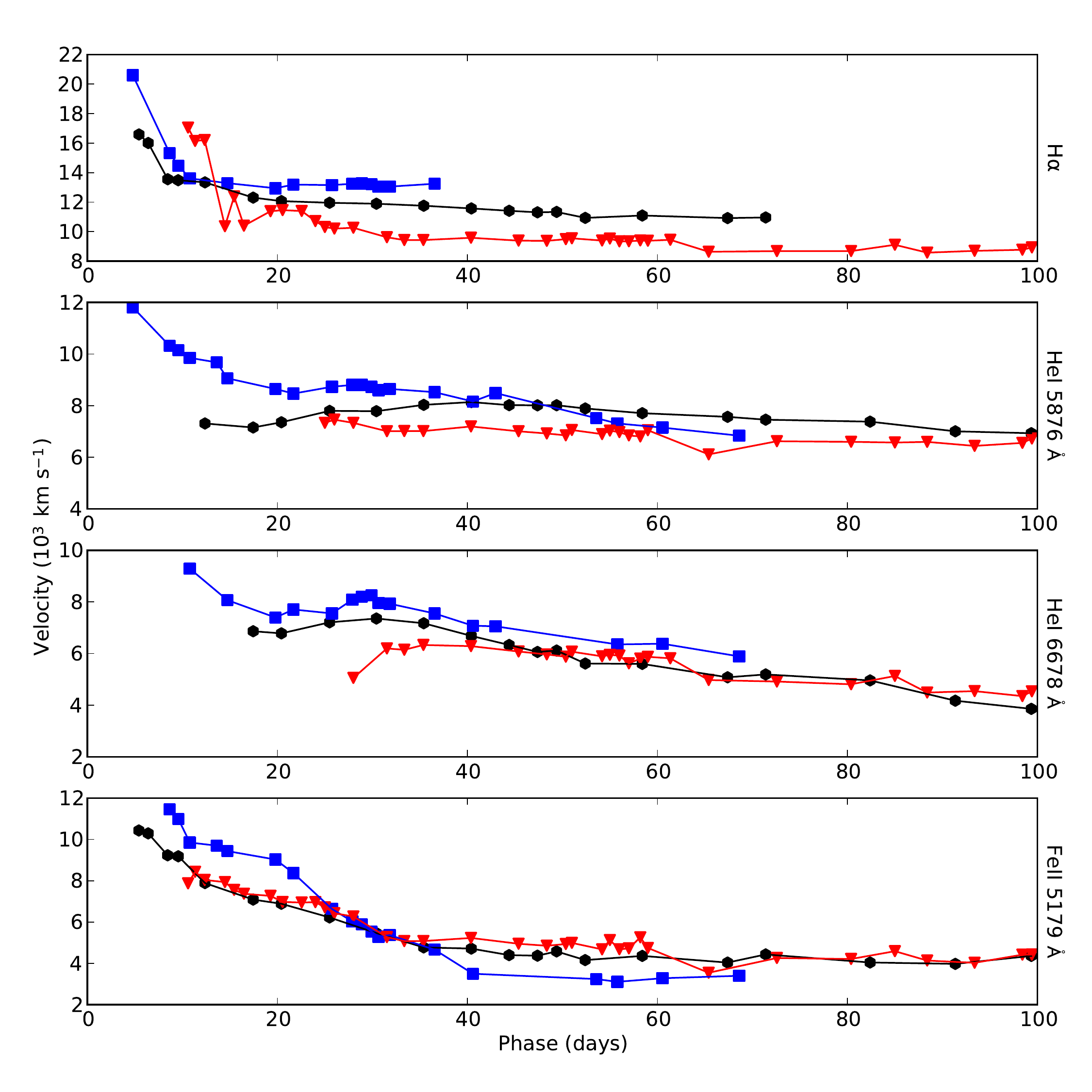}
\caption{Velocity evolution of the absorption minimum for the H$\alpha$ (upper panel), \ion{He}{i} 5876 \AA~(upper middle panel), \ion{He}{i} 6678 \AA~(lower middle panel) and  \ion{Fe}{ii} 5169 \AA~(lower panel) lines  for SNe 2011dh (black circles), 2008ax (blue squares) and 1993J (red triangles) measured as described in Sect. \ref{s_analysis_spec}.}
\label{f_vel_evo_p_cygni_comp}
\end{figure}

In Figures \ref{f_spec_evo_Ha_comp} and \ref{f_spec_evo_He_10833_comp} we show closeups of the spectral evolution centred on the H$\alpha$ and the \ion{He}{i} 10830 \AA~lines. The minimum velocity for the H$\alpha$ absorption minimum has been marked and occurs at $\sim$9000, $\sim$11000 and $\sim$13000 km s$^{-1}$ for SNe 1993J, 2011dh and 2008ax respectively. As discussed in Sect.~\ref{s_analysis_spec} this velocity likely corresponds to the interface between the helium core and the hydrogen envelope for SN 2011dh. The H$\alpha$ line disappears at $\sim$50 days for SN 2008ax, at $\sim$80 days for SN 2011dh and is still strong at 100 days for SN 1993J. Figure~\ref{f_vel_evo_p_cygni_comp} shows the evolution of the absorption minimum for the \ion{Fe}{ii} 5169 \AA, \ion{He}{i} 5876 and 6678 \AA~and H$\alpha$ lines measured as described in Sect. \ref{s_analysis_spec}. Interpreting the \ion{Fe}{ii} 5169 \AA~absorption minimum as the photosphere and the minimum velocity for the H$\alpha$ absorption minimum as the interface between the helium core and the hydrogen envelope the photosphere reaches the helium core at $\lesssim$10, $\sim$5 and $\lesssim$10 days for SNe 1993J, 2011dh and 2008ax respectively. The helium lines appear at $\sim$20, $\sim$10 and $\sim$5 days for SNe 1993J, 2011dh and 2008ax respectively, at lower velocities close to the region where we expect the continuum photosphere to be located. The initial evolution is different among the SNe but after $\sim$30 days the helium lines have increased in strength, moved outward as compared to the photosphere and show a quite similar evolution for all three SNe. The evolution of the \ion{Fe}{ii} 5169 \AA~line is very similar for SNe 1993J and 2011dh but a bit different for SN 2008ax. In general, lines originating closer to the photosphere seem to have similar velocities for the three SNe whereas lines originating further out in the ejecta seem to have progressively higher velocities for SNe 1993J, 2011dh and 2008ax respectively.

The differences in peak and tail luminosities suggest differences in the mass of ejected \element[ ][56]{Ni} (Sect.~\ref{s_physics_sn_IIb}). The differences in peak-to-tail luminosity ratios suggest differences in the ejecta mass, explosion energy and/or distribution of \element[ ][56]{Ni} (Sect.~\ref{s_physics_sn_IIb}). However, as seen in the left panels of Figures \ref{f_UK_bol_comp_comb} and \ref{f_colour_evo_comp_comb} the systematic errors in the luminosity and colour arising from the distance and extinction is large so similarity among the SNe cannot be excluded. \citetalias{Mar13} find both the luminosities and the colours to be similar, mainly due to differences in the adopted distances and extinctions. The similar velocities of lines originating closer to the photosphere and the times at which peak luminosity occurs, both of which are independent of the distance and extinction, suggests similar ejecta masses and explosion energies (Sect.~\ref{s_physics_sn_IIb}). Although the differences in the bolometric lightcurves could possibly be explained by differences in the mass and distribution of ejected \element[ ][56]{Ni} this is not fully satisfactory as the mass of ejected \element[ ][56]{Ni} is known from observations to be correlated with initial mass and expansion velocity \citep{Fra11,Mag12}. In all the observed characteristic of the SNe does not seem entirely consistent and we have to consider the possibility that the adopted distances and extinctions are in error.

Interestingly enough, it is possible to revise the extinctions alone, within the adopted error bars, in such a way that it brings the colour evolution, the bolometric luminosities and the peak-to-tail luminosity ratios in good agreement. This is shown in the right panels of Figures \ref{f_UK_bol_comp_comb} and \ref{f_colour_evo_comp_comb} where we have set $E$($B$-$V$)$_\mathrm{T}$ to 0.14, 0.09 and 0.27 mag for SNe 2011dh, 1993J and 2008ax respectively. Intrinsic differences among the SNe can not be excluded and the arguments used are only suggestive so we can not make a definite conclusion. It is clear, however, that a scenario where all three SNe have similar ejecta masses, explosion energies and ejected masses of \element[ ][56]{Ni} is possible. As shown in \citetalias{Ber12} the differences in the early evolution and the velocities of lines originating further out in the ejecta could be explained by differences in the mass and radius of the hydrogen envelope. The progressively higher minimum velocities for the H$\alpha$ absorption minimum, if interpreted as the interface between the helium core and the hydrogen envelope, would naively suggest progressively lower masses of this envelope for SNe 1993J, 2011dh and 2008ax respectively. Such a conclusion is supported by early photometric evolution, the strength and persistence of the H$\alpha$ line, the hydrodynamical modelling of SNe 1993J and 2011dh in \citetalias{Ber12}, the spectral modelling of SN 2011dh in this paper and by \citetalias{Arc11} and the spectral modelling of SN 2008ax by \citet{Mau10}. \citet{Mar13} reach a similar conclusion based on the progressively later times at which the helium lines appears although we do not find their physical argument convincing (Sect.~\ref{s_analysis_spec}).

\section{Discussion}
\label{s_discussion}

In Sect. \ref{s_extinction_rev} we revisit the issue of extinction and discuss constraints from the SN itself and the comparisons to SNe 1993J and 2008ax made in Sect.~\ref{s_sn_comp_comp}. In Sect.~\ref{s_physics_sn_IIb} we discuss the physics of Type IIb lightcurves as understood from approximate models, in particular in relation to the hydrodynamical modelling made in \citetalias{Ber12}. In Sect.~\ref{s_error_b12} we discuss the sensitivity of the SN and progenitor parameters derived in \citetalias{Ber12} to errors in the distance, extinction and photospheric velocity and also revise these parameters to agree with the distance and extinction adopted in this paper. In Sect.~\ref{s_prog_dis} we discuss the results on the disappearance of the progenitor star and what consequences this have for the results in \citetalias{Mau11} and \citetalias{Ber12} and our understanding of this star and Type IIb progenitors in general. Finally, in Sect.~\ref{s_45_micron_excess}, we discuss the excess in the Spitzer 4.5 $\mu$m band and possible explanations.

\subsection{Extinction revisited}
\label{s_extinction_rev}

In Sect.~\ref{s_extinction} we discussed different estimates of the extinction for SN 2011dh. Most estimates suggested a low extinction and we adopted $E$($B$-$V$)=0.07$^{+0.07}_{-0.04}$ mag as estimated from the equivalent widths of the \ion{Na}{i} D lines. The near simultaneous $V$ and $R$ band observations from day 1 presented in \citetalias{Arc11} and \citetalias{Tsv12} corresponds to an intrinsic $V$-$R$ colour of about $-$0.2 mag for the adopted extinction. The bluest $V$-$R$ colour allowed for a blackbody, which can be calculated from the Rayleigh-Jeans law, is $-$0.16 mag so this suggests a very high temperature. In higher extinction scenarios this colour would be even bluer and, even taking measurement and calibration errors into account, in conflict with the bluest $V$-$R$ colour allowed for a blackbody. Figure~\ref{f_bb_T_evo} shows the evolution of the blackbody temperature for two higher extinction scenarios where we have increased $E$($B$-$V$)$_\mathrm{T}$ in $\sim$0.1 steps to 0.2 and 0.3 mag. As seen the blackbody temperature would become quite high between 10 and 20 days and we would expect lines from low ionization potential ions such as \ion{Ca}{ii} and \ion{Fe}{ii} to be quite sensitive to this. As shown in Fig.~\ref{f_spec_model_comp}, the SN atmosphere code described in Sect.~\ref{s_analysis_spec} can neither reproduce the \ion{Ca}{ii} 8498/8542/8662 \AA~lines, nor the the \ion{Fe}{ii} lines, between 10 and 20 days for these higher extinction scenarios. Even though NLTE effects may change the ion fractions, this again suggests a low extinction scenario for SN 2011dh. Comparisons to SNe 2008ax and 1993J provides another source of information. As discussed in Sect.~\ref{s_sn_comp} an assumption of similarity in luminosity and colour among the SNe requires a revision of the extinctions adopted in this paper and suggest a revision of the extinction for SN 2011dh towards the upper error bar. However, as pointed out, intrinsic differences among the SNe cannot be excluded and as such a revision would be within our error bars we do not find this argument sufficient to revise our adopted value E(B-V)=0.07$^{+0.07}_{-0.04}$ mag.

\subsection{Physics of Type IIb SNe lightcurves}
\label{s_physics_sn_IIb}

The bolometric lightcurves of SN 2011dh and other Type IIb SNe can be divided in two distinct phases depending on the energy source powering the lightcurve. The first phase is powered by the thermal energy deposited in the ejecta by the explosion. The second phase is powered by the energy deposited in the ejecta by the $\gamma$-rays emitted in the radioactive decay chain of \element[ ][56]{Ni}. In \citetalias {Ber12} we used the \element[ ][56]{Ni} powered phase to estimate the ejecta mass, explosion energy and ejected mass of \element[ ][56]{Ni} whereas the explosion energy powered phase was used to estimate the radius of the progenitor star.

The explosion energy powered phase ends at $\sim$3 days when our observations begin but $V$ $R$ and $g$ band data have been published in \citetalias{Arc11} and \citetalias{Tsv12}. These data are insufficient to construct a bolometric lightcurve but it is clear that this phase corresponds to a strong decline of the bolometric luminosity. In the \citetalias{Ber12} modelling roughly half the explosion energy is deposited as thermal energy in the core but most of this is lost before shock breakout due to expansion. Only a small fraction of the explosion energy is deposited as thermal energy in the envelope and it is the cooling of this, both by expansion and radiative diffusion, that gives rise to the strong decline of the bolometric luminosity. The shape and extent of the bolometric lightcurve in the explosion energy powered phase depends on the mass, radius, density profile and composition of the envelope and, as discussed in \citetalias{Ber12}, requires detailed hydrodynamical modelling.

The subsequent \element[ ][56]{Ni} powered phase is well covered by our data and the bolometric lightcurve (Fig.~\ref{f_UV_MIR_bol}) shows the characteristics common to all Type I and IIb SNe; a rise to peak luminosity followed by a decline phase and a subsequent tail phase with a roughly linear decline rate. These characteristics can be qualitatively understood by approximate models such as the ones by \citet{Arn82} or \citet{Ims92}. The rising phase is caused by radiative diffusion of the energy deposited in the ejecta by the $\gamma$-rays. The radioactive heating decreases with time and so does the diffusion time because the ejecta are expanding. As shown by \citet{Arn82} the luminosity peak is reached when the radioactive heating equals the cooling by radiative diffusion. During the subsequent decline phase the diffusion time continues to decrease until the SN reaches the tail phase where the diffusion time is negligible and the luminosity equals the radioactive heating (instant diffusion). The shape of the tail is not exactly linear but is modulated by a term determined by the decreasing optical depth for $\gamma$-rays as the ejecta continue to expand.

From approximate models the qualitative dependence of the bolometric lightcurve in the \element[ ][56]{Ni} powered phase on basic parameters as the explosion energy, ejecta mass and mass of ejected \element[ ][56]{Ni} can be understood. Increasing the explosion energy will increase the expansion velocities which will decrease the diffusion time for thermal radiation and the optical depth for $\gamma$-rays. Increasing the ejecta mass will have the opposite effect but, as the optical depth $\tau \propto (M^{2}/E)$ and the diffusion time $t_{\mathrm{d}} \propto (M^{3}/E)^{1/4}$ \citep{Arn82}, the bolometric lightcurve depends stronger on the ejecta mass than on the explosion energy. Either an increase of the explosion energy or a decrease of the ejecta mass will result in an earlier and more luminous peak of the bolometric lightcurve whereas the tail luminosity will be decreased. Increasing the mass of \element[ ][56]{Ni} will increase the radioactive heating and thus result in an overall increase of the luminosity and in fact corresponds to a pure scaling in the approximate models. The distribution of \element[ ][56]{Ni} also affects the lightcurve and if the \element[ ][56]{Ni} is distributed further out in the ejecta the lightcurve will rise faster to the peak because of the decreased diffusion time for thermal radiation and have a lower luminosity on the tail because of the decreased optical depth for $\gamma$-rays. As shown in figures 2, 4, 5 and 6 in \citetalias{Ber12} all these qualitative dependencies are well followed by the hydrodynamical models.

If the optical depth for $\gamma$-rays in the tail phase is high the shape of the bolometric lightcurve in the \element[][56]{Ni} powered phase depends exclusively on the diffusion time for thermal radiation, which determines the quantity $(M^{3}/E)$, and the ejecta mass and explosion energy become degenerate. In this case knowledge of the expansion velocity, which determines the quantity $(M/E)$, is needed to determine the SN parameters. However, as seen in Fig.~\ref{f_UV_MIR_bol_model_comp}, the optical depth for $\gamma$-rays becomes $\lesssim$1 at $\sim$40 days for SN 2011dh. The bolometric lightcurve in the tail phase then depends on the optical depth for $\gamma$-rays, which determines the quantity $(M^{2}/E)$, and provides the constraint needed to break the degeneracy. However, as the bolometric lightcurve also depends on the distribution of \element[ ][56]{Ni} the problem is not necessarily well-conditioned. In our experience, knowledge of the expansion velocity, which corresponds to the fitting of photospheric velocities in \citetalias{Ber12}, is needed to robustly determine the SN parameters.

\subsection{Error sensitivity and revisions of the \citetalias{Ber12} modelling}
\label{s_error_b12}

What was not discussed in \citetalias{Ber12} was the sensitivity of the results to errors in the adopted distance and extinction.  A change in the distance corresponds to a scaling of the bolometric lightcurve whereas a change in the extinction is more complicated as the change in luminosity also depends on the colour. However, as seen in Fig.~\ref{f_UV_MIR_bol}, the change in luminosity for SN 2011dh due to the combined errors in distance and extinction does not differ significantly from a scaling. As the adopted distance and extinction have been revised as compared to \citetalias{Ber12} we also need to investigate the effect of this change on the derived quantities.

\begin{figure}[tb]
\includegraphics[width=0.48\textwidth,angle=0]{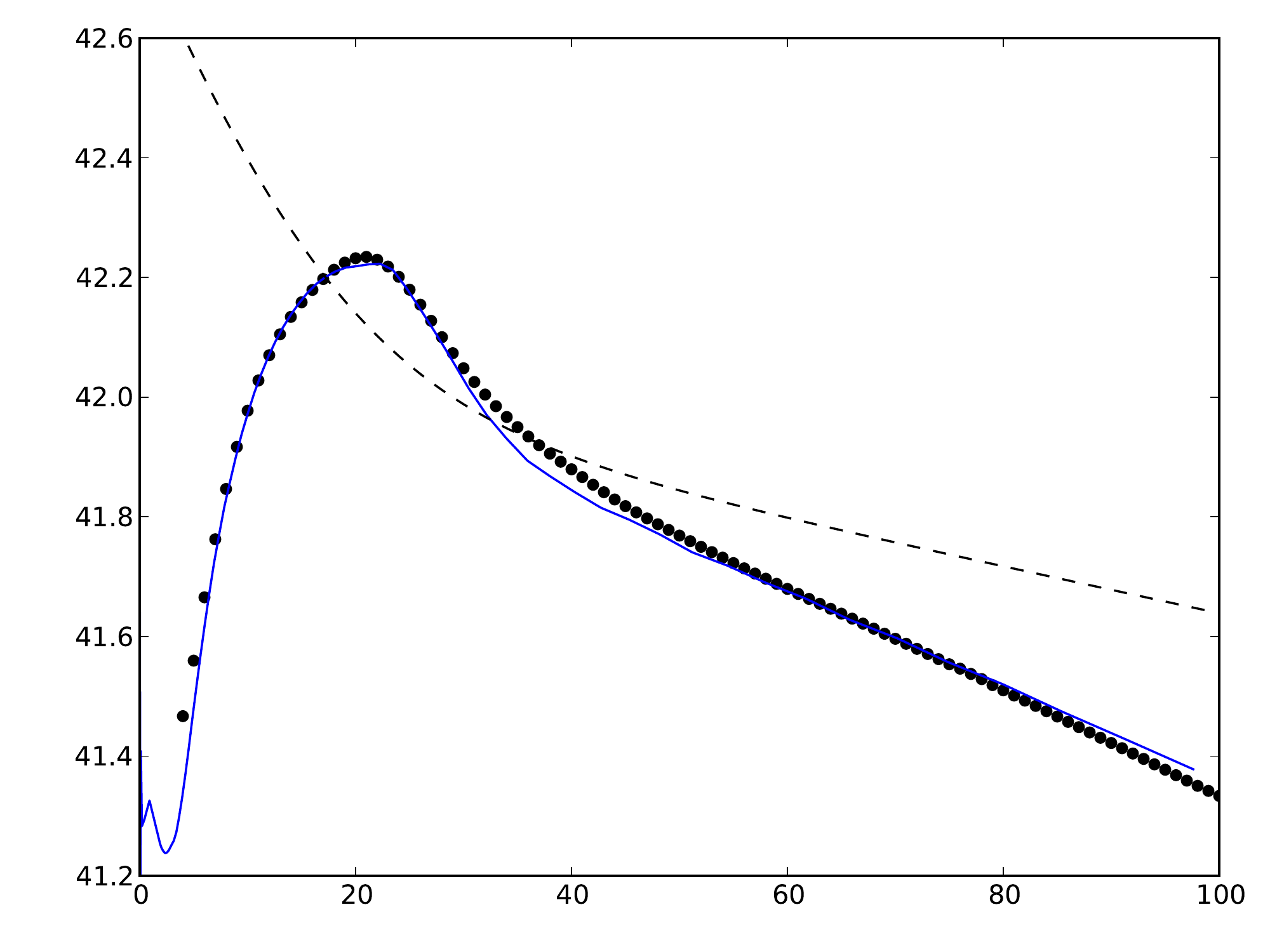}
\caption{Revised \citetalias{Ber12} He4 model bolometric lightcurve with the \element[ ][56]{Ni} mass increased to 0.075 M$_\odot$ (blue solid line) compared to the pseudo-bolometric UV to MIR lightcurve for SN 2011dh calculated with the spectroscopic method (black dots). For comparison we also show the total $\gamma$-ray luminosity corresponding to this amount of \element[ ][56]{Ni} (black dashed line).}
\label{f_UV_MIR_bol_model_comp}
\end{figure}

In the \element[ ][56]{Ni} powered phase, according to approximate models, the luminosity is proportional to the mass of ejected \element[ ][56]{Ni} (Sect.~\ref{s_physics_sn_IIb}). Therefore, ignoring possible degeneracy among the parameters, we expect the derived ejecta mass and explosion energy to be insensitive to the errors in the distance and extinction and the error in the derived mass of ejected \element[ ][56]{Ni} to be similar to the error in the luminosity. We have re-run the He4 model with the \element[ ][56]{Ni} mass increased to 0.075 M$_\odot$ to account for the revisions in the adopted distance and extinction. The model bolometric lightcurve is shown in Fig.~\ref{f_UV_MIR_bol_model_comp} and well reproduces the bolometric lightcurve presented in this paper. Estimating the errors arising from the distance and extinction as described the revised mass of ejected \element[ ][56]{Ni} becomes 0.05-0.10 M$_{\odot}$ whereas the ejecta mass and explosion energy remains the same as in \citetalias{Ber12}, 1.8-2.5 M$_{\odot}$ and 0.6-1.0$\times$10$^{51}$ erg respectively.

In the explosion energy powered phase, according to approximate models, the luminosity is proportional to the radius. However, the lightcurve in this phase is not well described by approximate models and detailed hydrodynamical modelling is needed (Sect.~\ref{s_physics_sn_IIb}). As discussed in \citetalias{Ber12} the sensitivity of the estimated radius to changes in the luminosity is modest. Examining the range of envelope models consistent with the \citetalias{Arc11} $g$ band observations, the distance and extinction adopted in this paper and the systematic error arising from these we find a revised progenitor radius of 200-300 R$_{\odot}$.

As discussed in Sect.~\ref{s_analysis_spec} the region where the \ion{Fe}{ii} 5169 \AA \ line is formed rather provides an upper limit for the location of the photosphere and the photospheric velocities used in the hydrodynamical modelling might therefore be overestimated. The dependence of the derived quantities on the photospheric velocity is complicated and a full scan of the model parameter space is probably needed to make a quantitative estimate. This is a potential problem and further work is needed to well constrain the photospheric velocities.

\subsection{Disappearance of the proposed progenitor star}
\label{s_prog_dis}

In \citet{Erg13} we presented $B$, $V$ and $r$ observations of the SN site obtained on Jan 20 2013 ($V$ and $r$) and Mar 19 2013 ($B$), 601 and 659 days past explosion respectively. An additional set of $B$, $V$ and $r$ band observations was obtained on Apr 14 2013, May 15 2013 and Jun 1 2013, 685, 715 and 732 days past explosion. In Appendix \ref{a_prog_obs} we give the details on these observations and the photometric measurements and calibration. Subtraction of pre-explosion images shows that the flux from the yellow supergiant proposed as the progenitor by \citetalias{Mau11} have been reduced with at least 74$\pm{9}$, 73$\pm{4}$ and 77$\pm{4}$ percent in the $B$, $V$ and $r$ bands respectively. The HST observations obtained on Mar 2 2013 and presented by \citetalias{Dyk13b} corresponds to a reduction of the flux of 71$\pm{1}$ and 70$\pm{1}$ in the $F555W$ and $F814W$ bands respectively. \citet{Szc12} find the progenitor to be variable at the five percent level so variability of the star is unlikely to explain the flux reduction. We find a 0.57, 0.76 and 0.64 mag decline corresponding to a decline rate of 0.0073, 0.0090 and 0.0053 mag day$^{-1}$ in the $B$, $V$ and $r$ bands respectively between the first and the second set of observations which is consistent with the remaining flux being emitted by the SN. As can be derived from the approximate models discussed in Sect. \ref{s_physics_sn_IIb}, in the limit of low optical depth for the $\gamma$-rays and if all positrons are trapped, the decline rate is $(1 + (1-f_{\mathrm{e^{+}}}(t)) \, 223/t) \, 0.0098$ mag day$^{-1}$, where $f_{\mathrm{e^{}+}}(t)$ is the fractional positron contribution to the luminosity and $t$ is given in days. This gives an expected decline rate of of 0.0098-0.0132 mag at 650 days, in rough agreement with the observed decline rates if the contribution from positrons is assumed to dominate. Given all this, although we can not exclude a minor contribution to the pre-explosion flux from other sources, we find that the yellow supergiant has disappeared. The only reasonable explanation is that the star was the progenitor of SN 2011dh as originally proposed in \citetalias{Mau11} which is also the conclusion reached by \citetalias{Dyk13b}.

The disappearance of the yellow supergiant confirms the results in \citetalias{Ber12} in which we showed that an extended progenitor with the observed properties of the yellow supergiant could well reproduce the early optical evolution. This shows that the duration of the initial cooling phase for a SN with an extended progenitor can be significantly shorter than commonly thought and that approximate models as the one by \citet{Rab11} used in \citetalias{Arc11} does not necessarily apply. It also indicates that the proposed division of Type IIb SNe progenitors in compact and extended ones and the relation between the speed of the shock and the type of progenitor \citep{Che10} needs to be revised. It is interesting to note that the two progenitors of Type IIb SNe (1993J and 2011dh) whose nature have been revealed were in both cases extended supergiants. The disappearance of the yellow supergiant in M51 was a major step in achieving one of our main goals, to determine the initial mass of the progenitor star. This mass has now been estimated to $\sim$13 M$_{\odot}$ by two different methods, the hydrodynamical modelling in \citetalias{Ber12} and the progenitor analysis in \citetalias{Mau11}, respectively. Both methods use results from stellar evolutionary modelling to relate the He core mass and the progenitor luminosity respectively to the initial mass but are otherwise independent. We note that, contrary to most other types, the initial mass of Type IIb SNe progenitors might be derived from hydrodynamical modelling without any assumptions of uncertain mass-loss rates as the star is essentially a bare He core (although with a thin and extended envelope).

\subsection{4.5 $\mu$m excess}
\label{s_45_micron_excess}

As mentioned in Sections \ref{s_analysis_phot} and \ref{s_bol_lightcurve} there is a flux excess in the Spitzer 4.5 $\mu$m band as compared to the 2MASS $JHK$ and Spitzer 3.6 $\mu$m bands developing during the first 100 days. This is most clearly seen in Fig. \ref{f_sed_evo}. Whereas other bands redward of $V$ are well approximated by the blackbody fits the flux at 4.5 $\mu$m is a factor of $\sim$5 in excess at 100 days. Warm dust or CO fundamental band emission are two possible explanations. For day 50-100 the excess (relative the blackbody fits discussed in Sect.~\ref{s_analysis_colour}) is well fitted by a blackbody with T$\simeq$400 K, R$\simeq$$5 \times 10^{16}$ cm and L$\simeq$$6 \times 10^{40}$ erg s$^{-1}$. \citet{Hel13} find a blackbody luminosity declining from $\sim$$7 \times 10^{40}$ to $\sim$$3 \times 10^{40}$ erg s$^{-1}$ and cooling from $\sim$1600 to $\sim$600 K between $\sim$20 and $\sim$90 days by fitting the MIR bands alone. Using a simple model for heated CSM dust they also find a thermal dust echo to be consistent with the observed MIR fluxes. Given the lack of MIR spectra we can neither confirm nor excluded CO fundamental band emission as the explanation of the excess. There is a possible excess (as compared to the continuum) developing near the location of the first overtone band at $\sim$23000 \AA~though. NIR spectra from later epochs may help to resolve this issue as we expect first overtone emission to grow stronger as compared to the continuum.

\section{Conclusions}
\label{s_conclusions}

We present extensive photometric and spectroscopic optical and NIR observations of SN 2011dh obtained during the first 100 days. The calibration of the photometry is discussed in some detail and we find it to be accurate to the five percent level in all bands. Using our observations as well as SWIFT UV and Spitzer MIR observations we calculate the bolometric UV to MIR lightcurve using both photometric and spectroscopic data. This bolometric lightcurve together with the photospheric velocity as estimated from the absorption minimum of the \ion{Fe}{ii} 5169 \AA~line provides the observational basis for the hydrodynamical modelling done in \citetalias{Ber12}.

We adopt a distance of 7.8$^{+1.1}_{-0.9}$ Mpc based on all estimates in the literature and find an extinction of $E$($B$-$V$)$_\mathrm{T}$=0.07$^{+0.07}_{-0.04}$ mag to be consistent with estimates and constraints presented in the literature and in this paper. The sensitivity of the results in \citetalias{Ber12} to these uncertainties is discussed and we find that only the derived mass of ejected \element[ ][56]{Ni} and radius is likely to be affected. We also revise the modelling made in \citetalias{Ber12} to agree with the values of the distance and extinction adopted in this paper and find that only the derived mass of ejected \element[ ][56]{Ni} and radius needs to be revised. The uncertainty in the photospheric velocity as estimated from the absorption minimum of the \ion{Fe}{ii} 5169 \AA~line is discussed and we find that we can not constrain this velocity very well. This is a potential problem as we are unable to quantify the sensitivity of the results in \citetalias{Ber12} to this uncertainty.

We present and discuss pre- and post-explosion observations which show that the yellow supergiant coincident with SN 2011dh has disappeared and indeed was the progenitor as proposed in \citetalias{Mau11}. Furthermore, the results from the progenitor analysis in \citetalias{Mau11} are consistent with those from the hydrodynamical modelling in \citetalias{Ber12}. Given the revisions in this paper, we find that an almost bare helium core with a mass of 3.3-4.0 M$_{\odot}$ surrounded by a thin hydrogen rich envelope extending to 200-300 R$_{\odot}$ exploded with an energy of 0.6$-$1.0 $\times10^{51}$ erg ejecting a mass of 1.8-2.5 M$_{\odot}$ of which 0.05-0.10 M$_{\odot}$ consisted of synthesised \element[ ][56]{Ni}.

The absorption minimum of the hydrogen lines is never seen below $\sim$11000 km s$^{-1}$ but approaches this value when the lines get weaker. Spectral modelling of the hydrogen lines using the \citetalias{Ber12} He4R270 ejecta model well reproduces this behaviour and the minimum velocity of the absorption minima coincides with the model interface between the helium core and the hydrogen rich envelope. The good agreement between the modelled and observed minimum velocities gives support to the \citetalias{Ber12} He4R270 ejecta model and we find it most likely that the observed minimum velocity of $\sim$11000 km s$^{-1}$ corresponds to the interface between the helium core and the hydrogen rich envelope. We note that the minimum velocity of the H$\alpha$ absorption minimum for SNe 1993J, 2011dh and 2008ax is $\sim$9000, $\sim$11000 and $\sim$13000 km s$^{-1}$ respectively which suggest that the interface between the helium core and the hydrogen rich envelope is located near these progressively higher velocities. By varying the fraction of hydrogen in the envelope we find a hydrogen mass of 0.01-0.04 M$_{\odot}$ to be consistent with the observed evolution of the hydrogen lines. This is in reasonable agreement with the 0.02 M$_{\odot}$ in the original model and the 0.024 M$_{\odot}$ estimated by \citetalias{Arc11} using spectral modelling similar to the one in this paper. We estimate that the photosphere reaches the interface between the helium core and the hydrogen rich envelope at 5-7 days. The helium lines appear between $\sim$10 days (\ion{He}{i} 10830 and 5876 \AA) and $\sim$15 days (\ion{He}{i} 6678, 7065 and 20581 \AA), close to the region where we expect the photosphere to be located and then move outward in velocity until $\sim$40 days. This suggests that the early evolution of these lines is driven by increasing non-thermal excitation due to decreasing optical depth for the $\gamma$-rays.

The photometric and spectral characteristics of SNe 2011dh, 1993J and 2008ax are compared and we find the colours and luminosities to differ significantly for the distances and extinctions adopted in this paper. However, the errors arising from the distance and extinction are large and a revision of the extinctions, just within the error bars, would bring the colours and luminosities in good agreement. Although a definite conclusion can not be made it is clear that a scenario where all three SNe have similar ejecta masses, explosion energies and ejected masses of $^{56}$Ni is possible. As shown in \citetalias{Ber12} the differences in the early evolution could be explained by differences in the mass and radius of the hydrogen envelope. Progressively higher velocities of the interface between the helium core and the hydrogen rich envelope, as proposed above, would naively correspond to progressively lower masses of this envelope for SNe 1993J, 2011dh and 2008ax respectively. Such a conclusion is supported by the early photometric evolution, the strength and persistence of the H$\alpha$ line and hydrodynamical as well as spectral modelling of these SNe.

We detect a flux excess in the 4.5 $\mu$m Spitzer band as compared to the NIR and the 3.6 $\mu$m Spitzer band. A thermal dust echo in the CSM as proposed by \citet{Hel13} or CO fundamental band emission are possible explanations but further work using late time observations is needed to resolve this issue.

The high quality dataset presented in this paper provides an ideal base for further modelling of the SN. One of the most interesting issues which remains unsolved is the possible existence of a bluer and more compact companion star as predicted by the binary evolutionary modelling in \citet{Ben12}. It is still not clear which of the single or binary star channels is dominating the production of Type IIb SNe. HST imaging, preferably in the UV, would have a good chance to detect such a companion.

\section{Acknowledgements}

This work is partially based on observations of the European supernova collaboration involved in the ESO-NTT and TNG large programmes led by Stefano Benetti. 

This work is partially based on observations made with the Nordic Optical Telescope, operated by the Nordic Optical Telescope Scientific Association at the Observatorio del Roque de los Muchachos, La Palma, Spain, of the Instituto de Astrofisica de Canarias. We acknowledge the exceptional support we got from the NOT staff throughout this campaign. 

This work is partially based on observations observations made with the Italian Telescopio Nazionale Galileo (TNG) operated by the Fundacio\'n Galileo Galilei of the INAF (Istituto Nazionale di Astrofisica) at the Spanish Observatorio del Roque de los Muchachos of the Instituto de Astrofisica de Canarias; the 1.82m Copernico and Schmidt 67/92 telescopes of INAF- Asiago Observatory; the 1.22m Galileo telescope of Dipartimento di Fisica e Astronomia (Universita' di Padova); the LBT, which is an international collaboration among institutions in the United States, Italy, and Germany. LBT Corporation partners are The Ohio State University, and The Research Corporation, on behalf of the University of Notre Dame, University of Minnesota and University of Virginia; the University of Arizona on behalf of the Arizona university system; INAF, Italy. 

We are in debt with S. Ciroi, A. Siviero and L. Aramyan for help with the Galileo 1.22m observations.

This work is partially based on observation made with the William Herschel Telescope, operated on the island of La Palma by the Isaac Newton Group in the Spanish Observatorio del Roque de los Muchachos of the Instituto de Astrofísica de Canarias and the Liverpool Telescope, operated on the island of La Palma by Liverpool John Moores University in the Spanish Observatorio del Roque de los Muchachos of the Instituto de Astrofisica de Canarias with financial support from the UK Science and Technology Facilities Council. 

This work is partially based on observations made with the Carlos S\'anchez Telescope operated on the island of Tenerife by the Instituto de Astrof\'{i}sica de Canarias in the Spanish Observatorio del Teide, and the Joan Or\'o Telescope of the Montsec Astronomical Observatory, which is owned by the Generalitat de Catalunya and operated by the Institute for Space Studies of Catalonia (IEEC).

L.T., A.P., S.B., E.C., and M.T. are partially supported by the PRIN-INAF 2011 with the project Transient Universe: from ESO Large to PESSTO. 

N.E.R. acknowledges financial support by the MICINN grant AYA08-1839/ESP, AYA2011-24704/ESP, and by the ESF EUROCORES Program EuroGENESIS (MINECO grants EUI2009-04170). 

S.T. acknowledges support by TRR 33 “The Dark Universe” of the German Research Foundation. J.S. and the OKC are supported by The Swedish Research Council.

R.K. and M.T. gratefully acknowledges the allocation of Liverpool Telescope time under the programmes ITP10-04 and PL11A-03 on which this study is partially-based.

F.B. acknowledges support from FONDECYT through grant 3120227 and by the Millennium Center for Supernova Science through grant P10-064-F (funded by "Programa Bicentenario de Ciencia y Tecnolog\'{i}a de CONICYT" and "Programa Iniciativa Cient\'{i}fica Milenio de MIDEPLAN").

Finally, we thank the referee Jozsef Vinko for his useful suggestions.

\begin{table*}[p]
\caption{Optical colour-corrected JC $U$ and S-corrected JC $BVRI$ magnitudes for SN 2011dh. Errors are given in parentheses.}
\begin{center}
\scalebox{1.00}{
\begin{tabular}{l l l l l l l l }
\hline\hline \\ [-1.5ex]
JD (+2400000) & Phase & $U$ & $B$ & $V$ & $R$ & $I$ & Telescope (Instrument)\\ [0.5ex]
(d) & (d) & (mag) & (mag) & (mag) & (mag) & (mag) & \\
\hline \\ [-1.5ex]
55716.43 & 3.43 & 14.99 (0.03) & 15.35 (0.02) & 14.92 (0.02) & 14.54 (0.01) & 14.41 (0.02) &  LT (RATCam) \\
55716.43 & 3.43 & 15.15 (0.08) & 15.39 (0.02) & 14.94 (0.02) & 14.57 (0.01) & 14.46 (0.01) &  TNG (LRS) \\
55717.43 & 4.43 & 15.03 (0.03) & 15.14 (0.02) & 14.67 (0.03) & 14.25 (0.01) & 14.26 (0.03) &  LT (RATCam) \\
55717.48 & 4.48 & 15.17 (0.09) & 15.21 (0.03) & 14.63 (0.03) & 14.24 (0.01) & 14.23 (0.02) &  AS-1.82m (AFOSC) \\
55717.48 & 4.48 & ... & 15.12 (0.03) & 14.63 (0.02) & 14.27 (0.01) & 14.28 (0.02) &  CANTAB (BIGST8) \\
55718.48 & 5.48 & ... & 14.84 (0.01) & 14.28 (0.02) & 13.94 (0.01) & 13.94 (0.01) &  LT (RATCam) \\
55718.57 & 5.57 & 14.68 (0.06) & 14.84 (0.02) & 14.24 (0.02) & 13.91 (0.01) & 14.04 (0.01) &  CA-2.2m (CAFOS) \\
55720.42 & 7.42 & 14.42 (0.02) & 14.25 (0.01) & 13.75 (0.03) & 13.41 (0.01) & 13.43 (0.02) &  LT (RATCam) \\
55721.42 & 8.42 & 14.28 (0.10) & 14.02 (0.01) & 13.48 (0.01) & 13.22 (0.01) & 13.24 (0.02) &  LT (RATCam) \\
55721.43 & 8.43 & 14.07 (0.07) & 14.06 (0.01) & 13.60 (0.04) & 13.27 (0.02) & 13.34 (0.02) &  NOT (ALFOSC) \\
55722.42 & 9.42 & ... & 13.86 (0.01) & 13.29 (0.01) & 13.05 (0.01) & 13.07 (0.01) &  LT (RATCam) \\
55723.41 & 10.41 & 13.98 (0.06) & 13.71 (0.01) & 13.16 (0.01) & 12.89 (0.01) & 12.90 (0.01) &  LT (RATCam) \\
55724.41 & 11.41 & 13.91 (0.08) & 13.62 (0.01) & 13.03 (0.01) & 12.79 (0.01) & 12.77 (0.01) &  LT (RATCam) \\
55725.39 & 12.39 & ... & ... & 12.94 (0.02) & 12.66 (0.01) & ... &  MONTCAB (BIGST8) \\
55725.43 & 12.43 & 13.88 (0.07) & 13.52 (0.02) & 12.92 (0.04) & 12.68 (0.01) & 12.68 (0.01) &  LT (RATCam) \\
55726.36 & 13.36 & ... & 13.52 (0.01) & 12.91 (0.02) & 12.59 (0.01) & ... &  MONTCAB (BIGST8) \\
55728.40 & 15.40 & ... & 13.39 (0.01) & 12.77 (0.01) & 12.44 (0.01) & ... &  MONTCAB (BIGST8) \\
55729.39 & 16.39 & 13.65 (0.01) & 13.35 (0.01) & 12.77 (0.06) & 12.39 (0.01) & 12.35 (0.02) &  LT (RATCam) \\
55730.40 & 17.40 & 13.64 (0.03) & 13.33 (0.01) & 12.66 (0.01) & 12.36 (0.01) & 12.32 (0.01) &  LT (RATCam) \\
55731.41 & 18.41 & 13.74 (0.09) & 13.30 (0.01) & 12.60 (0.02) & 12.31 (0.01) & 12.27 (0.01) &  LT (RATCam) \\
55731.82 & 18.82 & ... & ... & ... & 12.33 (0.02) & 12.25 (0.01) &  FTN (FS02) \\
55732.40 & 19.40 & ... & 13.35 (0.03) & 12.61 (0.01) & 12.27 (0.01) & 12.21 (0.01) &  CANTAB (BIGST8) \\
55732.41 & 19.41 & 13.44 (0.06) & 13.36 (0.02) & 12.64 (0.02) & 12.33 (0.02) & 12.31 (0.02) &  NOT (ALFOSC) \\
55732.46 & 19.46 & 13.71 (0.07) & 13.32 (0.01) & 12.58 (0.01) & 12.28 (0.02) & 12.22 (0.01) &  LT (RATCam) \\
55733.45 & 20.45 & 13.67 (0.07) & ... & ... & 12.26 (0.01) & 12.20 (0.02) &  LT (RATCam) \\
55734.52 & 21.52 & 13.37 (0.05) & 13.33 (0.01) & 12.58 (0.01) & 12.25 (0.01) & 12.29 (0.01) &  CA-2.2m (CAFOS) \\
55735.44 & 22.44 & 13.91 (0.04) & ... & ... & 12.26 (0.01) & 12.16 (0.01) &  LT (RATCam) \\
55736.44 & 23.44 & 14.13 (0.08) & ... & ... & 12.26 (0.01) & 12.16 (0.01) &  LT (RATCam) \\
55737.39 & 24.39 & ... & 13.65 (0.01) & 12.72 (0.01) & ... & ... &  LT (RATCam) \\
55738.42 & 25.42 & 14.50 (0.04) & 13.79 (0.02) & 12.81 (0.01) & 12.32 (0.02) & 12.22 (0.01) &  LT (RATCam) \\
55738.51 & 25.51 & 14.20 (0.04) & 13.77 (0.02) & 12.82 (0.01) & 12.38 (0.01) & 12.26 (0.01) &  NOT (ALFOSC) \\
55739.43 & 26.43 & 14.73 (0.04) & 13.95 (0.02) & 12.88 (0.01) & 12.38 (0.01) & 12.23 (0.01) &  LT (RATCam) \\
55740.36 & 27.36 & ... & 14.09 (0.04) & 12.93 (0.01) & 12.45 (0.01) & 12.29 (0.01) &  MONTCAB (BIGST8) \\
55740.43 & 27.43 & 14.91 (0.03) & 14.12 (0.01) & 12.97 (0.01) & 12.48 (0.01) & 12.30 (0.01) &  LT (RATCam) \\
55740.44 & 27.44 & ... & ... & 12.97 (0.01) & 12.47 (0.01) & ... &  TJO (MEIA) \\
55741.44 & 28.44 & ... & ... & ... & 12.54 (0.01) & 12.32 (0.01) &  LT (RATCam) \\
55742.49 & 29.49 & 15.33 (0.01) & ... & ... & 12.62 (0.01) & 12.40 (0.01) &  LT (RATCam) \\
55743.41 & 30.41 & ... & 14.53 (0.01) & 13.27 (0.02) & ... & ... &  LT (RATCam) \\
55743.42 & 30.42 & 15.18 (0.05) & 14.51 (0.02) & ... & 12.65 (0.01) & 12.53 (0.01) &  CA-2.2m (CAFOS) \\
55743.42 & 30.42 & 15.43 (0.05) & 14.53 (0.01) & 13.26 (0.03) & 12.68 (0.01) & 12.49 (0.01) &  NOT (ALFOSC) \\
55745.39 & 32.39 & 15.74 (0.03) & 14.74 (0.01) & 13.44 (0.01) & 12.77 (0.01) & 12.56 (0.01) &  NOT (ALFOSC) \\
55745.44 & 32.44 & 15.93 (0.04) & ... & ... & 12.81 (0.01) & 12.53 (0.01) &  LT (RATCam) \\
55745.80 & 32.80 & ... & ... & ... & 12.80 (0.01) & 12.51 (0.01) &  FTN (FS02) \\
55746.45 & 33.45 & 16.07 (0.04) & 14.87 (0.03) & 13.51 (0.01) & 12.83 (0.01) & 12.55 (0.02) &  LT (RATCam) \\
55747.44 & 34.44 & 16.12 (0.04) & ... & ... & 12.89 (0.01) & 12.59 (0.01) &  LT (RATCam) \\
55748.43 & 35.43 & 16.02 (0.02) & 14.97 (0.01) & 13.62 (0.01) & 12.88 (0.01) & 12.65 (0.01) &  NOT (ALFOSC) \\
55748.44 & 35.44 & 16.27 (0.04) & ... & ... & 12.94 (0.01) & 12.62 (0.01) &  LT (RATCam) \\
55750.40 & 37.40 & 16.20 (0.04) & 15.10 (0.01) & 13.73 (0.01) & 13.03 (0.01) & 12.73 (0.01) &  NOT (ALFOSC) \\
55750.42 & 37.42 & 16.41 (0.14) & 15.11 (0.02) & 13.78 (0.03) & 13.03 (0.01) & 12.73 (0.02) &  LT (RATCam) \\
55751.41 & 38.41 & ... & 15.14 (0.01) & 13.81 (0.01) & 13.08 (0.01) & 12.73 (0.01) &  TJO (MEIA) \\
55751.43 & 38.43 & ... & ... & ... & 13.11 (0.01) & 12.77 (0.01) &  LT (RATCam) \\
55752.45 & 39.45 & 16.54 (0.16) & ... & ... & 13.13 (0.01) & 12.75 (0.01) &  LT (RATCam) \\
\hline
\end{tabular}}

\end{center}
\label{t_jc}
\end{table*}

\setcounter{table}{2}
\begin{table*}[p]
\caption{Continued.}
\begin{center}
\scalebox{1.00}{
\begin{tabular}{l l l l l l l l }
\hline\hline \\ [-1.5ex]
JD (+2400000) & Phase & $U$ & $B$ & $V$ & $R$ & $I$ & Telescope (Instrument)\\ [0.5ex]
(d) & (d) & (mag) & (mag) & (mag) & (mag) & (mag) & \\
\hline \\ [-1.5ex]
55753.42 & 40.42 & ... & 15.29 (0.01) & 13.90 (0.02) & ... & ... &  LT (RATCam) \\
55753.46 & 40.46 & 16.45 (0.05) & 15.24 (0.01) & 13.86 (0.01) & 13.15 (0.01) & 12.81 (0.01) &  NOT (ALFOSC) \\
55755.40 & 42.40 & 16.42 (0.04) & 15.30 (0.01) & 13.96 (0.01) & 13.23 (0.01) & 12.89 (0.01) &  NOT (ALFOSC) \\
55756.44 & 43.44 & ... & 15.28 (0.02) & 13.98 (0.02) & 13.28 (0.02) & 12.86 (0.01) &  AS-Schmidt (SBIG) \\
55756.45 & 43.45 & ... & 15.38 (0.02) & 13.98 (0.01) & 13.27 (0.03) & 12.92 (0.01) &  LT (RATCam) \\
55757.43 & 44.43 & 16.42 (0.04) & 15.38 (0.01) & 14.05 (0.01) & 13.29 (0.01) & 12.97 (0.01) &  NOT (ALFOSC) \\
55759.45 & 46.45 & ... & 15.44 (0.01) & 14.06 (0.02) & ... & ... &  LT (RATCam) \\
55761.40 & 48.40 & ... & 15.44 (0.01) & 14.17 (0.01) & 13.44 (0.01) & 13.02 (0.01) &  AS-Schmidt (SBIG) \\
55762.41 & 49.41 & ... & 15.45 (0.01) & 14.16 (0.01) & 13.44 (0.01) & 13.06 (0.01) &  NOT (ALFOSC) \\
55762.78 & 49.78 & ... & ... & ... & 13.44 (0.01) & 13.03 (0.01) &  FTN (FS02) \\
55763.44 & 50.44 & ... & 15.47 (0.01) & 14.22 (0.01) & 13.47 (0.01) & 13.09 (0.01) &  AS-Schmidt (SBIG) \\
55765.43 & 52.43 & 16.44 (0.03) & 15.52 (0.01) & 14.26 (0.01) & 13.55 (0.01) & 13.17 (0.01) &  NOT (ALFOSC) \\
55767.43 & 54.43 & 16.50 (0.05) & ... & ... & 13.58 (0.01) & 13.16 (0.02) &  LT (RATCam) \\
55768.45 & 55.45 & 16.48 (0.04) & ... & ... & 13.60 (0.02) & 13.19 (0.02) &  LT (RATCam) \\
55771.40 & 58.40 & 16.37 (0.03) & 15.58 (0.01) & 14.32 (0.01) & 13.62 (0.01) & 13.28 (0.01) &  CA-2.2m (CAFOS) \\
55773.39 & 60.39 & 16.45 (0.04) & 15.60 (0.01) & 14.38 (0.01) & 13.71 (0.01) & 13.32 (0.01) &  NOT (ALFOSC) \\
55776.38 & 63.38 & 16.47 (0.04) & 15.64 (0.01) & 14.46 (0.01) & 13.77 (0.01) & 13.36 (0.01) &  NOT (ALFOSC) \\
55777.33 & 64.33 & ... & 15.52 (0.03) & 14.46 (0.02) & 13.78 (0.02) & 13.34 (0.02) &  AS-Schmidt (SBIG) \\
55780.40 & 67.40 & 16.42 (0.03) & 15.65 (0.01) & 14.50 (0.01) & 13.85 (0.01) & 13.43 (0.01) &  NOT (ALFOSC) \\
55783.43 & 70.43 & 16.41 (0.03) & 15.71 (0.01) & 14.58 (0.01) & 13.94 (0.01) & 13.51 (0.01) &  NOT (ALFOSC) \\
55784.33 & 71.33 & ... & 15.66 (0.02) & 14.59 (0.01) & ... & 13.43 (0.02) &  AS-Schmidt (SBIG) \\
55784.39 & 71.39 & 16.45 (0.04) & 15.66 (0.01) & 14.52 (0.02) & 13.90 (0.01) & 13.47 (0.02) &  CA-2.2m (CAFOS) \\
55784.77 & 71.77 & ... & ... & ... & 13.93 (0.02) & 13.45 (0.01) &  FTN (FS02) \\
55785.36 & 72.36 & ... & 15.70 (0.02) & 14.61 (0.01) & 13.96 (0.01) & 13.45 (0.01) &  AS-Schmidt (SBIG) \\
55788.41 & 75.41 & ... & ... & ... & 14.02 (0.02) & 13.52 (0.01) &  AS-Schmidt (SBIG) \\
55790.38 & 77.38 & 16.45 (0.09) & ... & ... & 14.03 (0.01) & 13.61 (0.01) &  LT (RATCam) \\
55793.37 & 80.37 & 16.55 (0.07) & 15.80 (0.01) & 14.74 (0.01) & 14.13 (0.01) & 13.67 (0.01) &  NOT (ALFOSC) \\
55795.35 & 82.35 & 16.40 (0.04) & 15.78 (0.01) & 14.76 (0.01) & 14.12 (0.01) & 13.68 (0.01) &  CA-2.2m (CAFOS) \\
55797.37 & 84.37 & ... & 15.83 (0.02) & 14.82 (0.01) & ... & ... &  AS-Schmidt (SBIG) \\
55797.76 & 84.76 & ... & ... & ... & 14.22 (0.01) & 13.68 (0.01) &  FTN (FS02) \\
55798.36 & 85.36 & 16.50 (0.03) & 15.84 (0.01) & 14.84 (0.01) & 14.25 (0.01) & 13.65 (0.02) &  NOT (ALFOSC) \\
55799.33 & 86.33 & ... & 15.82 (0.01) & 14.86 (0.01) & ... & ... &  AS-Schmidt (SBIG) \\
55801.36 & 88.36 & 16.44 (0.04) & 15.89 (0.01) & 14.90 (0.01) & 14.31 (0.01) & 13.80 (0.01) &  NOT (ALFOSC) \\
55801.40 & 88.40 & ... & 15.80 (0.02) & 14.90 (0.01) & ... & ... &  AS-Schmidt (SBIG) \\
55803.35 & 90.35 & ... & 15.88 (0.02) & 14.91 (0.01) & 14.32 (0.01) & 13.79 (0.01) &  AS-Schmidt (SBIG) \\
55805.33 & 92.33 & ... & 15.87 (0.02) & 14.97 (0.02) & 14.37 (0.01) & 13.83 (0.01) &  AS-Schmidt (SBIG) \\
55810.34 & 97.34 & 16.68 (0.06) & 16.00 (0.01) & 15.11 (0.01) & 14.52 (0.01) & 14.02 (0.01) &  NOT (ALFOSC) \\
55812.33 & 99.33 & 16.51 (0.03) & 16.02 (0.01) & 15.05 (0.01) & 14.49 (0.01) & 14.00 (0.01) &  CA-2.2m (CAFOS) \\
\hline
\end{tabular}}

\end{center}
\end{table*}

\begin{table*}
\caption{Optical S-corrected SWIFT JC $UBV$ magnitudes for SN 2011dh. Errors are given in parentheses.}
\begin{center}
\scalebox{1.00}{
\begin{tabular}{l l l l l l }
\hline\hline \\ [-1.5ex]
JD (+2400000) & Phase & $U$ & $B$ & $V$ & Telescope (Instrument)\\ [0.5ex]
(d) & (d) & (mag) & (mag) & (mag) & \\
\hline \\ [-1.5ex]
55716.01 & 3.01 & 14.92 (0.02) & 15.35 (0.02) & 14.96 (0.02) &  SWIFT (UVOT) \\
55716.68 & 3.68 & 15.05 (0.02) & 15.29 (0.02) & 14.90 (0.01) &  SWIFT (UVOT) \\
55717.82 & 4.82 & 15.03 (0.04) & 15.03 (0.03) & 14.57 (0.03) &  SWIFT (UVOT) \\
55719.03 & 6.03 & 14.77 (0.03) & 14.65 (0.02) & 14.14 (0.02) &  SWIFT (UVOT) \\
55720.83 & 7.83 & 14.31 (0.03) & 14.15 (0.02) & 13.65 (0.01) &  SWIFT (UVOT) \\
55721.84 & 8.84 & 14.11 (0.02) & 13.96 (0.02) & 13.43 (0.01) &  SWIFT (UVOT) \\
55723.18 & 10.18 & 13.93 (0.02) & 13.73 (0.02) & 13.20 (0.01) &  SWIFT (UVOT) \\
55723.98 & 10.98 & 13.85 (0.02) & 13.68 (0.02) & 13.12 (0.01) &  SWIFT (UVOT) \\
55725.13 & 12.13 & 13.77 (0.02) & 13.57 (0.02) & 12.97 (0.01) &  SWIFT (UVOT) \\
55726.66 & 13.66 & 13.78 (0.02) & 13.57 (0.02) & 12.91 (0.01) &  SWIFT (UVOT) \\
55727.79 & 14.79 & 13.71 (0.02) & 13.48 (0.02) & 12.80 (0.01) &  SWIFT (UVOT) \\
55727.87 & 14.87 & 13.76 (0.05) & ... & ... &  SWIFT (UVOT) \\
55729.25 & 16.25 & 13.69 (0.02) & 13.42 (0.02) & 12.77 (0.01) &  SWIFT (UVOT) \\
55729.60 & 16.60 & 13.63 (0.02) & 13.42 (0.02) & 12.72 (0.01) &  SWIFT (UVOT) \\
55730.53 & 17.53 & 13.66 (0.02) & 13.36 (0.02) & 12.69 (0.01) &  SWIFT (UVOT) \\
55731.65 & 18.65 & 13.66 (0.02) & 13.35 (0.02) & 12.64 (0.01) &  SWIFT (UVOT) \\
55732.67 & 19.67 & 13.69 (0.02) & 13.37 (0.02) & 12.62 (0.01) &  SWIFT (UVOT) \\
55733.89 & 20.89 & 13.74 (0.02) & 13.37 (0.02) & 12.63 (0.01) &  SWIFT (UVOT) \\
55734.75 & 21.75 & 13.82 (0.02) & 13.44 (0.02) & 12.63 (0.01) &  SWIFT (UVOT) \\
55735.95 & 22.95 & 13.99 (0.03) & 13.53 (0.02) & 12.68 (0.01) &  SWIFT (UVOT) \\
55736.55 & 23.55 & 14.05 (0.03) & 13.55 (0.02) & 12.71 (0.01) &  SWIFT (UVOT) \\
55737.55 & 24.55 & 14.30 (0.03) & 13.70 (0.02) & 12.77 (0.01) &  SWIFT (UVOT) \\
55738.76 & 25.76 & 14.54 (0.03) & 13.83 (0.02) & 12.84 (0.02) &  SWIFT (UVOT) \\
55740.28 & 27.28 & 14.85 (0.03) & 14.07 (0.02) & 13.01 (0.03) &  SWIFT (UVOT) \\
55741.37 & 28.37 & 15.08 (0.05) & 14.23 (0.03) & 13.09 (0.02) &  SWIFT (UVOT) \\
55741.77 & 28.77 & 15.20 (0.05) & 14.32 (0.03) & 13.16 (0.02) &  SWIFT (UVOT) \\
55742.84 & 29.84 & 15.43 (0.05) & 14.46 (0.03) & 13.26 (0.02) &  SWIFT (UVOT) \\
55743.84 & 30.84 & 15.63 (0.05) & 14.55 (0.03) & 13.33 (0.02) &  SWIFT (UVOT) \\
55745.25 & 32.25 & 15.76 (0.05) & 14.74 (0.03) & 13.43 (0.02) &  SWIFT (UVOT) \\
55746.12 & 33.12 & 15.91 (0.06) & 14.80 (0.03) & 13.50 (0.02) &  SWIFT (UVOT) \\
55750.60 & 37.60 & 16.30 (0.06) & 15.03 (0.03) & 13.74 (0.02) &  SWIFT (UVOT) \\
55754.62 & 41.62 & 16.46 (0.07) & 15.24 (0.03) & 13.97 (0.02) &  SWIFT (UVOT) \\
55758.55 & 45.55 & 16.54 (0.08) & 15.37 (0.03) & 14.07 (0.02) &  SWIFT (UVOT) \\
55762.57 & 49.57 & 16.75 (0.10) & 15.45 (0.04) & 14.19 (0.02) &  SWIFT (UVOT) \\
55766.52 & 53.52 & 16.60 (0.08) & 15.57 (0.04) & 14.30 (0.02) &  SWIFT (UVOT) \\
55770.80 & 57.80 & 16.48 (0.07) & 15.61 (0.04) & 14.35 (0.02) &  SWIFT (UVOT) \\
55775.69 & 62.69 & 16.42 (0.07) & 15.60 (0.04) & 14.48 (0.03) &  SWIFT (UVOT) \\
55780.50 & 67.50 & 16.60 (0.08) & 15.70 (0.04) & 14.52 (0.03) &  SWIFT (UVOT) \\
55784.80 & 71.80 & 16.46 (0.07) & 15.71 (0.04) & 14.61 (0.03) &  SWIFT (UVOT) \\
55788.74 & 75.74 & 16.43 (0.06) & 15.73 (0.04) & 14.64 (0.03) &  SWIFT (UVOT) \\ [0.5ex]
\hline
\end{tabular}}

\end{center}
\label{t_jc_swift}
\end{table*}

\begin{table*}[p]
\caption{Optical colour-corrected SDSS $u$ and S-corrected SDSS $griz$ magnitudes for SN 2011dh. Errors are given in parentheses.}
\begin{center}
\scalebox{1.00}{
\begin{tabular}{l l l l l l l l }
\hline\hline \\ [-1.5ex]
JD (+2400000) & Phase & $u$ & $g$ & $r$ & $i$ & $z$ & Telescope (Instrument)\\ [0.5ex]
(d) & (d) & (mag) & (mag) & (mag) & (mag) & (mag) & \\
\hline \\ [-1.5ex]
55716.47 & 3.47 & 15.90 (0.03) & 15.08 (0.01) & 14.68 (0.01) & 14.80 (0.01) & 14.76 (0.02) &  LT (RATCam) \\
55717.46 & 4.46 & 16.01 (0.03) & 14.80 (0.01) & 14.38 (0.01) & 14.61 (0.01) & 14.58 (0.02) &  LT (RATCam) \\
55718.53 & 5.53 & ... & 14.44 (0.04) & 14.06 (0.01) & 14.27 (0.01) & ... &  LT (RATCam) \\
55720.44 & 7.44 & 15.39 (0.02) & 13.97 (0.01) & 13.53 (0.01) & 13.73 (0.02) & 13.87 (0.01) &  LT (RATCam) \\
55721.44 & 8.44 & 15.09 (0.01) & 13.78 (0.01) & 13.33 (0.01) & 13.52 (0.01) & 13.64 (0.01) &  LT (RATCam) \\
55722.44 & 9.44 & ... & 13.59 (0.01) & 13.18 (0.01) & 13.35 (0.01) & 13.49 (0.01) &  LT (RATCam) \\
55723.41 & 10.41 & 14.82 (0.03) & ... & 13.02 (0.01) & 13.16 (0.01) & 13.34 (0.01) &  LT (RATCam) \\
55724.41 & 11.41 & 14.72 (0.02) & ... & 12.93 (0.01) & 13.05 (0.01) & 13.22 (0.01) &  LT (RATCam) \\
55725.43 & 12.43 & 14.74 (0.04) & ... & 12.83 (0.01) & 12.94 (0.01) & 13.09 (0.01) &  LT (RATCam) \\
55729.39 & 16.39 & 14.56 (0.03) & 13.10 (0.01) & 12.56 (0.01) & 12.62 (0.01) & 12.81 (0.01) &  LT (RATCam) \\
55730.40 & 17.40 & 14.45 (0.03) & 13.07 (0.01) & 12.51 (0.01) & 12.56 (0.01) & 12.77 (0.01) &  LT (RATCam) \\
55731.41 & 18.41 & 14.54 (0.03) & 13.02 (0.01) & 12.46 (0.01) & 12.51 (0.01) & 12.71 (0.01) &  LT (RATCam) \\
55731.82 & 18.82 & ... & 13.07 (0.01) & 12.46 (0.01) & 12.50 (0.01) & 12.65 (0.01) &  FTN (FS02) \\
55732.46 & 19.46 & 14.56 (0.01) & 13.00 (0.03) & 12.42 (0.01) & 12.48 (0.01) & 12.67 (0.01) &  LT (RATCam) \\
55733.45 & 20.45 & 14.52 (0.05) & 13.03 (0.01) & 12.41 (0.01) & 12.45 (0.01) & 12.65 (0.01) &  LT (RATCam) \\
55735.44 & 22.44 & 14.75 (0.04) & 13.12 (0.01) & 12.43 (0.01) & 12.41 (0.01) & 12.60 (0.01) &  LT (RATCam) \\
55736.44 & 23.44 & 14.96 (0.03) & 13.19 (0.02) & 12.45 (0.01) & 12.42 (0.01) & 12.59 (0.02) &  LT (RATCam) \\
55738.45 & 25.45 & 15.37 (0.02) & 13.43 (0.01) & 12.55 (0.01) & 12.47 (0.01) & 12.65 (0.01) &  LT (RATCam) \\
55739.44 & 26.44 & 15.55 (0.02) & 13.50 (0.03) & 12.59 (0.01) & 12.50 (0.01) & 12.65 (0.01) &  LT (RATCam) \\
55740.44 & 27.44 & 15.80 (0.01) & 13.66 (0.01) & 12.66 (0.01) & 12.55 (0.01) & 12.70 (0.01) &  LT (RATCam) \\
55741.44 & 28.44 & ... & 13.75 (0.02) & 12.76 (0.01) & 12.59 (0.02) & 12.76 (0.01) &  LT (RATCam) \\
55742.49 & 29.49 & 16.20 (0.02) & 13.92 (0.01) & 12.84 (0.01) & 12.65 (0.01) & 12.80 (0.01) &  LT (RATCam) \\
55745.44 & 32.44 & 16.71 (0.05) & 14.20 (0.02) & 13.04 (0.01) & 12.79 (0.01) & 12.87 (0.03) &  LT (RATCam) \\
55745.80 & 32.80 & ... & 14.35 (0.04) & 13.00 (0.01) & 12.79 (0.01) & 12.94 (0.01) &  FTN (FS02) \\
55746.45 & 33.45 & 16.83 (0.04) & 14.32 (0.01) & 13.09 (0.01) & 12.82 (0.01) & 12.94 (0.01) &  LT (RATCam) \\
55747.44 & 34.44 & 16.90 (0.04) & 14.40 (0.02) & 13.13 (0.01) & 12.86 (0.01) & 12.95 (0.01) &  LT (RATCam) \\
55748.44 & 35.44 & 17.09 (0.04) & 14.42 (0.02) & 13.19 (0.01) & 12.90 (0.01) & 13.01 (0.01) &  LT (RATCam) \\
55750.44 & 37.44 & 17.20 (0.10) & 14.55 (0.02) & 13.29 (0.01) & 13.02 (0.02) & 13.04 (0.04) &  LT (RATCam) \\
55751.43 & 38.43 & 17.14 (0.03) & 14.64 (0.03) & 13.36 (0.01) & 13.04 (0.01) & 13.11 (0.01) &  LT (RATCam) \\
55752.45 & 39.45 & 17.24 (0.07) & 14.66 (0.01) & 13.39 (0.01) & 13.04 (0.01) & 13.09 (0.01) &  LT (RATCam) \\
55756.46 & 43.46 & ... & 14.79 (0.01) & 13.55 (0.01) & 13.22 (0.01) & 13.19 (0.01) &  LT (RATCam) \\
55762.78 & 49.78 & ... & 15.00 (0.02) & 13.68 (0.01) & 13.37 (0.01) & 13.28 (0.01) &  FTN (FS02) \\
55767.43 & 54.43 & 17.30 (0.02) & 15.03 (0.01) & 13.84 (0.01) & 13.52 (0.01) & 13.38 (0.02) &  LT (RATCam) \\
55768.45 & 55.45 & 17.29 (0.02) & 15.03 (0.01) & 13.86 (0.01) & 13.56 (0.01) & 13.41 (0.01) &  LT (RATCam) \\
55773.39 & 60.39 & 17.27 (0.04) & 15.07 (0.01) & 13.99 (0.01) & 13.72 (0.01) & 13.54 (0.02) &  NOT (ALFOSC) \\
55776.38 & 63.38 & 17.36 (0.03) & 15.13 (0.01) & 14.03 (0.01) & 13.76 (0.01) & 13.56 (0.01) &  NOT (ALFOSC) \\
55780.41 & 67.41 & 17.33 (0.03) & 15.16 (0.01) & 14.09 (0.01) & 13.84 (0.01) & 13.61 (0.01) &  NOT (ALFOSC) \\
55783.44 & 70.44 & 17.26 (0.04) & 15.18 (0.01) & 14.19 (0.01) & 13.93 (0.01) & 13.67 (0.01) &  NOT (ALFOSC) \\
55784.77 & 71.77 & ... & 15.23 (0.02) & 14.16 (0.01) & 13.88 (0.01) & 13.64 (0.01) &  FTN (FS02) \\
55790.38 & 77.38 & 17.29 (0.03) & 15.35 (0.04) & 14.28 (0.01) & 14.04 (0.01) & 13.69 (0.02) &  LT (RATCam) \\
55793.37 & 80.37 & 17.32 (0.03) & 15.30 (0.01) & 14.39 (0.01) & 14.16 (0.01) & 13.84 (0.01) &  NOT (ALFOSC) \\
55797.76 & 84.76 & ... & 15.38 (0.01) & 14.42 (0.01) & 14.18 (0.01) & 13.82 (0.01) &  FTN (FS02) \\
55798.37 & 85.37 & 17.35 (0.03) & 15.38 (0.01) & 14.50 (0.01) & 14.26 (0.01) & 13.87 (0.01) &  NOT (ALFOSC) \\
55801.36 & 88.36 & 17.34 (0.01) & 15.42 (0.01) & 14.53 (0.01) & 14.31 (0.01) & 13.89 (0.01) &  NOT (ALFOSC) \\
55810.34 & 97.34 & 17.49 (0.02) & 15.55 (0.01) & 14.75 (0.01) & 14.56 (0.01) & 14.10 (0.02) &  NOT (ALFOSC) \\
\hline
\end{tabular}}

\end{center}
\label{t_sloan}
\end{table*}

\begin{table*}[p]
\caption{NIR S-corrected 2MASS $JHK$ magnitudes for SN 2011dh. Errors are given in parentheses.}
\begin{center}
\scalebox{1.00}{
\begin{tabular}{l l l l l l }
\hline\hline \\ [-1.5ex]
JD (+2400000) & Phase & $J$ & $H$ & $K$ & Telescope (Instrument)\\ [0.5ex]
(d) & (d) & (mag) & (mag) & (mag) & \\
\hline \\ [-1.5ex]
55716.51 & 3.51 & 14.09 (0.01) & 13.90 (0.01) & 13.68 (0.02) &  TNG (NICS) \\
55722.40 & 9.40 & 12.89 (0.01) & 12.87 (0.01) & 12.67 (0.01) &  TNG (NICS) \\
55725.50 & 12.50 & 12.61 (0.04) & 12.54 (0.01) & 12.43 (0.02) &  NOT (NOTCAM) \\
55730.51 & 17.51 & 12.12 (0.01) & 12.08 (0.01) & 11.94 (0.01) &  TNG (NICS) \\
55737.72 & 24.72 & 11.96 (0.01) & 11.90 (0.01) & 11.72 (0.03) &  LBT (LUCIFER) \\
55741.13 & 28.13 & 11.94 (0.01) & 11.90 (0.02) & 11.70 (0.05) &  TCS (CAIN) \\
55748.43 & 35.43 & 12.14 (0.01) & 12.00 (0.02) & 11.77 (0.01) &  TCS (CAIN) \\
55750.42 & 37.42 & 12.19 (0.01) & 12.00 (0.01) & 11.84 (0.04) &  TCS (CAIN) \\
55751.42 & 38.42 & 12.29 (0.01) & 12.01 (0.01) & 11.84 (0.03) &  TCS (CAIN) \\
55758.45 & 45.45 & 12.55 (0.01) & 12.22 (0.01) & 12.06 (0.01) &  TNG (NICS) \\
55759.41 & 46.41 & 12.49 (0.03) & 12.22 (0.03) & 12.11 (0.04) &  TCS (CAIN) \\
55762.41 & 49.41 & 12.57 (0.01) & 12.26 (0.01) & 12.17 (0.03) &  TCS (CAIN) \\
55763.42 & 50.42 & 12.62 (0.02) & 12.27 (0.04) & 12.25 (0.06) &  TCS (CAIN) \\
55765.45 & 52.45 & 12.79 (0.01) & 12.38 (0.01) & 12.23 (0.01) &  TNG (NICS) \\
55769.41 & 56.41 & 12.77 (0.01) & 12.48 (0.06) & 12.40 (0.03) &  TCS (CAIN) \\
55773.37 & 60.37 & 12.94 (0.03) & 12.58 (0.01) & 12.42 (0.02) &  TNG (NICS) \\
55774.40 & 61.40 & 12.90 (0.01) & 12.55 (0.03) & 12.43 (0.04) &  TCS (CAIN) \\
55776.40 & 63.40 & 13.00 (0.01) & 12.64 (0.01) & 12.53 (0.02) &  TCS (CAIN) \\
55781.41 & 68.41 & 13.23 (0.01) & 12.76 (0.01) & 12.66 (0.01) &  WHT (LIRIS) \\
55787.44 & 74.44 & 13.56 (0.03) & 13.03 (0.02) & 12.95 (0.02) &  NOT (NOTCAM) \\
55794.41 & 81.41 & ... & 13.18 (0.01) & ... &  TNG (NICS) \\
55801.36 & 88.36 & 13.90 (0.02) & 13.41 (0.02) & 13.17 (0.01) &  TNG (NICS) \\
55804.34 & 91.34 & 14.10 (0.01) & 13.50 (0.01) & 13.26 (0.01) &  CA-3.5m (O2000) \\
\hline
\end{tabular}}

\end{center}
\label{t_nir}
\end{table*}

\begin{table*}[p]
\caption{MIR Spitzer 3.6 $\mu$m and 4.5 $\mu$m magnitudes for SN 2011dh. Errors are given in parentheses.}
\begin{center}
\scalebox{1.00}{
\begin{tabular}{l l l l l }
\hline\hline \\ [-1.5ex]
JD (+2400000) & Phase & 3.6 $\mu$m & 4.5 $\mu$m & Telescope (Instrument)\\ [0.5ex]
(d) & (d) & (mag) & (mag) & \\
\hline \\ [-1.5ex]
55731.21 & 18.21 & 11.83 (0.02) & 11.48 (0.02) &  SPITZER (IRAC) \\
55737.06 & 24.06 & 11.66 (0.02) & 11.31 (0.02) &  SPITZER (IRAC) \\
55744.32 & 31.32 & 11.66 (0.02) & 11.30 (0.02) &  SPITZER (IRAC) \\
55751.46 & 38.46 & 11.68 (0.02) & 11.30 (0.02) &  SPITZER (IRAC) \\
55758.75 & 45.75 & 11.79 (0.02) & 11.32 (0.02) &  SPITZER (IRAC) \\
55766.45 & 53.45 & 11.96 (0.02) & 11.34 (0.02) &  SPITZER (IRAC) \\
55772.33 & 59.33 & 12.11 (0.03) & 11.38 (0.02) &  SPITZER (IRAC) \\
55779.12 & 66.12 & 12.30 (0.03) & 11.43 (0.02) &  SPITZER (IRAC) \\
55785.60 & 72.60 & 12.50 (0.03) & 11.50 (0.02) &  SPITZER (IRAC) \\
55798.28 & 85.28 & 12.84 (0.04) & 11.66 (0.03) &  SPITZER (IRAC) \\
\hline
\end{tabular}}

\end{center}
\label{t_mir}
\end{table*}

\begin{table*}[p]
\caption{UV SWIFT $UVW1$, $UVM2$ and $UVW2$ magnitudes for SN 2011dh. Errors are given in parentheses.}
\begin{center}
\scalebox{1.00}{
\begin{tabular}{l l l l l l }
\hline\hline \\ [-1.5ex]
JD (+2400000) & Phase & $UVW1$ & $UVM2$ & $UVW2$ & Telescope (Instrument)\\ [0.5ex]
(d) & (d) & (mag) & (mag) & (mag) & \\
\hline \\ [-1.5ex]
55716.01 & 3.01 & 15.40 (0.04) & 15.99 (0.04) & 16.40 (0.04) &  SWIFT (UVOT) \\
55716.68 & 3.68 & 15.56 (0.04) & 16.27 (0.05) & 16.69 (0.04) &  SWIFT (UVOT) \\
55717.55 & 4.55 & 15.80 (0.05) & 16.59 (0.08) & 16.91 (0.08) &  SWIFT (UVOT) \\
55718.90 & 5.90 & 15.85 (0.04) & 16.77 (0.04) & 16.98 (0.05) &  SWIFT (UVOT) \\
55720.60 & 7.60 & 15.75 (0.04) & 16.89 (0.09) & ... &  SWIFT (UVOT) \\
55721.30 & 8.30 & ... & ... & 16.95 (0.07) &  SWIFT (UVOT) \\
55722.50 & 9.50 & 15.62 (0.04) & 17.28 (0.08) & ... &  SWIFT (UVOT) \\
55723.10 & 10.10 & ... & ... & 17.02 (0.13) &  SWIFT (UVOT) \\
55724.80 & 11.80 & 15.66 (0.04) & 17.37 (0.08) & 17.00 (0.05) &  SWIFT (UVOT) \\
55727.10 & 14.10 & 15.59 (0.04) & 17.61 (0.09) & 17.11 (0.05) &  SWIFT (UVOT) \\
55730.66 & 17.66 & 15.56 (0.03) & 17.72 (0.06) & 17.00 (0.04) &  SWIFT (UVOT) \\
55733.70 & 20.70 & 15.65 (0.03) & 17.93 (0.07) & 17.14 (0.04) &  SWIFT (UVOT) \\
55736.57 & 23.57 & 15.87 (0.04) & 18.16 (0.09) & 17.32 (0.05) &  SWIFT (UVOT) \\
55739.95 & 26.95 & 16.42 (0.04) & ... & 17.82 (0.08) &  SWIFT (UVOT) \\
55740.70 & 27.70 & ... & 18.49 (0.19) & ... &  SWIFT (UVOT) \\
55742.27 & 29.27 & 16.67 (0.06) & 19.06 (0.26) & 18.36 (0.13) &  SWIFT (UVOT) \\
55747.73 & 34.73 & 17.18 (0.06) & 19.11 (0.20) & 19.03 (0.19) &  SWIFT (UVOT) \\
55758.17 & 45.17 & 17.64 (0.08) & 19.37 (0.25) & 19.47 (0.25) &  SWIFT (UVOT) \\
55770.83 & 57.83 & 17.67 (0.08) & 19.74 (0.33) & 19.97 (0.40) &  SWIFT (UVOT) \\
55784.97 & 71.97 & 17.77 (0.09) & 20.22 (0.58) & 19.91 (0.40) &  SWIFT (UVOT) \\ [0.5ex]
\hline
\end{tabular}}

\end{center}
\label{t_uv}
\end{table*}

\begin{table*}[p]
\caption{List of optical and NIR spectroscopic observations.}
\begin{center}
\scalebox{1.0}{
\begin{tabular}{l l l l l l l}
\hline\hline \\ [-1.5ex]
JD (+2400000) & Phase & Grism & Range & Resolution & Resolution & Telescope (Instrument)\\ [0.5ex]
(d) & (d) & & (\AA) & & (\AA) &\\ [0.5ex]
\hline \\ [-1.5ex]
55716.41 & 3.41 & LRB & 3300-8000 & 585 & 10.0 & TNG (LRS) \\
55716.41 & 3.41 & LRR & 5300-9200 & 714 & 10.4 & TNG (LRS) \\
55716.47 & 3.47 & IJ & 9000-14500 & 333 & ... & TNG (NICS) \\
55716.49 & 3.49 & HK & 14000-25000 & 333 & ... & TNG (NICS) \\
55717.37 & 4.37 & b200 & 3300-8700 & .... & 12.0 & CA-2.2m (CAFOS) \\
55717.37 & 4.37 & r200 & 6300-10500 & ... & 12.0 & CA-2.2m (CAFOS) \\
55717.49 & 4.49 & Grism 4 & 3500-8450 & 613 & ... & AS 1.82m (AFOSC) \\
55718.42 & 5.42 & Grism 4 & 3200-9100 & 355 & 16.2 & NOT (ALFOSC) \\
55718.44 & 5.44 & Grism 5 & 5000-10250 & 415 & 16.8 & NOT (ALFOSC) \\
55719.40 & 6.40 & Grism 4 & 3200-9100 & 355 & 16.2 & NOT (ALFOSC) \\
55719.42 & 6.42 & Grism 5 & 5000-10250 & 415 & 16.8 & NOT (ALFOSC) \\
55719.47 & 6.47 & VPH4 & 6350-7090 & ... & 3.7 & AS-1.82m (AFOSC) \\
55721.39 & 8.39 & Grism 4 & 3200-9100 & 355 & 16.2 & NOT (ALFOSC) \\
55721.40 & 8.40 & Grism 5 & 5000-10250 & 415 & 16.8 & NOT (ALFOSC) \\
55721.45 & 8.45	& R300B & 3200-5300 & ... & 4.1 & WHT (ISIS) \\
55721.45 & 8.45	& R158R & 5300-10000 & ... & 7.7 & WHT(ISIS) \\
55722.57 & 9.57 & R300B & 3200-5300 & ... & 4.1 & WHT (ISIS) \\
55722.57 & 9.57 & R158R & 5300-10000 & ... & 7.7 & WHT (ISIS) \\
55722.42 & 9.42 & IJ & 9000-14500 & 333 & ... & TNG (NICS) \\
55722.46 & 9.48 & HK & 14000-25000 & 333 & ... & TNG (NICS) \\
55723.61 & 10.61 & VHRV & 4752-6698 & 2181 & 2.6 & TNG (LRS) \\
55725.38 & 12.38 & R300B & 3200-5300 & ... & 4.1 & WHT (ISIS) \\
55725.38 & 12.38 & R158R & 5300-10000 & ... & 7.7 & WHT (ISIS) \\
55730.45 & 17.45 & Grism 4 & 3200-9100 & 355 & 16.2 & NOT (ALFOSC) \\
55730.46 & 17.46 & Grism 5 & 5000-10250 & 415 & 16.8  & NOT (ALFOSC) \\
55730.52 & 17.52 & IJ & 9000-14500 & 333 & ... & TNG (NICS) \\
55730.57 & 17.57 & HK & 14000-25000 & 333 & ... & TNG (NICS) \\
55733.37 & 20.37 & b200 & 3300-8700 & ... & 12.0 & CA-2.2m (CAFOS) \\
55733.37 & 20.37 & r200 & 6300-10500 & ... & 12.0 & CA-2.2m (CAFOS) \\
55733.42 & 20.42 & Grism 4 & 3200-9100 & 355 & 16.2 & NOT (ALFOSC) \\
55733.43 & 20.43 & Grism 5 & 5000-10250 & 415 & 16.8 & NOT (ALFOSC) \\
55737.68 & 24.68 & 200 H+K & 14900-24000 & 1881(H)/2573(K) & ... & LBT (LUCIFER) \\
55738.49 & 25.49 & Grism 4 & 3200-9100 & 355 & 16.2 & NOT (ALFOSC) \\
55738.50 & 25.50 & Grism 5 & 5000-10250 & 415 & 16.8 & NOT (ALFOSC) \\
55738.41 & 25.41 & IJ & 9000-14500 & 333 & ... & TNG (NICS) \\
55743.40 & 30.40 & Grism 4 & 3200-9100 & 355 & 16.2 & NOT (ALFOSC) \\
55743.44 & 30.44 & Grism 5 & 5000-10250 & 415 & 16.8 & NOT (ALFOSC) \\
55743.43 & 30.43 & b200 & 3300-8700 & ... & 12.0 & CA-2.2m (CAFOS) \\
55743.43 & 30.43 & r200 & 6300-10500 & ... & 12.0 & CA-2.2m (CAFOS) \\
55748.40 & 35.40 & Grism 4 & 3200-9100 & 355 & 16.2 & NOT (ALFOSC) \\
55748.41 & 35.41 & Grism 5 & 5000-10250 & 415 & 16.8 & NOT (ALFOSC) \\
55748.39 & 35.39 & IJ & 9000-14500 & 333 & ... & TNG (NICS) \\
55748.42 & 35.42 & HK & 14000-25000 & 333 & ... & TNG (NICS) \\
55753.41 & 40.41 & Grism 4 & 3200-9100 & 355 & 16.2 & NOT (ALFOSC) \\
55753.43 & 40.43 & Grism 5 & 5000-10250 & 415 & 16.8 & NOT (ALFOSC) \\
55757.39 & 44.39 & Grism 4 & 3200-9100 & 355 & 16.2 & NOT (ALFOSC) \\
55757.41 & 44.41 & Grism 5 & 5000-10250 & 415 & 16.8 & NOT (ALFOSC) \\
55757.41 & 44.41 & gt300 & 3200-7700 & 555 & 9.0 & AS-1.22m (DU440) \\
55758.39 & 45.39 & IJ & 9000-14500 & 333 & ... & TNG (NICS) \\
55758.42 & 45.42 & HK & 14000-25000 & 333 & ... & TNG (NICS) \\
55760.38 & 47.38 & b200 & 3300-8700 & ... & 12.0 & CA-2.2m (CAFOS) \\
55762.39 & 49.39 & Grism 4 & 3200-9100 & 355 & 16.2 & NOT (ALFOSC) \\
55762.40 & 49.40 & Grism 5 & 5000-10250 & 415 & 16.8 & NOT (ALFOSC) \\
55765.40 & 52.40 & Grism 4 & 3200-9100 & 355 & 16.2 & NOT (ALFOSC) \\
55765.42 & 52.42 & Grism 5 & 5000-10250 & 415 & 16.8 & NOT (ALFOSC) \\
\hline
\end{tabular}}
\end{center}
\label{t_speclog}
\end{table*}

\setcounter{table}{8}
\begin{table*}[p]
\caption{Continued.}
\begin{center}
\scalebox{1.0}{
\begin{tabular}{l l l l l l l}
\hline\hline \\ [-1.5ex]
JD (+2400000) & Phase & Grism & Range & Resolution & Resolution & Telescope (Instrument)\\ [0.5ex]
(d) & (d) & & (\AA) & & (\AA) &\\ [0.5ex]
\hline \\ [-1.5ex]
55765.39 & 52.39 & IJ & 9000-14500 & 333 & ... & TNG (NICS) \\
55765.42 & 52.42 & HK & 14000-25000 & 333 & ... & TNG (NICS) \\
55771.41 & 58.41 & b200 & 3300-8700 & ... & 12.0 & CA-2.2m (CAFOS) \\
55771.41 & 58.41 & r200 & 6300-10500 & ... & 12.0 & CA-2.2m (CAFOS) \\
55780.39 & 67.39 & Grism 4 & 3200-9100 & 355 & 16.2 & NOT (ALFOSC) \\
55780.43 & 67.43 & zJ & 8900-15100 & 700 & ... & WHT (LIRIS) \\
55780.40 & 67.40 & HK & 14000-23800 & 700 & ... & WHT (LIRIS) \\
55784.40 & 71.40 & b200 & 3300-8700 & ... & 12.0 & CA-2.2m (CAFOS) \\
55784.40 & 71.40 & r200 & 6300-10500 & ... & 12.0 & CA-2.2m (CAFOS) \\
55795.39 & 82.39 & b200 & 3300-8700 & ... & 12.0 & CA-2.2m (CAFOS) \\
55795.39 & 82.39 & r200 & 6300-10500 & ... & 12.0 & CA-2.2m (CAFOS) \\
55801.37 & 88.37 & IJ & 9000-14500 & 333 & ... & TNG (NICS) \\
55801.40 & 88.40 & HK & 14000-25000 & 333 & ... & TNG (NICS) \\
55802.37 & 89.37 & gt300 & 3200-7700 & 396 & 12.6 & AS-1.22m (DU440) \\
55804.36 & 91.36 & R300B & 3200-5300 & ... & 4.1 & WHT (ISIS) \\
55804.36 & 91.36 & R158R & 5300-10000 & ... & 7.7 & WHT (ISIS) \\
55812.36 & 99.36 & b200 & 3300-8700 & ... & 12.0 & CA-2.2m (CAFOS) \\
55812.36 & 99.36 & r200 & 6300-10500 & ... & 12.0 & CA-2.2m (CAFOS) \\
\hline
\end{tabular}}
\end{center}
\end{table*}

\begin{table*}[p]
\caption{Pseudo-bolometric UV to MIR lightcurve for SN 2011dh calculated from spectroscopic and photometric data. Random errors are given in the first parentheses and systematic lower and upper errors (arising from the distance and extinction) respectively in the second parentheses.}
\begin{center}
\scalebox{1.00}{
\begin{tabular}{llllll}
\hline\hline \\ [-1.5ex]
JD (+2400000) & Phase & L & JD (+2400000) & Phase & L \\ [0.5ex]
(d) & (d) & (10$^{41}$ erg s$^{-1}$) & (d) & (d) & (10$^{41}$ erg s$^{-1}$) \\ [0.5ex]
\hline \\ [-1.5ex]
55717.00 & 4.00 & 2.90 (0.01)  (0.66,1.24) & 55765.00 & 52.00 & 5.53 (0.02)  (1.18,2.02) \\
55718.00 & 5.00 & 3.57 (0.01)  (0.80,1.49) & 55766.00 & 53.00 & 5.42 (0.02)  (1.16,1.98) \\
55719.00 & 6.00 & 4.55 (0.01)  (1.02,1.88) & 55767.00 & 54.00 & 5.30 (0.02)  (1.13,1.94) \\
55720.00 & 7.00 & 5.67 (0.02)  (1.27,2.33) & 55768.00 & 55.00 & 5.19 (0.02)  (1.11,1.90) \\
55721.00 & 8.00 & 6.86 (0.02)  (1.53,2.81) & 55769.00 & 56.00 & 5.09 (0.02)  (1.09,1.86) \\
55722.00 & 9.00 & 8.05 (0.02)  (1.80,3.29) & 55770.00 & 57.00 & 4.99 (0.02)  (1.07,1.83) \\
55723.00 & 10.00 & 9.25 (0.03)  (2.06,3.77) & 55771.00 & 58.00 & 4.89 (0.02)  (1.05,1.79) \\
55724.00 & 11.00 & 10.38 (0.03)  (2.31,4.21) & 55772.00 & 59.00 & 4.79 (0.02)  (1.03,1.76) \\
55725.00 & 12.00 & 11.43 (0.03)  (2.54,4.62) & 55773.00 & 60.00 & 4.70 (0.02)  (1.01,1.73) \\
55726.00 & 13.00 & 12.39 (0.03)  (2.75,5.00) & 55774.00 & 61.00 & 4.61 (0.02)  (0.99,1.69) \\
55727.00 & 14.00 & 13.25 (0.03)  (2.94,5.32) & 55775.00 & 62.00 & 4.52 (0.02)  (0.97,1.66) \\
55728.00 & 15.00 & 14.01 (0.03)  (3.10,5.62) & 55776.00 & 63.00 & 4.44 (0.02)  (0.95,1.63) \\
55729.00 & 16.00 & 14.69 (0.04)  (3.25,5.87) & 55777.00 & 64.00 & 4.35 (0.02)  (0.93,1.60) \\
55730.00 & 17.00 & 15.31 (0.05)  (3.39,6.11) & 55778.00 & 65.00 & 4.27 (0.02)  (0.91,1.57) \\
55731.00 & 18.00 & 15.86 (0.05)  (3.51,6.32) & 55779.00 & 66.00 & 4.19 (0.02)  (0.90,1.54) \\
55732.00 & 19.00 & 16.30 (0.05)  (3.60,6.48) & 55780.00 & 67.00 & 4.11 (0.02)  (0.88,1.52) \\
55733.00 & 20.00 & 16.57 (0.05)  (3.66,6.57) & 55781.00 & 68.00 & 4.03 (0.02)  (0.86,1.49) \\
55734.00 & 21.00 & 16.67 (0.05)  (3.67,6.59) & 55782.00 & 69.00 & 3.95 (0.02)  (0.85,1.46) \\
55735.00 & 22.00 & 16.52 (0.07)  (3.63,6.49) & 55783.00 & 70.00 & 3.88 (0.02)  (0.83,1.43) \\
55736.00 & 23.00 & 16.12 (0.06)  (3.54,6.30) & 55784.00 & 71.00 & 3.80 (0.02)  (0.82,1.41) \\
55737.00 & 24.00 & 15.53 (0.06)  (3.40,6.03) & 55785.00 & 72.00 & 3.73 (0.02)  (0.80,1.38) \\
55738.00 & 25.00 & 14.81 (0.05)  (3.23,5.71) & 55786.00 & 73.00 & 3.66 (0.02)  (0.78,1.35) \\
55739.00 & 26.00 & 14.00 (0.04)  (3.05,5.36) & 55787.00 & 74.00 & 3.58 (0.02)  (0.77,1.33) \\
55740.00 & 27.00 & 13.17 (0.04)  (2.86,5.01) & 55788.00 & 75.00 & 3.51 (0.02)  (0.75,1.30) \\
55741.00 & 28.00 & 12.38 (0.03)  (2.68,4.68) & 55789.00 & 76.00 & 3.45 (0.02)  (0.74,1.28) \\
55742.00 & 29.00 & 11.65 (0.04)  (2.52,4.38) & 55790.00 & 77.00 & 3.38 (0.01)  (0.73,1.26) \\
55743.00 & 30.00 & 11.00 (0.03)  (2.37,4.11) & 55791.00 & 78.00 & 3.32 (0.01)  (0.71,1.23) \\
55744.00 & 31.00 & 10.44 (0.03)  (2.24,3.88) & 55792.00 & 79.00 & 3.24 (0.01)  (0.70,1.20) \\
55745.00 & 32.00 & 9.94 (0.03)  (2.14,3.69) & 55793.00 & 80.00 & 3.18 (0.01)  (0.68,1.18) \\
55746.00 & 33.00 & 9.51 (0.03)  (2.04,3.52) & 55794.00 & 81.00 & 3.11 (0.01)  (0.67,1.16) \\
55747.00 & 34.00 & 9.12 (0.03)  (1.96,3.37) & 55795.00 & 82.00 & 3.05 (0.01)  (0.66,1.14) \\
55748.00 & 35.00 & 8.77 (0.03)  (1.88,3.23) & 55796.00 & 83.00 & 2.99 (0.01)  (0.64,1.11) \\
55749.00 & 36.00 & 8.46 (0.03)  (1.81,3.11) & 55797.00 & 84.00 & 2.93 (0.01)  (0.63,1.09) \\
55750.00 & 37.00 & 8.18 (0.03)  (1.75,3.00) & 55798.00 & 85.00 & 2.87 (0.01)  (0.62,1.07) \\
55751.00 & 38.00 & 7.92 (0.03)  (1.69,2.90) & 55799.00 & 86.00 & 2.81 (0.01)  (0.61,1.05) \\
55752.00 & 39.00 & 7.68 (0.02)  (1.64,2.81) & 55800.00 & 87.00 & 2.75 (0.01)  (0.59,1.03) \\
55753.00 & 40.00 & 7.46 (0.02)  (1.59,2.73) & 55801.00 & 88.00 & 2.70 (0.01)  (0.58,1.01) \\
55754.00 & 41.00 & 7.24 (0.02)  (1.55,2.65) & 55802.00 & 89.00 & 2.64 (0.01)  (0.57,0.99) \\
55755.00 & 42.00 & 7.02 (0.02)  (1.50,2.57) & 55803.00 & 90.00 & 2.59 (0.01)  (0.56,0.97) \\
55756.00 & 43.00 & 6.82 (0.02)  (1.46,2.49) & 55804.00 & 91.00 & 2.54 (0.01)  (0.55,0.95) \\
55757.00 & 44.00 & 6.64 (0.02)  (1.42,2.43) & 55805.00 & 92.00 & 2.49 (0.01)  (0.54,0.93) \\
55758.00 & 45.00 & 6.47 (0.02)  (1.38,2.36) & 55806.00 & 93.00 & 2.44 (0.01)  (0.53,0.91) \\
55759.00 & 46.00 & 6.32 (0.02)  (1.35,2.31) & 55807.00 & 94.00 & 2.39 (0.01)  (0.51,0.90) \\
55760.00 & 47.00 & 6.17 (0.02)  (1.32,2.25) & 55808.00 & 95.00 & 2.34 (0.01)  (0.50,0.88) \\
55761.00 & 48.00 & 6.03 (0.02)  (1.29,2.20) & 55809.00 & 96.00 & 2.29 (0.01)  (0.49,0.86) \\
55762.00 & 49.00 & 5.90 (0.02)  (1.26,2.16) & 55810.00 & 97.00 & 2.24 (0.01)  (0.48,0.84) \\
55763.00 & 50.00 & 5.78 (0.02)  (1.23,2.11) & 55811.00 & 98.00 & 2.20 (0.01)  (0.47,0.83) \\
55764.00 & 51.00 & 5.65 (0.02)  (1.21,2.07) & 55812.00 & 99.00 & 2.15 (0.01)  (0.47,0.81) \\ [0.5ex]
\hline
\end{tabular}}

\end{center}
\label{t_UV_MIR_bol}
\end{table*}

\appendix

\section{Photometric calibration}
\label{a_phot_cal}

The optical photometry was tied to the Johnson-Cousins (JC) and Sloan Digital Sky Survey (SDSS) systems. The NIR photometry was tied to the 2 Micron All Sky Survey (2MASS) system. Table \ref{t_sys_map} lists the filters used at each instrument and the mapping of these to the standard systems. Note that we have used JC-like $UBVRI$ filters and SDSS-like $gz$ filters at NOT whereas we have used JC-like $BV$ filters and SDSS-like $ugriz$ filters at LT and FTN. The JC-like $URI$ and SDSS-like $uri$ photometry were then tied to both the JC and SDSS systems to produce full sets of JC and SDSS photometry. The SWIFT photometry was tied to the natural (photon count based) Vega system although the SWIFT $UBV$ photometry was also tied to the JC system for comparison. The Spitzer photometry was tied to the natural (energy flux based) Vega system.

\begin{table*}[p]
\caption{Mapping of natural systems to standard systems.}
\begin{center}
\begin{tabular}{l l l}
\hline\hline \\ [-1.5ex]
Telescope (Instrument) & Natural system & Standard system \\ [0.5ex]
\hline \\ [-1.5ex]
NOT (ALFOSC) & Bessel U \#7 & JC U, SDSS u \\
NOT (ALFOSC) & Bessel B \#74 & JC B \\
NOT (ALFOSC) & Bessel V \#75 & JC V \\
NOT (ALFOSC) & Bessel R \#76 & JC R, SDSS r \\
NOT (ALFOSC) & Interference i \#12 & JC I, SDSS i \\
NOT (ALFOSC) & SDSS g \#120 & SDSS g \\
NOT (ALFOSC) & SDSS z \#112 & SDSS z \\
LT (RATCam) & Sloan u' & SDSS u, JC U \\ 
LT (RATCam) & Sloan g' & SDSS g \\
LT (RATCam) & Sloan r' & SDSS r, JC R \\
LT (RATCam) & Sloan i' & SDSS i, JC I \\
LT (RATCam) & Sloan z' & SDSS z \\
LT (RATCam) & Bessel B & JC B \\
LT (RATCam) & Bessel V & JC V \\
CA (CAFOS) & Johnson U 370/47b & JC U \\
CA (CAFOS) & Johnson B 451/73 & JC B \\
CA (CAFOS) & Johnson V 534/97b & JC V \\
CA (CAFOS) & Cousins R 641/158 & JC R \\
CA (CAFOS) & Johnson I 850/150b & JC I \\
ASIAGO (AFOSC) & Bessel U & JC U \\
ASIAGO (AFOSC) & Bessel B & JC B \\
ASIAGO (AFOSC) & Bessel V & JC V \\
ASIAGO (AFOSC) & Bessel R & JC R \\
ASIAGO (AFOSC) & Gunn i & JC I \\
ASIAGO (SCHMIDT) & B & JC B \\
ASIAGO (SCHMIDT) & V & JC V \\
ASIAGO (SCHMIDT) & R & JC R \\
ASIAGO (SCHMIDT) & I & JC I \\
FTN (FS02) & SDSS G & SDSS g \\
FTN (FS02) & SDSS R & SDSS r, JC R \\
FTN (FS02) & SDSS I & SDSS i, JC I \\
FTN (FS02) & Pann Starrs Z & SDSS z \\
MONTSEC (CCD) & B & JC B \\
MONTSEC (CCD) & V & JC V \\
MONTSEC (CCD) & R & JC R \\
MONTSEC (CCD) & I & JC I \\
TNG (LRS) & Johnson U & JC U \\
TNG (LRS) & Johnson B & JC B \\
TNG (LRS) & Johnson V & JC V \\
TNG (LRS) & Cousins R & JC R \\
TNG (LRS) & Cousins I & JC I \\
TNG (NICS) & J & 2MASS J \\
TNG (NICS) & H & 2MASS H \\
TNG (NICS) & K & 2MASS K \\
TCS (CAIN) & J & 2MASS J \\
TCS (CAIN) & H & 2MASS H \\
TCS (CAIN) & Kshort & 2MASS K \\
WHT (LIRIS) & j & 2MASS J \\
WHT (LIRIS) & h & 2MASS H \\
WHT (LIRIS) & ks & 2MASS K \\
NOT (NOTCAM) & J & 2MASS J \\
NOT (NOTCAM) & H & 2MASS H \\
NOT (NOTCAM) & Ks & 2MASS K \\
CA (O2000) & J & 2MASS J \\
CA (O2000) & H & 2MASS H \\
CA (O2000) & KS & 2MASS K \\
LBT (LUCIFER) & J & 2MASS J \\
LBT (LUCIFER) & H & 2MASS H \\
LBT (LUCIFER) & Ks & 2MASS K \\
\hline
\end{tabular}
\end{center}
\label{t_sys_map}
\end{table*}

\subsection{Calibration method}

The SN photometry was calibrated using reference stars within the SN field. These reference stars, in turn, were calibrated using standard fields. The calibration was performed using the {\sc sne} pipeline. To calibrate the SN photometry we fitted transformation equations of the type $m_{i}^{\mathrm{sys}}=m_{i}^{\mathrm{ins}}+C_{i,jk} (m_{j}^{\mathrm{sys}}-m_{k}^{\mathrm{sys}})+Z_{i}$, where $m_{i}^{\mathrm{sys}}$, $m_{i}^{\mathrm{ins}}$ and $Z_{i}$ are the system and instrumental magnitudes and the zeropoint for band $i$ respectively and $C_{i,jk}$ is the colour-term coefficient for band $i$ using the colour $jk$. The magnitudes of the SN was evaluated both using these transformation equations and by the use of S-corrections. In the latter case the linear colour-terms are replaced by the S-corrections as determined from the SN spectra and the filter response functions of the natural systems of the instruments and the standard systems. We will discuss S-corrections in Appendix~\ref{a_scorr} and compare the results from the two methods. In the end we have decided to use the S-corrected photometry for all bands except the JC $U$ and SDSS $u$ bands. To calibrate the reference star photometry we fitted transformation equations of the type $m_{i}^{\mathrm{sys}}=m_{i}^{\mathrm{ins}}+C_{i,jk} (m_{j}^{\mathrm{sys}}-m_{k}^{\mathrm{sys}})+Z_{i}+e_{i} X_{i}$, where $X_{i}$ and $e_{i}$ are the airmass and the extinction coefficient for band $i$ respectively. The magnitudes of the reference stars were evaluated using these transformation equations and averaged using a magnitude error limit (0.05 mag) and mild (3 $\sigma$) rejection. Both measurement errors and calibration errors in the fitted quantities were propagated using standard methods. The errors in the reference star magnitudes were calculated as the standard deviation of all measurements corrected for the degrees of freedom.

The coefficients of the linear colour terms ($C_{i,jk}$) used to transform from the natural system of the instruments to the JC and SDSS standard systems were determined separately. For each instrument, system and band we determined the coefficient by least-square fitting of a common value to a large number of observations. For the NOT and the LT we also fitted the coefficients of a cross-term between colour and airmass for $U$ and $B$ to correct for the change in the filter response functions due to the variation of the extinction with airmass. However, given the colour and airmass range spanned by our observations, the correction turned out to be at the few-percent level and we decided to drop it. Because of the lower precision in the 2MASS catalogue as compared to the Landolt and SDSS catalogues we could not achieve the desired precision in measured 2MASS colour-term coefficients. Therefore we have used synthetic colour terms computed for a blackbody SED using the NIR filter response functions described in Sect. \ref{a_scorr}. The measured JC and SDSS and synthetic 2MASS colour-term coefficients determined for each instrument are listed in Tables \ref{t_cconst_jc}-\ref{t_cconst_2mass}.

\begin{table*}[p]
\caption{JC $UBVRI$ colour terms for all telescope/instrument combinations. Errors are given in parentheses.}
\begin{center}
\scalebox{0.70}{
\begin{tabular}{l l l l l l l l l l l l l l l}
\hline\hline \\ [-1.5ex]
CT & NOT (ALFOSC) & LT (RATCam) & FTN (FS02) & CA-2.2m (CAFOS) & AS-Schmidt (SBIG) & AS-1.82m (AFOSC) & TNG (LRS) & TJO (MEIA) \\ [0.5ex]
\hline \\ [-1.5ex]
C$_{U,UB}$ & 0.127 (0.005) & 0.018 (0.006) & ...& 0.186 (0.025) & ...& 0.179 (0.048) & 0.181 (0.020) & ...\\ 
C$_{U,UR}$ & ...& 0.009 (0.003) & ...& 0.089 (0.012) & ...& ...& ...& ...\\ 
C$_{U,UV}$ & ...& ...& ...& ...& ...& 0.106 (0.028) & 0.107 (0.011) & ...\\ 
C$_{B,BV}$ & 0.039 (0.005) & 0.059 (0.005) & ...& 0.135 (0.010) & 0.133 (0.041) & 0.035 (0.014) & 0.071 (0.004) & 0.237 (0.016) \\ 
C$_{B,UB}$ & ...& 0.039 (0.004) & ...& 0.095 (0.007) & ...& 0.024 (0.010) & 0.050 (0.006) & ...\\ 
C$_{B,BR}$ & ...& ...& ...& ...& 0.086 (0.026) & ...& ...& ...\\ 
C$_{B,BI}$ & ...& ...& ...& ...& 0.065 (0.020) & ...& ...& ...\\ 
C$_{V,BV}$ & -0.048 (0.007) & -0.058 (0.006) & ...& -0.036 (0.009) & -0.075 (0.043) & 0.034 (0.016) & -0.053 (0.009) & -0.038 (0.008) \\ 
C$_{V,VR}$ & ...& -0.103 (0.009) & ...& -0.060 (0.016) & -0.083 (0.069) & 0.064 (0.029) & -0.096 (0.018) & -0.072 (0.014) \\ 
C$_{R,VR}$ & -0.064 (0.010) & -0.180 (0.006) & ...& 0.051 (0.014) & -0.054 (0.050) & 0.043 (0.032) & -0.031 (0.015) & -0.061 (0.020) \\ 
C$_{R,RI}$ & ...& -0.183 (0.011) & -0.324 (0.014) & 0.043 (0.011) & -0.050 (0.036) & 0.048 (0.032) & -0.029 (0.015) & -0.066 (0.020) \\ 
C$_{I,VI}$ & -0.021 (0.004) & -0.085 (0.006) & ...& 0.072 (0.020) & 0.024 (0.018) & -0.012 (0.015) & 0.061 (0.018) & 0.041 (0.010) \\ 
C$_{I,RI}$ & ...& -0.168 (0.012) & -0.231 (0.019) & 0.128 (0.038) & 0.039 (0.031) & -0.021 (0.030) & 0.122 (0.034) & 0.082 (0.022) \\ 
C$_{I,BI}$ & ...& ...& ...& ...& 0.013 (0.011) & ...& ...& ...\\ 
\hline
\end{tabular}}

\end{center}
\label{t_cconst_jc}
\end{table*}

\begin{table*}[p]
\caption{SDSS $ugriz$ colour terms for all telescope/instrument combinations. Errors are given in parentheses.}
\begin{center}
\scalebox{1.00}{
\begin{tabular}{l l l l l l l l l l l}
\hline\hline \\ [-1.5ex]
CT & NOT (ALFOSC) & LT (RATCam) & FTN (FS02) \\ [0.5ex]
\hline \\ [-1.5ex]
C$_{u,ug}$ & 0.043 (0.027) & 0.032 (0.024) & ...\\ 
C$_{u,ur}$ & 0.029 (0.018) & 0.021 (0.015) & ...\\ 
C$_{g,gr}$ & 0.009 (0.011) & 0.102 (0.012) & 0.200 (0.038) \\ 
C$_{g,ug}$ & 0.007 (0.008) & 0.056 (0.010) & ...\\ 
C$_{r,gr}$ & 0.077 (0.019) & 0.023 (0.016) & 0.032 (0.014) \\ [0.5ex]
C$_{r,ri}$ & 0.162 (0.016) & 0.057 (0.020) & 0.045 (0.014) \\ [0.5ex]
C$_{i,ri}$ & 0.185 (0.027) & 0.094 (0.009) & 0.045 (0.011) \\ 
C$_{i,iz}$ & 0.310 (0.038) & 0.159 (0.015) & 0.082 (0.018) \\ 
C$_{z,iz}$ & -0.094 (0.046) & -0.081 (0.065) & -0.119 (0.023) \\ 
C$_{z,rz}$ & -0.027 (0.019) & -0.024 (0.028) & ...\\ 
\hline
\end{tabular}}

\end{center}
\label{t_cconst_sloan}
\end{table*}

\begin{table*}[p]
\caption{Synthetic 2MASS $JHK$ colour terms for all telescope/instrument combinations.}
\begin{center}
\scalebox{1.00}{
\begin{tabular}{l l l l l l l}
\hline\hline \\ [-1.5ex]
CT & TNG (NICS) & NOT (NOTCAM) & TCS (CAIN) & LBT (LUCIFER) & WHT (LIRIS) & CA-3.5m (O2000) \\ [0.5ex]
\hline \\ [-1.5ex]
C$_{J,JH}$ & 0.028 & 0.039 & 0.026 & -0.032 & 0.056 & -0.092 \\ 
C$_{J,JK}$ & 0.017 & 0.024 & 0.016 & -0.020 & 0.034 & -0.055 \\ 
C$_{H,JH}$ & -0.006 & -0.008 & 0.051 & 0.021 & -0.011 & 0.019 \\ 
C$_{H,HK}$ & -0.008 & -0.011 & 0.077 & 0.032 & -0.016 & 0.028 \\ 
C$_{K,JK}$ & 0.025 & -0.005 & 0.027 & 0.001 & -0.003 & -0.007 \\ 
C$_{K,HK}$ & 0.062 & -0.012 & 0.067 & 0.001 & -0.007 & -0.018 \\ 
\hline
\end{tabular}}

\end{center}
\label{t_cconst_2mass}
\end{table*}

\subsection{JC calibration}

\begin{figure}[tb]
\includegraphics[width=0.48\textwidth,angle=0]{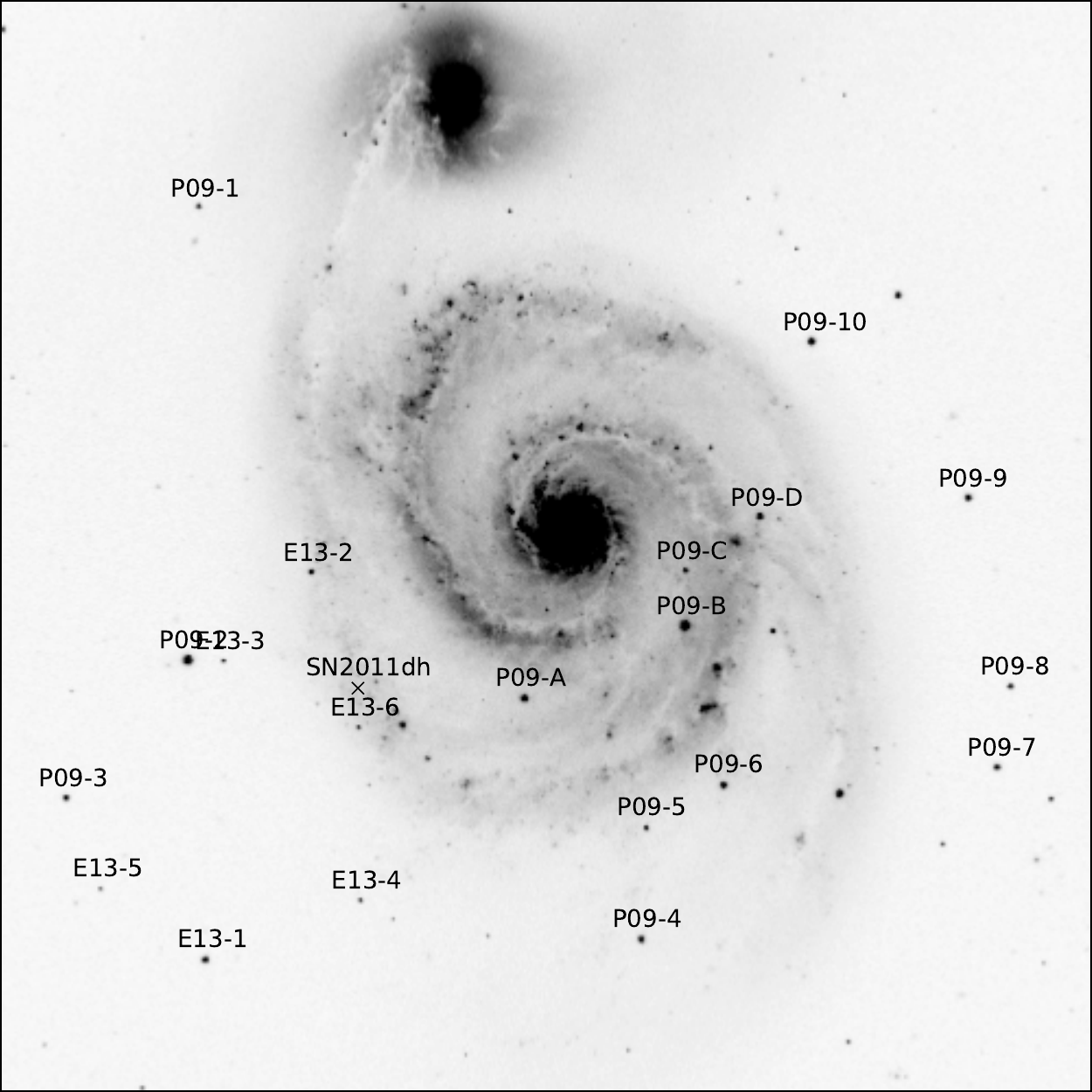}
\caption{Reference stars used for calibration of the optical photometry marked on a SDSS $r$ band image.}
\label{f_refstar_opt}
\end{figure}

The optical photometry was tied to the JC system using the reference stars presented in \citet[hereafter \citetalias{Pas09}]{Pas09} as well as a number of additional fainter stars close to the SN. In the following the reference stars from \citetalias{Pas09} will be abbreviated as P09-N and those added in this paper as E13-N. Those reference stars, in turn, have been tied to the JC system using standard fields from \citet{Lan83,Lan92}. Taking advantage of the large number of standard star observations obtained with the LT we have re-measured the magnitudes of the P09 reference stars within the LT field of view (FOV). The mean and root mean square (RMS) of the difference was at the few-percent level for the $B$, $V$, $R$ and $I$ bands and at the 10-percent level for the $U$ band except for P09-3 which differed considerably. We have also re-measured the magnitudes of the Landolt standard stars and the mean and RMS of the difference was at the few-percent level in all bands. This shows that, in spite of the SDSS-like nature of the LT $u$, $r$ and $i$ filters, the natural LT photometry transform to the JC system with good precision. In the end we chose to keep the \citetalias{Pas09} magnitudes, except for P09-3, and use the LT observations as confirmation. The magnitudes of the additional reference stars and P09-3 were determined using the remaining \citetalias{Pas09} reference stars and large number of deep, high quality NOT images of the SN field. The coordinates and magnitudes of the JC reference stars are listed Table~\ref{t_refstar_jc} and their positions marked in Fig.~\ref{f_refstar_opt}.

\begin{table*}[p]
\caption{JC $UBVRI$ magnitudes of local reference stars used to calibrate the photometry. Errors are given in parentheses.}
\begin{center}
\scalebox{1.00}{
\begin{tabular}{l l l l l l l l}
\hline\hline \\ [-1.5ex]
ID & RA & DEC & $U$ & $B$ & $V$ & $R$ & $I$  \\ [0.5ex]
& (h m s) & (\degr~\arcmin~\arcsec) & (mag) &(mag) &(mag) &(mag) &(mag)  \\
\hline \\ [-1.5ex]
P09-1 & $13~30~14.2$ & $+47~14~55$ & ...& 17.01 (0.18) & 16.27 (0.03) & 15.81 (0.08) & 15.44 (0.06)  \\
P09-2 & $13~30~14.9$ & $+47~10~27$ & 14.62 (0.04) & 14.32 (0.02) & 13.60 (0.01) & 13.19 (0.01) & 12.81 (0.01)  \\
P09-3 & $13~30~21.9$ & $+47~09~05$ & 16.63 (0.01) & 16.55 (0.01) & 15.92 (0.01) & 15.48 (0.01) & 15.10 (0.01)  \\
P09-4 & $13~29~48.5$ & $+47~07~42$ & 16.15 (0.05) & 16.23 (0.02) & 15.66 (0.02) & 15.23 (0.01) & 14.91 (0.03)  \\
P09-5 & $13~29~48.2$ & $+47~08~48$ & 18.88 (0.07) & 18.25 (0.07) & 17.36 (0.01) & 16.75 (0.02) & 16.28 (0.01)  \\
P09-6 & $13~29~43.7$ & $+47~09~14$ & 15.51 (0.02) & 15.85 (0.01) & 15.39 (0.02) & 15.05 (0.01) & 14.74 (0.01)  \\
P09-7 & $13~29~27.9$ & $+47~09~23$ & 17.13 (0.19) & 16.73 (0.01) & 16.06 (0.02) & 15.54 (0.01) & 15.10 (0.05)  \\
P09-8 & $13~29~27.1$ & $+47~10~12$ & 19.76 (0.02) & 18.72 (0.05) & 17.41 (0.04) & 15.82 (0.03) & 14.38 (0.03)  \\
P09-9 & $13~29~29.5$ & $+47~12~03$ & 18.01 (0.23) & 16.66 (0.03) & 15.68 (0.01) & 14.97 (0.01) & 14.36 (0.04)  \\
P09-10 & $13~29~38.6$ & $+47~13~36$ & 16.74 (0.04) & 16.20 (0.03) & 15.29 (0.01) & 14.75 (0.02) & 14.20 (0.02)  \\
P09-A & $13~29~55.3$ & $+47~10~05$ & 17.77 (0.03) & 16.34 (0.01) & 15.11 (0.01) & 14.33 (0.01) & 13.68 (0.01)  \\
P09-B & $13~29~45.9$ & $+47~10~47$ & 14.12 (0.02) & 14.01 (0.01) & 13.43 (0.01) & 13.06 (0.01) & 12.73 (0.01)  \\
P09-C & $13~29~45.9$ & $+47~11~20$ & 17.19 (0.04) & 17.15 (0.02) & 16.67 (0.01) & 16.29 (0.01) & 15.94 (0.01)  \\
P09-D & $13~29~41.6$ & $+47~11~52$ & 15.76 (0.03) & 15.77 (0.01) & 15.24 (0.01) & 14.85 (0.01) & 14.52 (0.01)  \\
E13-1 & $13~30~13.8$ & $+47~07~30$ & 17.35 (0.01) & 16.59 (0.01) & 15.70 (0.01) & 15.12 (0.01) & 14.67 (0.01)  \\
E13-2 & $13~30~07.6$ & $+47~11~19$ & 17.76 (0.01) & 17.26 (0.01) & 16.48 (0.01) & 16.02 (0.01) & 15.63 (0.01)  \\
E13-3 & $13~30~12.8$ & $+47~10~27$ & 18.94 (0.02) & 18.58 (0.01) & 17.82 (0.01) & 17.29 (0.01) & 16.84 (0.01)  \\
E13-4 & $13~30~04.8$ & $+47~08~05$ & 20.57 (0.03) & 19.46 (0.01) & 17.99 (0.01) & 16.98 (0.01) & 15.61 (0.01)  \\
E13-5 & $13~30~19.9$ & $+47~08~12$ & 19.79 (0.03) & 19.17 (0.01) & 18.34 (0.01) & 17.76 (0.01) & 17.31 (0.01)  \\
E13-6 & $13~30~04.9$ & $+47~09~47$ & 20.98 (0.06) & 19.61 (0.01) & 18.20 (0.01) & 17.28 (0.01) & 16.41 (0.01)  \\
\hline
\end{tabular}}

\end{center}
\label{t_refstar_jc}
\end{table*}

\subsection{SDSS calibration}

The optical photometry was tied to the SDSS system using the subset of the reference stars within the LT FOV. Those reference stars, in turn, have been tied to the SDSS system using fields covered by the SDSS DR8 catalogue \citep{Aih11}. The calibration was not straightforward as the SN field and a number of the LT standard fields were not well covered by the catalogue and many of the brighter stars in the fields covered by the catalogue were marked as saturated in the $g$, $r$ and $i$ bands. The procedure used to calibrate the reference star magnitudes was as follows. First we re-measured the $g$, $r$ and $i$ magnitudes for all stars marked as saturated using the remaining stars in the LT standard fields covered by the catalogue. We then measured the magnitudes for the stars in the fields not covered by the catalogue and finally we measured the reference star magnitudes using all the LT standard field observations. For reference, our measured SDSS magnitudes for the LT standard fields are listed in Table~\ref{t_standard_sloan}. Magnitudes of stars covered by the catalogue and not marked as saturated were adopted from the catalogue. The coordinates and the magnitudes of the SDSS reference stars are listed in Table~\ref{t_refstar_sloan} and their positions marked in Fig.~\ref{f_refstar_opt}. The magnitudes of P09-3, E13-1 and E13-5 were adopted from the catalogue. The mean and RMS of the difference between measured and catalogue magnitudes for stars with measured and catalogue errors less than 5 percent was less than 5 percent in all bands.

\begin{table*}[p]
\caption{SDSS $ugriz$ magnitudes for the standard fields PG0231+051, PG1047+003, PG1525-071, PG2331+046 and Mark-A. Errors are given in parentheses.}
\begin{center}
\scalebox{1.00}{
\begin{tabular}{l l l l l l l l}
\hline\hline \\ [-1.5ex]
ID & RA & DEC & $u$ & $g$ & $r$ & $i$ & $z$  \\ [0.5ex]
& (h m s) & (\degr~\arcmin~\arcsec) & (mag) &(mag) &(mag) &(mag) &(mag)  \\
\hline \\ [-1.5ex]
PG0231+051E & $02~33~28.8$ & $+05~19~48$ & 15.56 (0.01) & ...& ...& ...& 13.41 (0.02)  \\
PG0231+051D & $02~33~34.0$ & $+05~19~30$ & 17.00 (0.02) & 14.59 (0.02) & 13.62 (0.01) & 13.24 (0.02) & 13.04 (0.02)  \\
PG0231+051A & $02~33~40.0$ & $+05~17~40$ & 14.63 (0.02) & 13.07 (0.02) & 12.55 (0.01) & 12.39 (0.01) & 12.31 (0.02)  \\
PG0231+051 & $02~33~41.3$ & $+05~18~43$ & 15.36 (0.01) & 15.80 (0.02) & 16.30 (0.02) & 16.61 (0.02) & 16.96 (0.02)  \\
PG0231+051B & $02~33~45.5$ & $+05~17~33$ & 18.28 (0.02) & 15.54 (0.02) & 14.16 (0.01) & 13.36 (0.01) & 12.93 (0.02)  \\
PG0231+051C & $02~33~48.1$ & $+05~20~26$ & 15.33 (0.01) & 13.97 (0.02) & 13.49 (0.01) & 13.33 (0.01) & 13.29 (0.02)  \\
PG1047+003 & $10~50~02.8$ & $-00~00~36$ & ...& 13.23 (0.02) & 13.68 (0.01) & 14.06 (0.02) & 14.40 (0.03)  \\
PG1047+003A & $10~50~05.6$ & $-00~01~10$ & 15.22 (0.01) & 13.86 (0.02) & 13.32 (0.01) & 13.13 (0.01) & 13.00 (0.03)  \\
PG1047+003B & $10~50~07.9$ & $-00~02~04$ & 16.45 (0.01) & 15.02 (0.02) & 14.54 (0.02) & 14.45 (0.01) & 14.36 (0.03)  \\
PG1047+003C & $10~50~13.6$ & $-00~00~31$ & 14.04 (0.01) & 12.72 (0.02) & 12.27 (0.01) & 12.14 (0.01) & 12.20 (0.03)  \\
PG1525-071 & $15~28~11.5$ & $-07~16~32$ & 14.44 (0.01) & 14.84 (0.01) & 15.24 (0.01) & 15.56 (0.01) & 15.90 (0.01)  \\
PG1525-071D & $15~28~12.0$ & $-07~16~39$ & 17.84 (0.03) & 16.64 (0.01) & 16.12 (0.01) & 15.96 (0.01) & 15.89 (0.01)  \\
PG1525-071A & $15~28~13.4$ & $-07~16~01$ & 15.38 (0.01) & 13.88 (0.01) & 13.29 (0.01) & 13.09 (0.01) & 13.01 (0.01)  \\
PG1525-071B & $15~28~14.3$ & $-07~16~13$ & 18.01 (0.04) & 16.73 (0.01) & 16.17 (0.01) & 15.98 (0.01) & 15.88 (0.01)  \\
PG1525-071C & $15~28~16.5$ & $-07~14~30$ & 16.57 (0.02) & 14.10 (0.01) & 13.20 (0.01) & 12.90 (0.01) & 12.74 (0.01)  \\
PG2331+055 & $23~33~44.4$ & $+05~46~39$ & 15.43 (0.02) & 15.04 (0.02) & 15.30 (0.02) & 15.52 (0.02) & 15.70 (0.02)  \\
PG2331+055A & $23~33~49.3$ & $+05~46~52$ & 14.85 (0.02) & 13.36 (0.01) & 12.85 (0.01) & 12.65 (0.01) & 12.56 (0.02)  \\
PG2331+055B & $23~33~51.1$ & $+05~45~08$ & 16.79 (0.02) & 15.10 (0.02) & 14.47 (0.02) & 14.21 (0.02) & 14.07 (0.02)  \\
MarkA4 & $20~43~53.5$ & $-10~45~05$ & 16.53 (0.02) & 15.14 (0.01) & 14.53 (0.01) & 14.25 (0.01) & 14.10 (0.01)  \\
MarkA2 & $20~43~54.9$ & $-10~45~31$ & 16.10 (0.01) & 14.82 (0.01) & 14.37 (0.01) & 14.21 (0.01) & 14.15 (0.01)  \\
MarkA1 & $20~43~58.4$ & $-10~47~12$ & 17.26 (0.02) & 16.16 (0.01) & 15.73 (0.01) & 15.57 (0.01) & 15.51 (0.01)  \\
MarkA & $20~43~59.2$ & $-10~47~41$ & 12.61 (0.01) & 12.97 (0.01) & 13.47 (0.01) & 13.78 (0.01) & 14.12 (0.01)  \\
MarkA3 & $20~44~03.8$ & $-10~45~37$ & 17.20 (0.02) & 15.28 (0.01) & 14.48 (0.01) & 14.17 (0.01) & 13.99 (0.01)  \\ [0.5ex]
\hline
\end{tabular}}

\end{center}
\label{t_standard_sloan}
\end{table*}

\begin{table*}[p]
\caption{SDSS $ugriz$ magnitudes of local reference stars used to calibrate the photometry. Errors are given in parentheses.}
\begin{center}
\scalebox{1.00}{
\begin{tabular}{l l l l l l l l}
\hline\hline \\ [-1.5ex]
ID & RA & DEC & $u$ & $g$ & $r$ & $i$ & $z$  \\ [0.5ex]
& (h m s) & (\degr~\arcmin~\arcsec) & (mag) &(mag) &(mag) &(mag) &(mag)  \\
\hline \\ [-1.5ex]
P09-1 & $13~30~14.2$ & $+47~14~55$ & 17.99 (0.02) & 16.64 (0.02) & 16.09 (0.01) & 15.86 (0.02) & 15.77 (0.02)  \\
P09-2 & $13~30~14.9$ & $+47~10~27$ & 15.59 (0.01) & 13.96 (0.02) & 13.42 (0.01) & 13.31 (0.02) & 13.21 (0.02)  \\
P09-3 & $13~30~21.9$ & $+47~09~05$ & ...& 16.15 (0.02) & 15.69 (0.02) & 15.52 (0.01) & 15.39 (0.02)  \\
P09-4 & $13~29~48.5$ & $+47~07~42$ & 17.07 (0.02) & 15.90 (0.01) & 15.45 (0.01) & 15.30 (0.01) & 15.25 (0.01)  \\
P09-5 & $13~29~48.2$ & $+47~08~48$ & 19.33 (0.17) & 17.84 (0.02) & 17.00 (0.01) & 16.71 (0.01) & 16.53 (0.02)  \\
P09-6 & $13~29~43.7$ & $+47~09~14$ & 16.41 (0.01) & 15.55 (0.01) & 15.24 (0.01) & 15.13 (0.01) & 15.10 (0.01)  \\
P09-7 & $13~29~27.9$ & $+47~09~23$ & ...& ...& ...& ...& ... \\
P09-8 & $13~29~27.1$ & $+47~10~12$ & ...& ...& ...& ...& ... \\
P09-9 & $13~29~29.5$ & $+47~12~03$ & ...& ...& ...& ...& ... \\
P09-10 & $13~29~38.6$ & $+47~13~36$ & ...& ...& ...& ...& ... \\
P09-A & $13~29~55.3$ & $+47~10~05$ & 18.52 (0.03) & 15.86 (0.01) & 14.65 (0.01) & 14.22 (0.01) & 13.99 (0.01)  \\
P09-B & $13~29~45.9$ & $+47~10~47$ & 14.95 (0.02) & 13.71 (0.01) & 13.26 (0.01) & 13.15 (0.01) & 13.12 (0.01)  \\
P09-C & $13~29~45.9$ & $+47~11~20$ & 17.88 (0.11) & 16.88 (0.01) & 16.48 (0.01) & 16.38 (0.01) & 16.33 (0.05)  \\
P09-D & $13~29~41.6$ & $+47~11~52$ & 16.64 (0.04) & 15.49 (0.01) & 15.09 (0.01) & 14.99 (0.01) & 14.95 (0.02)  \\
E13-1 & $13~30~13.8$ & $+47~07~30$ & 18.03 (0.02) & 16.12 (0.02) & 15.32 (0.02) & 15.08 (0.01) & 14.93 (0.02)  \\
E13-2 & $13~30~07.6$ & $+47~11~19$ & ...& 16.86 (0.03) & 16.23 (0.03) & 16.04 (0.01) & 15.96 (0.03)  \\
E13-3 & $13~30~12.8$ & $+47~10~27$ & ...& 18.23 (0.02) & 17.53 (0.01) & 17.32 (0.04) & 17.04 (0.02)  \\
E13-4 & $13~30~04.8$ & $+47~08~05$ & ...& 18.76 (0.02) & 17.37 (0.01) & 16.27 (0.01) & 15.61 (0.01)  \\
E13-5 & $13~30~19.9$ & $+47~08~12$ & 20.54 (0.06) & 18.67 (0.02) & 17.95 (0.02) & 17.72 (0.02) & 17.56 (0.03)  \\
E13-6 & $13~30~04.9$ & $+47~09~47$ & ...& 18.96 (0.03) & 17.59 (0.01) & 16.90 (0.01) & 16.48 (0.02)  \\
\hline
\end{tabular}}

\end{center}
\label{t_refstar_sloan}
\end{table*}

\subsection{2MASS calibration}

The NIR photometry was tied to the 2MASS system using all stars within 7 arc minutes distance from the SN with $J$ magnitude brighter than 18.0 detected in deep UKIRT imaging of the SN field. This includes the optical reference stars as well as $\sim$50 additional stars, although for most observations the small FOV prevented use of more than about 10 of these. Those reference stars, in turn, have been tied to the 2MASS system using all stars from the 2MASS Point Source catalogue \citep{Skr06} with $J$ magnitude error less than 0.05 mag within the 13.65$\times$13.65 arc minute FOV. The coordinates and magnitudes for the 2MASS reference stars are listed in Table~\ref{t_refstar_2mass}. Magnitudes of stars covered by the catalogue and with errors less than 0.05 were adopted from the catalogue. The mean and RMS of the difference between measured and catalogue magnitudes for stars with measured and catalogue errors less than 5 percent was less than 5 percent in all bands.

\begin{table*}[p]
\caption{2MASS $JHK$ magnitudes of local reference stars used to calibrate the photometry. Errors are given in parentheses.}
\begin{center}
\scalebox{1.00}{
\begin{tabular}{l l l l l l}
\hline\hline \\ [-1.5ex]
ID & RA & DEC & $J$ & $H$ & $K$  \\ [0.5ex]
& (h m s) & (\degr~\arcmin~\arcsec) & (mag) &(mag) &(mag)  \\
\hline \\ [-1.5ex]
P09-1 & $13~30~14.3$ & $+47~14~55$ & 14.87 (0.01) & 14.43 (0.01) & 14.36 (0.01)  \\
P09-2 & $13~30~14.8$ & $+47~10~27$ & 12.35 (0.01) & 12.02 (0.01) & 11.96 (0.01)  \\
P09-3 & $13~30~21.9$ & $+47~09~05$ & 14.55 (0.01) & 14.13 (0.01) & 14.09 (0.01)  \\
P09-4 & $13~29~48.5$ & $+47~07~42$ & 14.38 (0.01) & 14.04 (0.01) & 13.99 (0.01)  \\
P09-5 & $13~29~48.2$ & $+47~08~48$ & 15.52 (0.01) & 14.96 (0.01) & 14.87 (0.01)  \\
P09-6 & $13~29~43.7$ & $+47~09~13$ & 14.28 (0.01) & 13.99 (0.01) & 13.93 (0.01)  \\
P09-7 & $13~29~27.8$ & $+47~09~23$ & 14.46 (0.01) & 14.00 (0.01) & 13.91 (0.01)  \\
P09-8 & $13~29~27.0$ & $+47~10~10$ & 12.90 (0.01) & 12.38 (0.01) & 12.12 (0.01)  \\
P09-9 & $13~29~29.4$ & $+47~12~03$ & 13.52 (0.01) & 12.91 (0.01) & 12.79 (0.01)  \\
P09-10 & $13~29~38.6$ & $+47~13~35$ & 13.46 (0.01) & 12.93 (0.01) & 12.79 (0.01)  \\
P09-A & $13~29~55.3$ & $+47~10~04$ & 12.84 (0.01) & 12.24 (0.01) & 12.12 (0.01)  \\
P09-B & $13~29~46.0$ & $+47~10~47$ & 12.25 (0.01) & 11.96 (0.01) & 11.90 (0.01)  \\
P09-C & $13~29~45.9$ & $+47~11~20$ & 15.49 (0.01) & 15.16 (0.01) & 15.10 (0.01)  \\
P09-D & $13~29~41.6$ & $+47~11~52$ & 14.06 (0.01) & 13.72 (0.01) & 13.66 (0.01)  \\
E13-1 & $13~30~13.8$ & $+47~07~30$ & 13.96 (0.01) & 13.46 (0.01) & 13.39 (0.01)  \\
E13-2 & $13~30~07.7$ & $+47~11~19$ & 15.00 (0.01) & 14.59 (0.01) & 14.52 (0.01)  \\
E13-3 & $13~30~12.8$ & $+47~10~26$ & 16.14 (0.01) & 15.64 (0.01) & 15.56 (0.01)  \\
E13-4 & $13~30~04.8$ & $+47~08~05$ & 14.30 (0.01) & 13.64 (0.01) & 13.41 (0.01)  \\
E13-5 & $13~30~19.9$ & $+47~08~11$ & 16.62 (0.01) & 16.09 (0.01) & 16.03 (0.01)  \\
E13-6 & $13~30~04.9$ & $+47~09~47$ & 15.31 (0.01) & 14.66 (0.01) & 14.48 (0.01)  \\
E13-7 & $13~29~31.0$ & $+47~08~38$ & 16.54 (0.01) & 16.12 (0.03) & 16.05 (0.01)  \\
E13-8 & $13~29~31.8$ & $+47~09~20$ & 17.47 (0.01) & 16.85 (0.03) & 16.71 (0.03)  \\
E13-9 & $13~29~33.6$ & $+47~14~03$ & 12.73 (0.01) & 12.13 (0.01) & 11.89 (0.01)  \\
E13-10 & $13~29~36.5$ & $+47~06~41$ & 15.31 (0.01) & 14.68 (0.01) & 14.47 (0.01)  \\
E13-11 & $13~29~36.9$ & $+47~09~07$ & 11.96 (0.01) & 11.39 (0.01) & 11.15 (0.01)  \\
E13-12 & $13~29~39.5$ & $+47~14~30$ & 15.43 (0.01) & 14.85 (0.01) & 14.58 (0.01)  \\
E13-13 & $13~29~40.9$ & $+47~10~44$ & 14.22 (0.01) & 13.57 (0.01) & 13.37 (0.01)  \\
E13-14 & $13~29~42.3$ & $+47~15~00$ & 17.61 (0.02) & 17.07 (0.02) & 16.99 (0.01)  \\
E13-15 & $13~29~44.4$ & $+47~12~32$ & 13.57 (0.01) & 13.09 (0.01) & 12.80 (0.01)  \\
E13-16 & $13~29~49.2$ & $+47~14~21$ & 17.19 (0.01) & 16.72 (0.03) & 16.48 (0.02)  \\
E13-17 & $13~29~49.3$ & $+47~06~36$ & 15.45 (0.01) & 14.78 (0.01) & 14.54 (0.01)  \\
E13-18 & $13~29~50.2$ & $+47~10~28$ & 17.91 (0.01) & 17.38 (0.01) & 17.21 (0.10)  \\
E13-19 & $13~29~51.6$ & $+47~14~14$ & 17.28 (0.01) & 16.67 (0.01) & 16.27 (0.01)  \\
E13-20 & $13~29~51.6$ & $+47~09~08$ & 17.92 (0.01) & 17.34 (0.04) & 17.05 (0.02)  \\
E13-21 & $13~29~54.0$ & $+47~10~57$ & 16.19 (0.01) & 15.50 (0.01) & 15.37 (0.02)  \\
E13-22 & $13~29~55.5$ & $+47~13~05$ & 17.88 (0.02) & 17.25 (0.01) & 17.12 (0.06)  \\
E13-23 & $13~29~56.2$ & $+47~14~53$ & 16.11 (0.01) & 15.73 (0.01) & 15.67 (0.01)  \\
E13-24 & $13~29~57.4$ & $+47~07~44$ & 16.37 (0.01) & 15.78 (0.01) & 15.55 (0.01)  \\
E13-25 & $13~29~58.0$ & $+47~04~54$ & 16.84 (0.01) & 16.46 (0.02) & 16.45 (0.01)  \\
E13-26 & $13~29~59.7$ & $+47~08~38$ & 17.00 (0.01) & 16.40 (0.01) & 16.20 (0.03)  \\
E13-27 & $13~30~01.7$ & $+47~14~30$ & 17.64 (0.02) & 17.02 (0.01) & 16.84 (0.05)  \\
E13-28 & $13~30~02.9$ & $+47~07~54$ & 17.19 (0.01) & 16.57 (0.01) & 16.48 (0.03)  \\
E13-29 & $13~30~04.3$ & $+47~09~13$ & 17.84 (0.01) & 17.37 (0.06) & 17.27 (0.03)  \\
E13-30 & $13~30~04.4$ & $+47~09~50$ & 17.20 (0.01) & 16.63 (0.07) & 16.38 (0.01)  \\
E13-31 & $13~30~05.7$ & $+47~15~38$ & 14.70 (0.01) & 14.13 (0.01) & 13.88 (0.01)  \\
E13-32 & $13~30~06.1$ & $+47~14~25$ & 17.31 (0.02) & 16.68 (0.02) & 16.45 (0.02)  \\
E13-33 & $13~30~07.3$ & $+47~10~01$ & 17.72 (0.02) & 17.24 (0.06) & 17.12 (0.03)  \\
E13-34 & $13~30~07.9$ & $+47~05~18$ & 13.67 (0.01) & 13.30 (0.01) & 13.27 (0.01)  \\
E13-35 & $13~30~08.0$ & $+47~04~09$ & 17.75 (0.01) & 17.13 (0.02) & 16.80 (0.04)  \\
E13-36 & $13~30~10.0$ & $+47~14~21$ & 17.94 (0.07) & 17.46 (0.03) & 17.21 (0.04)  \\
E13-37 & $13~30~11.4$ & $+47~05~11$ & 16.04 (0.01) & 15.50 (0.01) & 15.18 (0.01)  \\
\hline
\end{tabular}}

\end{center}
\label{t_refstar_2mass}
\end{table*}

\setcounter{table}{7}
\begin{table*}[p]
\caption{Continued.}
\begin{center}
\scalebox{1.00}{
\begin{tabular}{l l l l l l}
\hline\hline \\ [-1.5ex]
ID & RA & DEC & $J$ & $H$ & $K$  \\ [0.5ex]
& (h m s) & (\degr~\arcmin~\arcsec) & (mag) &(mag) &(mag)  \\
\hline \\ [-1.5ex]
E13-38 & $13~30~15.4$ & $+47~11~01$ & 17.68 (0.01) & 17.11 (0.04) & 16.88 (0.02)  \\
E13-39 & $13~30~17.4$ & $+47~06~14$ & 15.79 (0.01) & 15.16 (0.01) & 14.98 (0.01)  \\
E13-40 & $13~30~18.4$ & $+47~12~14$ & 17.81 (0.03) & 17.29 (0.02) & 17.04 (0.01)  \\
E13-41 & $13~30~21.6$ & $+47~10~09$ & 16.41 (0.01) & 15.78 (0.01) & 15.53 (0.01)  \\
E13-42 & $13~30~23.1$ & $+47~06~39$ & 16.75 (0.01) & 16.12 (0.01) & 15.99 (0.01)  \\
E13-43 & $13~30~26.7$ & $+47~07~26$ & 15.39 (0.01) & 14.79 (0.01) & 14.57 (0.01)  \\
E13-44 & $13~30~30.2$ & $+47~12~34$ & 16.34 (0.01) & 15.70 (0.01) & 15.44 (0.01)  \\
E13-45 & $13~30~31.8$ & $+47~12~44$ & 17.61 (0.01) & 16.99 (0.01) & 16.71 (0.06)  \\
E13-46 & $13~30~33.2$ & $+47~06~39$ & 15.50 (0.01) & 14.93 (0.01) & 14.67 (0.01)  \\
E13-47 & $13~30~33.8$ & $+47~14~03$ & 15.78 (0.01) & 15.10 (0.01) & 15.01 (0.01)  \\
E13-48 & $13~30~34.8$ & $+47~13~35$ & 16.12 (0.01) & 15.72 (0.01) & 15.68 (0.03)  \\
E13-49 & $13~30~34.8$ & $+47~12~29$ & 15.08 (0.01) & 14.47 (0.01) & 14.26 (0.01)  \\
E13-50 & $13~30~35.9$ & $+47~07~19$ & 17.98 (0.03) & 17.63 (0.04) & 17.50 (0.01)  \\
E13-51 & $13~30~36.6$ & $+47~09~17$ & 15.42 (0.01) & 14.80 (0.01) & 14.58 (0.01)  \\
E13-52 & $13~30~39.8$ & $+47~11~19$ & 17.96 (0.01) & 17.43 (0.04) & 17.21 (0.04)  \\
E13-53 & $13~30~42.4$ & $+47~10~11$ & 14.71 (0.01) & 14.03 (0.01) & 13.93 (0.01)  \\
E13-54 & $13~30~46.2$ & $+47~10~04$ & 16.58 (0.01) & 15.96 (0.01) & 15.72 (0.03)  \\
\hline
\end{tabular}}

\end{center}
\end{table*}

\subsection{S-corrections}
\label{a_scorr}

From the above and the fact that a fair (usually 5-10) number of reference stars were used we conclude that the calibration of the JC, SDSS and 2MASS photometry, with the possible exception of the $U$ band, should be good to the few-percent level as long as a linear colour-correction is sufficient to transform to the standard systems. This is known to work well for stars, but is not necessarily true for SNe. Photometry for well monitored SNe as 1987A and 1993J shows significant differences between different datasets and telescopes, in particular at late times, as might be expected by the increasingly line-dominated nature of the spectrum. A more elaborate method to transform from the natural system to the standard system is S-corrections \citep{Str02}. Using this method we first transform the reference star magnitudes to the natural system using linear colour-terms and then transform the SN magnitudes to the standard system by replacing the linear colour-terms with S-corrections calculated as the difference of the synthetic magnitudes in the standard and natural systems. Note that this definition differs from the one by \citet{Str02} but is the same as used by \citet{Tau11}. Success of the method depends critically on the accuracy of the filter response functions and a well sampled spectral sequence.

For all telescopes we constructed optical filter response functions from filter and CCD data provided by the observatory or the manufacturer and extinction data for the site. Extinction data for Roque de los Muchachos at La Palma where NOT, LT, TNG and WHT are located where obtained from the Isaac Newton Group of Telescopes (ING). A typical telluric absorption profile, as determined from spectroscopy, was also added to the filter response functions. We have assumed that the optics response functions vary slowly enough not to affect the S-corrections. To test this we constructed optics response functions for a number of telescopes from filter zeropoints, measured from standard star data or provided by the observatories. Except below $\sim$4000 \AA\ they vary slowly as expected and applying them to our data the S-corrections vary at the percent level or less so the assumption seems to be justified. Below $\sim$4000 \AA\ the optics response functions may vary rapidly and because of this and other difficulties in this wavelength region wave have not applied S-corrections to the JC $U$ and SDSS $u$ bands. To test the constructed JC and SDSS filter response functions we compared synthetic colour-term coefficients derived from the STIS NGSL\footnote{http://archive.stsci.edu/prepds/stisngsl/} spectra with observed colour-term coefficients and the agreement was generally good. However, for those filters were the synthetic and observed colour-term coefficients deviated the most, we adjusted the response functions by small wavelength shifts (typically between 50 and 100 \AA). This method is similar to the one applied by S11 and, in all cases, improved the agreement between S-corrected photometry from different telescopes. As we could not measure the 2MASS colour-term coefficients with sufficient precision this method could not be applied to the NIR filters.

\subsection{Systematic errors}

\begin{figure}[tb]
\includegraphics[width=0.48\textwidth,angle=0]{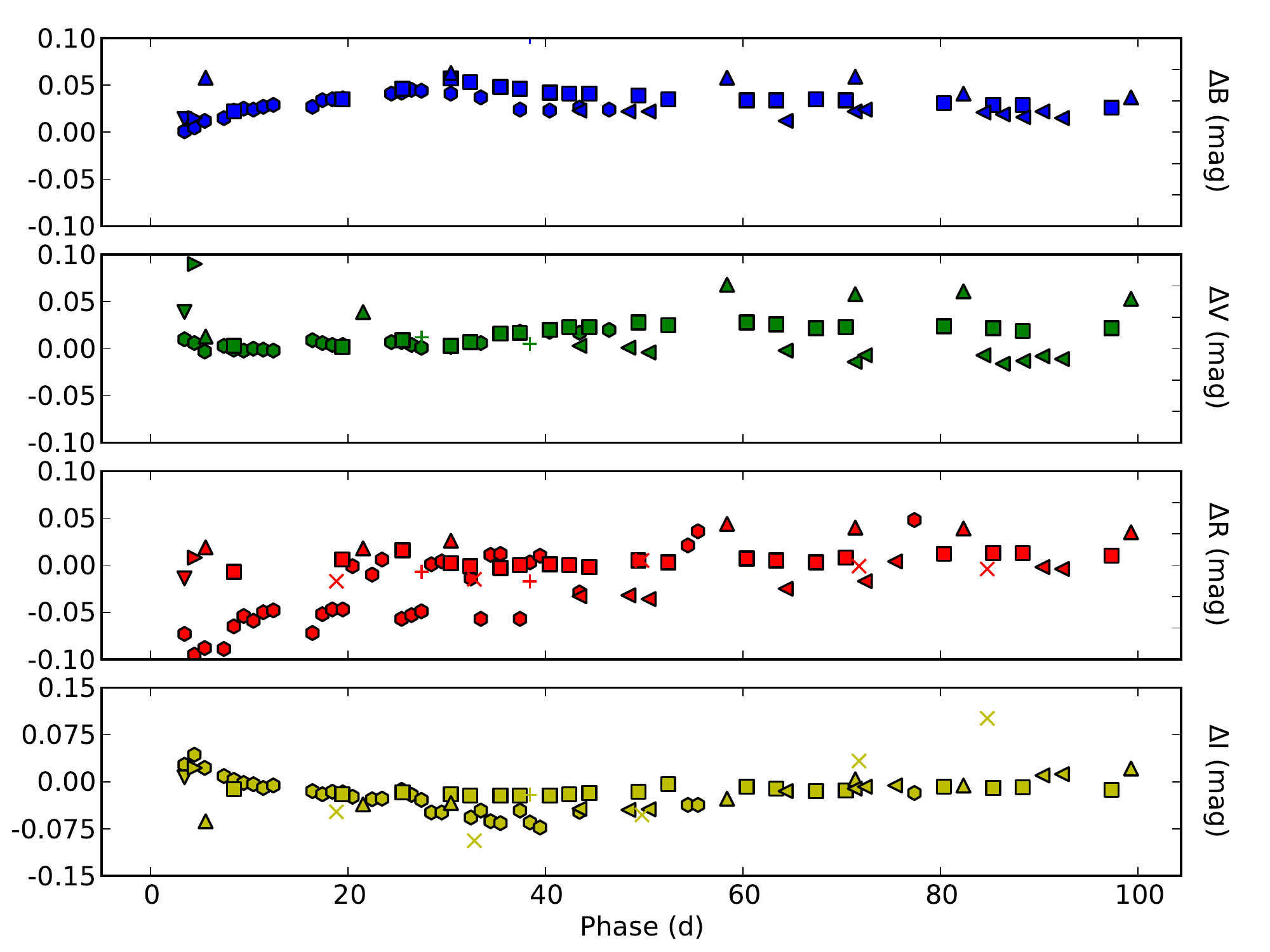}
\caption{Difference between JC colour- and S-corrected photometry for NOT (squares), LT (circles), CA 2.2m (upward triangles), TNG (downward triangles), AS 1.82m (rightward triangles), AS Schmidt (leftward triangles), TJO (pluses) and FTN (crosses).}
\label{f_scorr_jc}
\end{figure}

\begin{figure}[tb]
\includegraphics[width=0.48\textwidth,angle=0]{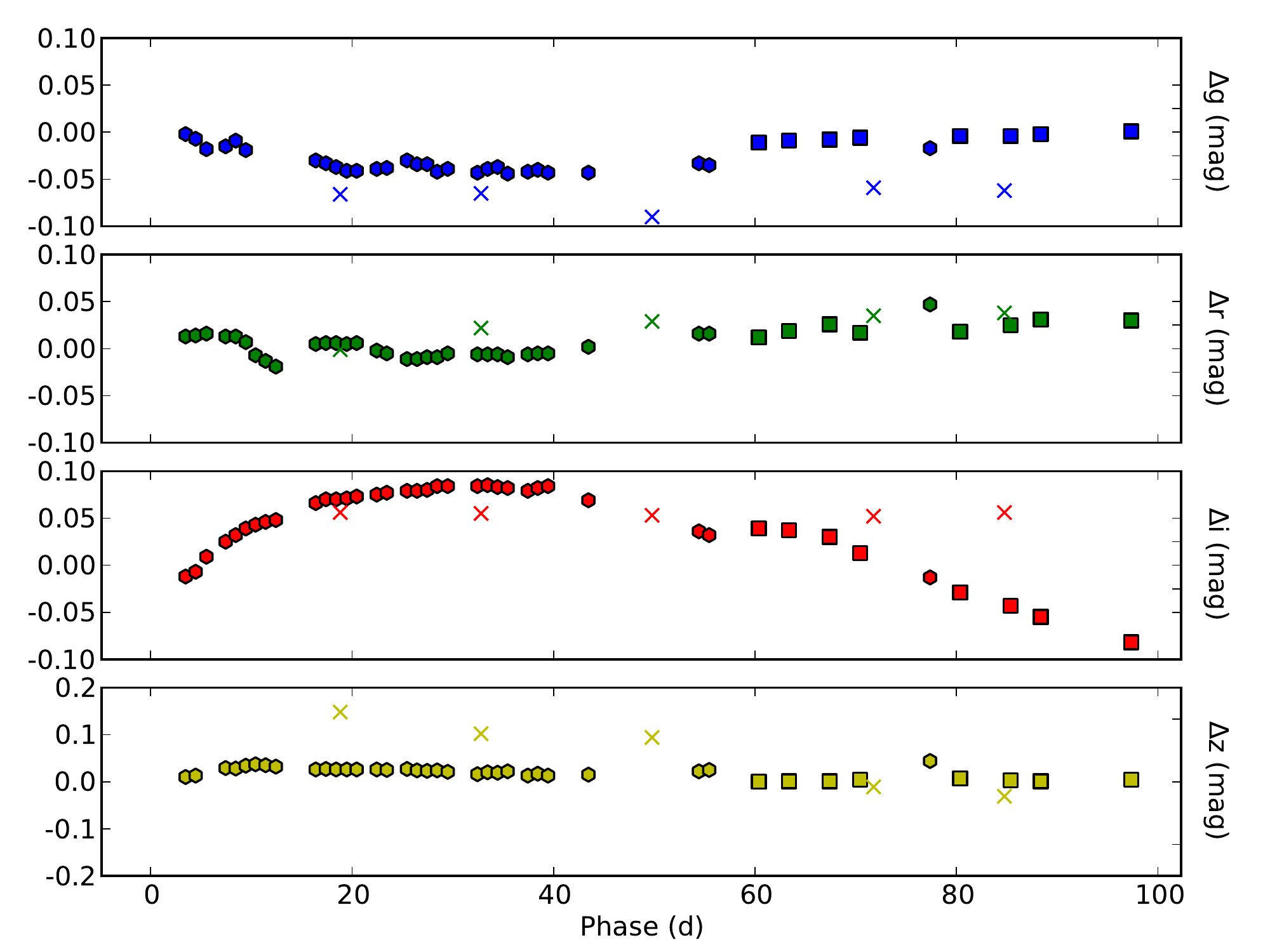}
\caption{Difference between SDSS colour- and S-corrected photometry for NOT (squares), LT (circles) and FTN (crosses).}
\label{f_scorr_sloan}
\end{figure}

\begin{figure}[tb]
\includegraphics[width=0.48\textwidth,angle=0]{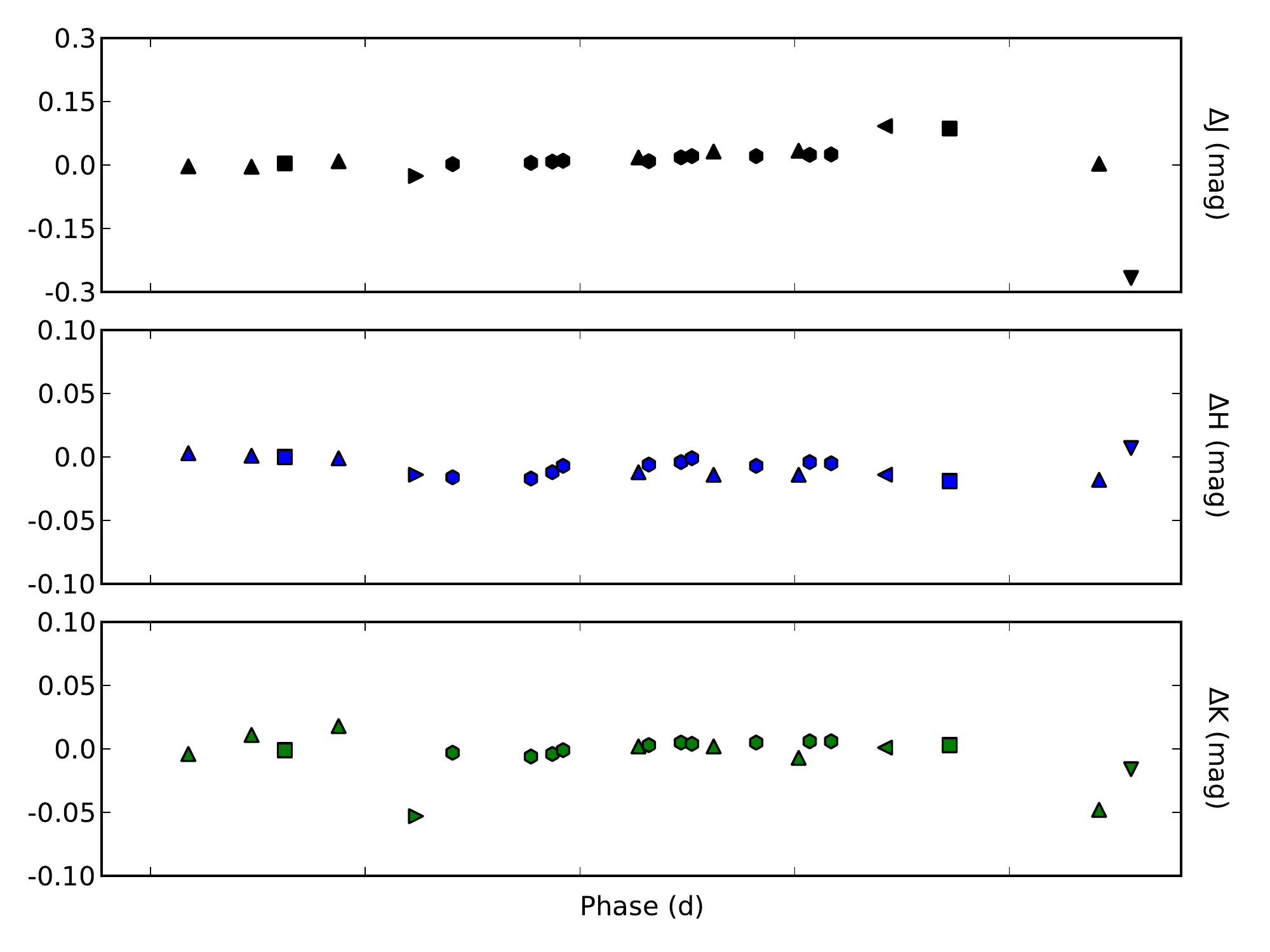}
\caption{Difference between 2MASS colour- and S-corrected photometry for NOT (squares), TCS (circles), TNG (upward triangles), CA 3.5m (downward triangles), WHT (rightward triangles) and LBT (leftward triangles).}
\label{f_scorr_2mass}
\end{figure}

The difference between the S-corrected and colour-corrected JC, SDSS and 2MASS photometry is shown in Figures \ref{f_scorr_jc}-\ref{f_scorr_2mass}. The differences are mostly less than 5 percent but approaches 10 percent in some cases. Most notably, the difference for the late CA 3.5m $J$ band observation is $\sim$30 percent because of the strong He 10830 \AA\ line. So even if the differences are mostly small, S-corrections seems to be needed to achieve 5 percent accuracy in the photometry.

\begin{figure}[tb]
\includegraphics[width=0.48\textwidth,angle=0]{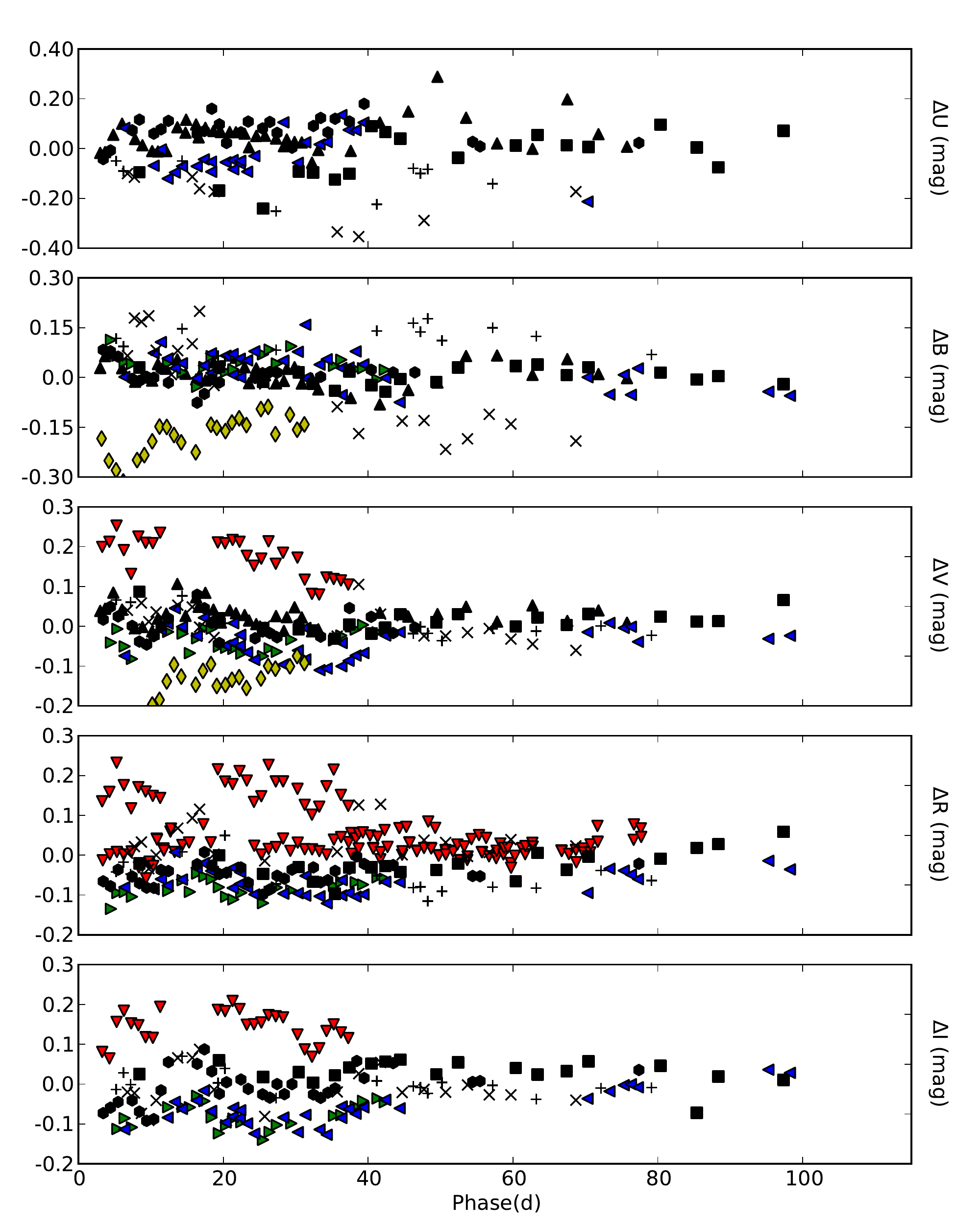}
\caption{Difference between JC photometry for the NOT (black squares), LT (black circles), SWIFT (black upward triangles), \citetalias{Arc11} (red downward triangles), \citetalias{Vin12} (green rightward triangles), \citetalias{Tsv12} (blue leftward triangles), \citetalias{Mar13} (yellow diamonds), \citetalias{Dyk13b} (black crosses) and \citetalias{Sah13} (black pluses) datasets and a cubic spline fit to all these datasets.}
\label{f_jc_ext_diff}
\end{figure}

A further check of the precision in the photometry is provided by comparison to the \citetalias{Arc11}, \citetalias{Vin12}, \citetalias{Tsv12}, \citetalias{Mar13}, \citetalias{Dyk13b}, \citetalias{Sah13} and, in particular, the SWIFT photometry. The SWIFT photometry was transformed to the JC system using S-corrections calculated from the SWIFT filter response functions, which are well known and not affected by the atmosphere. The differences between the JC photometry for all datasets and a spline fit to these observations is shown in Fig.~\ref{f_jc_ext_diff}. The systematic (mean) difference for the NOT and LT datasets (which constitutes the bulk of our data) is less than $\sim$5 percent in the $B$, $V$, $R$ and $I$ bands and less than $\sim$10 percent in the $U$ band. Considerably larger differences, in the 15-30 percent range, are seen in some datasets, like the \citetalias{Arc11} WISE 1m $V$, $R$ and $I$, the \citetalias{Mar13} $B$ and $V$, the \citetalias{Dyk13b} $U$ and the \citetalias{Sah13} $B$. The difference between the NOT and LT and the WISE 1m photometry seems to arise partly from differences in the reference stars magnitudes (Iair Arcavi, private communication). As seen in Fig.~\ref{f_jc_ext_diff} as well as in Fig.~\ref{f_uv_opt_nir_mir}, except for the early (0-40 days) NOT $U$ band observations, the agreement between the NOT, LT and SWIFT $U$, $B$ and $V$ band photometry is excellent and the systematic (mean) differences are only a few percent. The early (0-40 days) NOT $U$ band photometry shows a systematic (mean) difference of $\sim$20 percent as compared to the LT and SWIFT photometry which could be explained by the lack of S-corrections. As the LT filter, in spite of its SDSS nature, is more similar to the JC $U$ band than the NOT filter we favour the LT and SWIFT photometry in this phase. The good agreement between the NOT and LT JC photometry and the S-corrected SWIFT JC photometry as well as the bulk of the published JC photometry, gives confidence in the method used. The limited amount of SDSS and 2MASS observations published so far prevents a similar comparison to be done for the SDSS and 2MASS photometry.

\subsection{Synthetic photometry}

Synthetic photometry (used for e.g. S-corrections) was calculated as energy flux based magnitudes in the form $m_i = \int F_{\mathrm{\lambda}} S_{\mathrm{\lambda},i} d\lambda / \int S_{\mathrm{\lambda},i} d\lambda + Z_i$, where $S_{\mathrm{\lambda},i}$ and $Z_i$ is the energy response function \citepalias{Bes12} and zeropoint of band $i$ respectively. Note that filter response functions are commonly given as photon response functions \citepalias{Bes12} and then have to be multiplied with wavelength to give the energy response functions. JC filter response functions have been adopted from \citetalias{Bes12} and zeropoints calculated using the Vega spectrum and JC magnitudes. SDSS filter response functions have been adopted from \citet{Doi10} and zeropoints calculated using the definition of AB magnitudes \citep{Oke83} and small corrections following the instructions given at the SDSS site. 2MASS filter response functions have been adopted from \citet{Coh03} as provided by the Explanatory Supplement to the 2MASS All Sky Data Release and Extended Mission Products. SWIFT filter response functions have been adopted from \citet{Poo08} as provided by the SWIFT calibration database. To transform the photon count based SWIFT system into an energy flux based system we have multiplied the response functions with wavelength and re-normalized. The zeropoints were then calculated using the Vega spectrum and SWIFT magnitudes \citep{Poo08}. Spitzer filter response functions have been adopted from \citet{Hor08} as provided at the Spitzer web site and zeropoints calculated using the Vega spectrum and Spitzer magnitudes.

\subsection{SWIFT UV read leak}

\begin{figure}[tb]
\includegraphics[width=0.48\textwidth,angle=0]{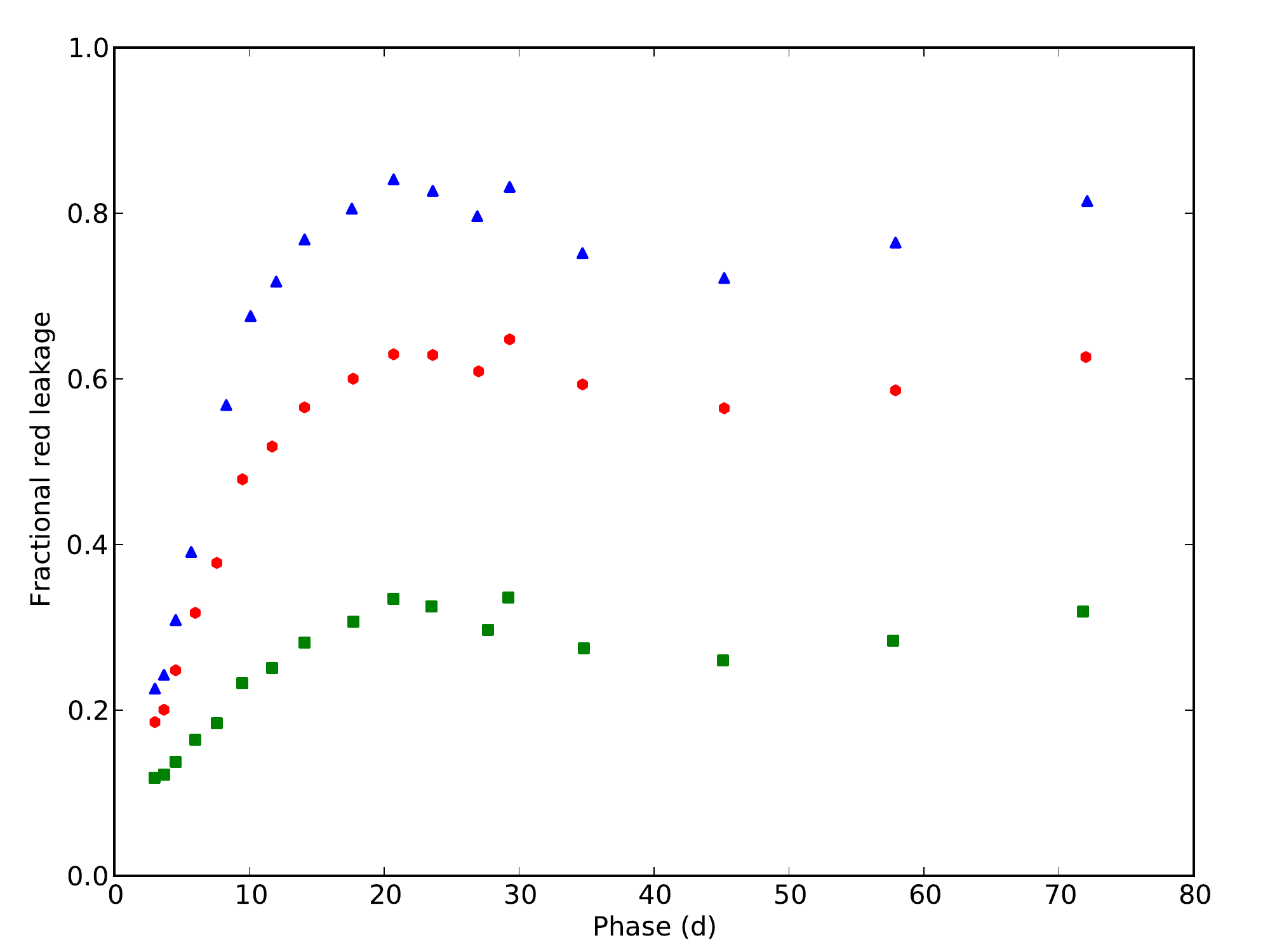}
\caption{Fractional red leakage in the SWIFT $UVW1$ (red circles), $UVM2$ (green squares) and $UVW2$ (blue triangles) filters.}
\label{f_swift_red_leak}
\end{figure}

Finally we note that the response functions of the SWIFT $UVW1$ and $UVW2$ filters have a quite strong red tail. If, as is often the case for SNe, there is a strong blueward slope of the spectrum in the UV region this will result in a red leakage that might even dominate the flux in these filters. In Fig.~\ref{f_swift_red_leak} we quantify this by showing the fractional red leakage defined as the fractional flux more than half the equivalent width redwards of the mean energy wavelengths of the filters. The spectrum was interpolated from the photometry as explained in Sect.~\ref{s_bol_lightcurve} excluding the $UVW1$ and $UVW2$ filters. After $\sim$20 days the leakage is $\sim$80 and $\sim$60 percent in the $UVW1$ and $UVW2$ filters respectively. Given this the $UVW1$ and $UVW2$ lightcurves do not reflect the evolution of the spectrum at their mean energy wavelengths and we will therefore exclude these from the analysis and when calculating the bolometric lightcurve in Sect.~\ref{s_bol_lightcurve}.

\section{Progenitor observations}
\label{a_prog_obs}

\begin{figure}[tb]
\includegraphics[width=0.48\textwidth,angle=0]{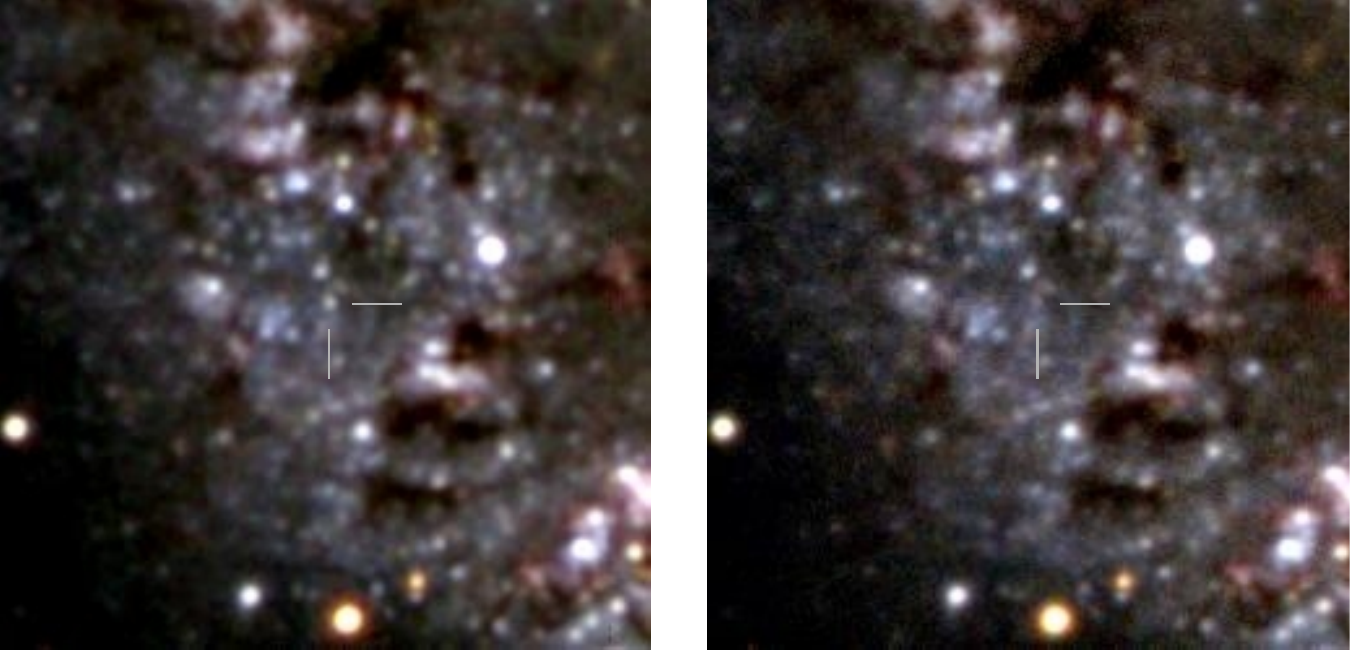}
\caption{Colour composite of the pre- (left panel) and post- (right panel) explosion NOT imaging. The RGB values have been scaled to match the number of photons.}
\label{f_prog_pre_post}
\end{figure}

We have obtained high quality pre- and post-explosion $B$, $V$ and $r$ band images of the SN site with the NOT. The pre-explosion images were obtained on May 26 2008 ($B$) and May 29 2011 ($V$ and $r$), the latter just 2 days before the explosion. Two sets of post-explosion images were obtained, the first on Jan 20 2013 ($V$ and $r$) and Mar 19 2013 ($B$), 601 and 659 days post explosion respectively, and the second on Apr 14 2013 ($V$), May 15 2013 ($r$) and June 1 2013 ($B$), 685, 715 and 732 days post explosion respectively. In Fig. \ref{f_prog_pre_post} we show a colour composite of the pre-explosion and the second set of post-explosion $B$, $V$ and $r$ band images where the RGB values have been scaled to match the number of photons. The photometry presented below have been calibrated to the natural Vega ($BV$) and AB ($r$) systems of the NOT using the reference star magnitudes and colour constants presented in this paper (Tables \ref{t_refstar_jc}, \ref{t_refstar_sloan}, \ref{t_cconst_jc} and \ref{t_cconst_sloan}). 

We have used the {\sc hotpants} package to perform subtractions of the pre- and post-explosion images and aperture photometry to measure the magnitudes of the residuals to $B$=23.00$\pm{0.10}$, $V$=22.73$\pm{0.07}$ and $r$=22.22$\pm{0.05}$ mag for the first set of post-explosion observations and $B$=22.73$\pm{0.06}$, $V$=22.23$\pm{0.05}$ and $r$=21.95$\pm{0.04}$ mag for the second set of post-explosion observations. The positions of the residuals in all bands are within 0.15 arcsec from the position of the SN. The two fainter nearby stars, seen in pre-explosion HST images, that could possibly contaminate the result are $\sim$0.5 arcsec away from the SN so their contribution (due to variability) to the residuals is likely to be small. Using PSF photometry where we have iteratively fitted the PSF subtracted background we measure the magnitudes of the yellow supergiant in the pre-explosion images to $B$=22.41$\pm{0.12}$, $V$=21.89$\pm{0.04}$ and $r$=21.67$\pm{0.03}$ mag. The residuals for the second set of post-explosion observations then corresponds to a reduction of the flux with 74$\pm{9}$, 73$\pm{4}$ and 77$\pm{4}$ percent in the $B$, $V$ and $r$ bands respectively. The remaining flux, at least partly emitted by the SN, corresponds to $B$=23.35$\pm{0.32}$, $V$=22.56$\pm{0.10}$ and $r$=22.67$\pm{0.11}$ mag for the first set of post-explosion observations and $B$=23.89$\pm{0.50}$, $V$=23.32$\pm{0.20}$ and $r$=23.28$\pm{0.19}$ mag for the second set of post-explosion observations.

\bibliographystyle{aa}
\bibliography{sn2011dh}

\label{lastpage}

\end{document}